\documentclass{phd}	

\usepackage{amssymb, amsmath, amsfonts, latexsym, graphicx, tabularx, tabulary, appendix} 

\usepackage{ccaption, rotfloat}
\usepackage{listings, subfig, multicol}
\usepackage{rotfloat} 

\usepackage[sort&compress]{natbib} 

\usepackage{subfig}
\usepackage{float}

\theoremstyle{plain}

\theoremstyle{definition}

\numberwithin{equation}{section} \numberwithin{theorem}{chapter}
\numberwithin{corollary}{chapter}
\numberwithin{definition}{chapter} \numberwithin{remark}{chapter}
\numberwithin{lemma}{chapter} \numberwithin{proposition}{chapter}
\numberwithin{example}{chapter}

\def\sin{{\rm sin}}
\def\cos{{\rm cos}}

\def\dsum{\displaystyle\sum}

\def\mod#1{{\rm \,\, mod}(#1)}

\def\IZ{\relax\ifmmode\mathchoice
 {\hbox{\cmss Z\kern-.4em Z}}{\hbox{\cmss Z\kern-.4em Z}}
 {\lower.9pt\hbox{\cmsss Z\kern-.4em Z}}
 {\lower1.2pt\hbox{\cmsss Z\kern-.4em Z}}\else{\cmss Z\kern-.4em Z}\fi}

\def\Io{\relax\ifmmode\mathchoice
 {\hbox{\cmss 1\kern-.4em 1}}{\hbox{\cmss 1\kern-.4em 1}}
 {\lower.9pt\hbox{\cmsss 1\kern-.4em 1}}
{\lower1.2pt\hbox{\cmsss 1\kern-.4em 1}}\else{\cmss 1\kern-.4em 1}\fi}

\def\half{{\textstyle{1\over 2}}}
\def\third{{\textstyle {1\over3}}}
\def\fourth{{\textstyle {1\over4}}}

\def\twothird{{\textstyle {2\over3}}}

\def\I{{\bf I}}
\def\bone{{\mathbf 1}}

\def\bo{{\mathbf 0}}
\def\mF{{\mathbf F}}
\def\mQ{{\mathbf Q}}

\def\bS{{\mathbf S}}

\def\mk{{\mathbf k}}
\def\mV{{\mathbf V}}
\def\bV{{\mathbf V}}

\def\balpha{{\mathbf \alpha}}

\def\malpha{{\mathbf \alpha}}
\def\mbeta{{\mathbf \beta}}

\def\bone{{\mathbf 1}}

\def\p4{\Phi_4}
\def\pp4{\Phi^{'}_4}
\def\pb4{\bar{\Phi}_4}
\def\ppb4{\bar{\Phi}^{'}_4}
\def\p#1{{\Phi_{#1}}}

\def\pp#1{{\Phi^{'}_{#1}}}
\def\pb#1{{{\overline{\Phi}}_{#1}}}

\def\ppb#1{{{\overline{\Phi}}^{'}_{#1}}}

\def\FD2pv{FD2$^{'}$V }
\def\FD2p{FD2$^{'}$ }

\def\beq{\begin{equation}}
\def\eeq{\end{equation}}
\def\beqn{\begin{eqnarray}}
\def\eeqn{\end{eqnarray}}

\def\half{{\textstyle{1\over 2}}}
\def\third{{\textstyle {1\over3}}}
\def\fourth{{\textstyle {1\over4}}}

\def\twothird{{\textstyle {2\over3}}}

\def\mod#1{{\rm \,\, (mod\, #1)}}

\def\I{{\bf I}}
\def\mbf{\mathbf}
\def\bone{{\mathbf 1}}

\def\bo{{\mathbf 0}}

\def\mF{{\mathbf F}}
\def\mQ{{\mathbf Q}}

\def\bS{{\mathbf S}}
\def\bV{{\mathbf V}}

\def\mk{{\mathbf k}}

\def\balpha{{\mathbf \alpha}}

\def\inbar{\,\vrule height1.5ex width.4pt depth0pt}
\def\IT{\relax\hbox{$\inbar\kern-.3em{\rm T}$}}
\def\IS{\relax\hbox{$\inbar\kern-.3em{\rm S}$}}
\def\IC{\relax\hbox{$\inbar\kern-.3em{\rm C}$}}
\def\IQ{\relax\hbox{$\inbar\kern-.3em{\rm Q}$}}
\def\IR{\relax{\rm I\kern-.18em R}}
 \font\cmss=cmss10 \font\cmsss=cmss10 at 7pt

\def\IZ{\relax\ifmmode\mathchoice
 {\hbox{\cmss Z\kern-.4em Z}}{\hbox{\cmss Z\kern-.4em Z}}
 {\lower.9pt\hbox{\cmsss Z\kern-.4em Z}}
 {\lower1.2pt\hbox{\cmsss Z\kern-.4em Z}}\else{\cmss Z\kern-.4em Z}\fi}

\def\Io{\relax\ifmmode\mathchoice
 {\hbox{\cmss 1\kern-.4em 1}}{\hbox{\cmss 1\kern-.4em 1}}
 {\lower.9pt\hbox{\cmsss 1\kern-.4em 1}}
{\lower1.2pt\hbox{\cmsss 1\kern-.4em 1}}\else{\cmss 1\kern-.4em 1}\fi}

\title{Investigation into Compactified Dimensions: Casimir Energies and Phenomenological Aspects}
\author{Richard K. Obousy}
\degrees{M.Phys.}
\mentor{Gerald B. Cleaver, Ph.D.}
\reader{Anzhong Wang, Ph.D.}
\readerThree{Dwight P. Russell, Ph.D.}
\readerFour{Gregory A. Benesh, Ph.D.}
\readerFive{Qin Sheng, Ph.D.}
\confDate{December 2008}

\makeCopyrightPage

\graduateDean{J. Larry Lyon, Ph.D.}
\deptChair{Gregory A. Benesh, Ph.D., Chairperson}

\abstract{The primary focus of this dissertation is the study of the Casimir effect and the possibility that this phenomenon may serve as a mechanism to mediate higher dimensional stability, and also as a possible mechanism for creating a small but non-zero vacuum energy density. In chapter one we review the nature of the quantum vacuum and discuss the different contributions to the vacuum energy density arising from different sectors of the standard model. Next, in chapter two, we discuss cosmology and the introduction of the cosmological constant into Einstein's field equations. In chapter three we explore the Casimir effect and study a number of mathematical techniques used to obtain a finite physical result for the Casimir energy. We also review the experiments that have verified the Casimir force. In chapter four we discuss the introduction of extra dimensions into physics. We begin by reviewing Kaluza Klein theory, and then discuss three popular higher dimensional models: bosonic string theory, large extra dimensions and warped extra dimensions. Chapter five is devoted to an original derivation of the Casimir energy we derived for the scenario of a higher dimensional vector field coupled to a scalar field in the fifth dimension. In chapter six we explore a range of vacuum scenarios and discuss research we have performed regarding moduli stability. Chapter seven explores a novel approach to spacecraft propulsion\pagestyle{empty} we have proposed based on the idea of manipulating the extra dimensions of string/M theory. Finally, in chapter 8, we discuss some issues in heterotic string phenomenology derived from the free fermionic approach.}

\acknowledgements{First I would like to thank my beautiful wife and best friend, Annie, for her constant love, support and patience during these long and challenging years of graduate school. Annie read this dissertation in its entirety and advised me on many aspects of readability, for which I am extremely grateful. Next, I would like to thank my supervisor, Dr. Gerald Cleaver, for his guidance, supervision, and support, and for giving me the academic freedom and encouragement to explore ideas beyond the usual sphere of academic research. I have benefitted deeply from his instruction, and have learned many of the subtle aspects of field theory from his seemingly endless knowledge and profound grasp of the subject.

I would also like to thank my mother, Elizabeth Obousy, for her belief in me and for her dedication and commitment to helping me achieve my ambitions. Without her love and support, I know I would not be here today. It is also a pleasure to thank my parents-in-law, Ross and Wendy Whitney, for welcoming me into their delightful family, and for their interest in my work and their eagerness to see me succeed.

I have been privileged to learn physics from the many talented professors of the Baylor physics department, but I would especially like to thank Dr. Gregory Benesh and Dr. Anzhong Wang for their invaluable support over the years. I would also like to thank Dr. Tibra Ali for helping me to understand a number of complicated aspects regarding higher dimensions, and for his insight into physics in general.

I am especially grateful to all the astonishing and brilliant graduate students I have gotten to know over the years. First, I'd like to thank my closest friend,  Andreas Tziolas, with whom I have shared countless insightful and compelling conversations and whose company I have enjoyed during the otherwise lonely nocturnal hours that I favor. I'd also like to thank Zoi Maroudas, his lovely wife, for her friendship and kindness. And finally, my office companions over the years: Ben Dundee, Mick Whitaker, Martin Frank, and Tim Renner, who have been forced to endure my quirky humor on a daily basis. In addition, Sammy Joseph, Victor Guerrero, Kristin Pechan, Qianyu Ma, John Perkins, Michael Devin, Qiang Wu, David Katz, Matt Robinson and James Creel all deserve a special mention.}

\dedication{\begin{center} For Elizabeth, Nadia and Annie\end{center}}

\begin{document}
\pagenumbering{arabic}


\chapter{Introduction}
\label{chap:intro}

\section{The Vacuum}
\label{sec:tv}
A central theme in this disseration is the notion of the quantum vacuum. To a particle physicist, the term `vacuum' means the ground state of a theory. In general, this ground state must obey Lorentz invariance, at least with regards to 3 spatial dimensions, meaning that the vacuum must look identical to all observers. At all energies probed by experiments to date, the universe is accurately described as a set of quantum fields. If we take the Fourier transform of a free quantum field,\footnote{By `free' we mean that the field does not interact with other fields.} each mode of a fixed wavelength behaves like a simple harmonic oscillator. A quantum mechanical property of a simple harmonic oscillator is that the ground state exhibits zero-point fluctuations as a consequence of the Heisenberg Uncertainty Principle, with energy $E=\frac{1}{2}\hbar\omega$. 

These fluctuations give rise to a number of phenomena; however, two are particularly striking. First, the Casimir Effect \cite{cas}, which will be examined in detail in this dissertation is arguably the most salient manifestation of the quantum vacuum. In its most basic form it is realized through the interaction of a pair of neutral parallel conducting plates. The presence of the plates modifies the quantum vacuum, and this modifcation causes the plates to be pulled toward each other. Second is the prediction of a vacuum energy density, which is an intrinsic feature of space itself. Many attempts have been made to relate this vacuum energy to the cosmological constant $\Lambda$, which is a common feature in modern cosmology \cite{Weinberg}; however, calculations are typically plagued either by divergences or by ridiculously high predictions which are far removed from observation. This chapter will first provide a brief historical review of the vacuum and then discuss in detail some of the attempts to explain the vacuum in the language of Quantum Field Theory (QFT).

\subsection{Early Studies of The Vacuum}
\label{sec:esotv}
One fundamental feature of QFT is the notion that \textit{empty} space is not really empty. Emptiness has been replaced by the concept of the vacuum (i.e. the ground state). It is this ground state which is responsible for a ubiquitous energy density that is ultimately believed to act as a contribution to the cosmological constant $\Lambda$, that appears in Einstein's field equations from 1917 onward \cite{E},
\begin{equation}
R_{\mu\nu}-\frac{1}{2}Rg_{\mu\nu}=8\pi G T_{\mu\nu}+\Lambda g_{\mu\nu}
\label{eq1}
\end{equation}
where $g_{\mu\nu}$ is the spacetime metric, $R_{\mu\nu}$ is the Ricci curvature tensor, $R$ is the Ricci scalar or scalar curvature and $T_{\mu\nu}$ is the energy momentum tensor. We will follow the convention of using natural units throughout this dissertation and set $c=\hbar=1$. The left-hand side of the equation characterizes the geometry of spacetime and the right-hand side encodes the matter and energy sources. $T_{\mu\nu}$ acts as a source for the gravitational field; its role is analogous to the electro-magnetic currents $J_\mu$, which acts as a source for the electromagnetic field $A_\mu$ in Maxwell's equations.

In 1916 Nernst, who was originally inspired by the new ideas of quantum theory and Plancks law for the radiation
from a blackbody \cite{Ner}, put forward the proposition that the vacuum of spacetime is not empty but is, in fact, a medium filled with radiation containing a large amount of energy. One feature of this model was that the energy density of the vacuum was infinite, and even when a modest cutoff was proposed, the total energy content was still large. Nernst's ideas about the vacuum were never used for any cosmological models, as his interests were in chemistry and in forming a model of the water molecule. \newpage

One problem that frequently arises in the calculations of vacuum ground state energies is the huge energies that are found. Pauli was concerned with the \textit{gravitational} effects of the zero-point energy \cite{ET}.  Pauli's calculation in the mid-late 1920s demonstrated that if the gravitational effect of the zero-point energies was taken into account, the radius of the universe would be smaller than the distance from the Earth to the Moon. Pauli's calculation invloved applying a cut-off energy at the classical electron radius, which was considered to be a natural cut-off at the time. Pauli's concern was readdressed by Straumann \cite{St} in 1999, who found that, indeed, the radius of the universe would be $\approx 31 \ km$. Setting $\hbar =c=1$ the calculation reads 
\begin{eqnarray}
\left< \rho_{\rm eff} \right>  = \frac{8\pi}{(2\pi)^3}\int_{0}^{\omega_{\rm max}} \frac{\omega}{2}\omega^2 d\omega  
  =  \frac{1}{8\pi^2}\omega_{\rm max}^4 . 
\label{eq2}
\end{eqnarray}
Inserting the appropriate cutoff,
\begin{equation}
\omega_{\rm max}=\frac{2\pi}{\lambda_{\rm max}}=\frac{2\pi m_{\rm e}}{\alpha},
\label{eq3}
\end{equation}
and plugging into
\begin{equation}
8\pi G \rho = \frac{1}{a^2}=\Lambda
\label{eq4}
\end{equation}
where $a$ is the radius of curvature obtained from solving Einstein's equation for a static dust filled universe. Thus, one obtains
\begin{equation}
a=\frac{\alpha^2}{(2\pi)^{2/3}}\frac{M_{\rm pl}}{m_{\rm e}^2}\approx 31\ {\rm km}.
\label{eq5}
\end{equation} 
One way to reconcile this inconsistency with the known size of the universe, as noted in Pauli's \textit{Handbuch der Physik}, is to begin from the ansatz that the zero-point energy does not interact with the gravitational field. Indeed, the speculations of Dirac regarding the huge vacuum energy and also the final version of quantum electrodynamics (QED): constructed by Schwinger, Feynman and others never prompted any interest in the \textit{gravitational} consequences of these theories. This is not surprising when one considers the theoretical landscape of QED; it was plagued with divergences in higher order calculations that preoccupied the community.

\subsection{Modern Studies of the Vacuum}
\label{sec:msotv}

A more solid foundation for speculations about the energy density of the vacuum became available with the development of QFT, in which all the fields in nature are treated as a collection of quantized harmonic oscillators. The various amplitudes and frequencies of oscillation represent the different boson and fermion species that are observed in nature. The vacuum contains all the quantum properties a particle may acquire: energy, spin and polarization. These quantities, on average, cancel each other out, with the exception of the vacuum expectation value of the energy $\left<E_{\rm vac} \right> $. A consequence of the Heisenberg Uncertainty Principle is that no field oscillator can ever be completely at rest \cite{Carroll2004}; there will always be some residual `zero-point energy.' Plainly, one can see this in $E=(n+\frac{1}{2})\hbar \omega $. For $n=0$ we are left with $ E=\frac{1}{2}\hbar\omega $.

The first published discussion relating the cosmological constant to quantum vacuum energy was contained in a 1967 paper by Zel'dovich \cite{Z}, who assumed that the zero-point energies and their higher order electromagnetic corrections cancel. After cancellation, the only remaining components are the higher order corrections where gravity is involved, and this remaining vacuum energy is the cosmological constant. In a more detailed article, Zel'dovich derived a QED-inspired \cite{Z1} vacuum energy; however, his result exceeded observational bounds by 46 orders of magnitude, leading to the statement that ``...such an estimate has nothing in common with reality.''

The `serious worry' about the vacuum energy dates from the early and mid-1970s as noted by Weinberg \cite{Weinberg}. Around this time, it was realized that the spontaneous symmetry-breaking mechanism invoked in the electroweak theory might have cosmological consequences. In the next three sections we review modern derivations of the vacuum of QFT.

\section{The Vacuum of QFT}
\label{sec:tvoqft}

Typically, one sees in the literature discussions of the the \textit{electromagnetic} vacuum. However, the ground state of the vacuum contains field contributions from numerous sectors of the standard model (SM) of particle physics, of which the electromagnetic vacuum is but one. These components include, but are not limited to, the QED vacuum, the electoweak (Higgs) vacuum and the Quantum Chromo Dynamical (QCD) vacuum. In the next section we will review the contributions from each of these sectors, and we will understand why we typically discuss the electromagnetic contribution.

\subsection{QED Vacuum}
\label{sec:qedv}

The electromagnetic force can be accurately described using the theory of Quantum Electro\-Dynamics (QED), which  describes how light and matter  interact. The quantization procedure for the classical fields of electromagnetism, $\vec{E}({\bf{x}},t)$ and $\vec{B}({\bf{x}},t),$\footnote{Throughout this dissertation we will follow the convention that a bold font character (e.g. ${\mathbf k}$) represents a 3-vector, and an arrowed character 
(e.g. $\vec{E}$) represents a 4-vector} involves replacing these fields by quantum operators that are defined at all locations in spacetime. To construct quantum theory with the correct classical limit, the Hamiltonian density of quantum theory is taken to be the same function as in classical theory, $\mathcal{H}=\frac{1}{2}(\vec{E}^2+\vec{B}^2)$. The total zero-point energy of QED is 
\begin{equation}
E=\left<0|H|0\right>=\frac{1}{2}\left<0|\int(\tilde{E}^2+\tilde{B}^2)|0\right>,
\label{eq6}
\end{equation} 
and the ground state energy density is given by \cite{Mandl, Peskin, Ryder}
\begin{equation}
\rho_{vac}=\frac{1}{V}\sum_{\mathbf{k}}\frac{1}{2}\hbar \omega_{\mathbf{k}},
\label{eq7}
\end{equation} 
where the wave vector ${\mathbf{k}}$ refers to the normal modes of the electromagnetic field.
Using the well known relation
\begin{equation}
\frac{1}{V}\sum_k \rightarrow\frac{1}{(2\pi)^3}\int d^3{\mathbf{k}}
\label{eq8}
\end{equation}
we can express eq.\ (\ref{eq7}) as an integral equation:
\begin{eqnarray}
\rho_{\rm vac} &=& \frac{1}{16\pi^2}\int_0^\Lambda \omega^3 d\omega \label{eq8a} \nonumber \\
           &=& \frac{\Lambda^4}{64\pi^2}
\label{eq9}
\end{eqnarray} 
where $\Lambda$ represents some cutoff that we impose to ensure convergence of the integral. Using eq.\ (\ref{eq9}) formula, we can now estimate the value of the QED zero-point energy. It is generally accepted that the SM can be believed up to $100 \ {\rm GeV}$, which is set by the electroweak scale. This is the energy in which the electromagnetic interaction is unified with the weak forces. Thus a rough estimate of the zero-point energy is
\begin{equation}
\rho_{\rm vac}^{\rm EW}\approx(100 \ GeV)^4 \approx 10^8 {\rm GeV}^4
\label{eq9a}
\end{equation}
Typically, observational estimates that come from cosmology give 
\begin{equation}
\rho_{\rm vac}^{\rm obs}\approx 10^{-47}({\rm GeV})^4;
\label{eq10}
\end{equation}
so estimates of the QED vacuum exceed the observational bound by $55$ orders of magnitude, even when using the most conservative cut-offs. It has been customary to assume that QFT is valid up to the Planck energy scale of $10^{19}GeV$, which when inserted as the cutoff yields 
\begin{equation}
\rho_{\rm vac}^{\rm Planck}\approx 10^{76}({\rm GeV})^4.
\label{eq11}
\end{equation}
This value exceeds the observational bound by about 120 orders of magnitude. Thus, modification of the cut-off to the Planck scale does nothing to reconcile the discrepancy between observation and theory.

\subsection{The Electroweak Theory and the Higgs Vacuum}
\label{sec:tetathv}

Electroweak theory is the framework for describing weak interactions such as $\beta$ decay. If electroweak theory is to describe particles with mass, the Higgs field must be introduced. This field gives mass to particles via the process of `spontaneous symmetry breaking,' which reflects the fact that the Lagrangian of the theory contains a symmetry that is not shared by the vacuum state. All massive particles couple to the Higgs field by a `Yukawa coupling' and their masses are proportional to the vacuum expectation value (VEV) of the Higgs field. The vacuum resulting from the Higgs field is expected to take the form
\begin{equation}
V(\phi)=V_0-\mu^2\phi^2+g\phi^4
\label{eq12}
\end{equation}
where g is the self coupling of the Higgs, and $\mu$ is related to the VEV of the Higgs field. The VEV of the Higgs field is found experimentally from the Fermi coupling constant, which itself is found from the muon lifetime. Taking the Higgs coupling constant $g$ to be the electromagnetic fine structure constant squared, we obtain the Higgs vacuum energy density (after minimizing) to be
\begin{equation}
\rho_{\rm vac}^{\rm Higgs}=\frac{\mu}{4g} \approx 10^5({\rm GeV})^4
\label{eq13}
\end{equation}
which is 52 orders of magnitude larger than $\rho_{\rm vac}^{\rm obs}$ when the absolute value is taken.
This estimate of the vacuum energy density is model-dependent, but clearly, a high degree of fine-tuning would be required to reduce $\rho_{\rm vac}^{\rm Higgs}$ to that set by observational bounds.

\subsection{The QCD Vacuum}
\label{sec:tqcdv}

QCD is a theory that describes the forces which bind the constituents of an atomic nucleus. It explains the interactions of quarks and gluons. QCD is both non-perturbative and non-linear at low energies, and its ground state is not well described in terms of harmonic oscillators. Thus, the vacuum structure of QCD is not a currently settled issue. With this said, it is given that the non-perturbative sector of QCD forms gluon and quark `condensates' to which estimates of the vacuum energy density can be given \cite{Sh}. One typically finds 
\begin{equation}
\rho_{\rm vac}^{\rm QCD}\approx 10^{-3}({\rm GeV})^4.
\label{eq14}
\end{equation}
A comparison of $\rho_{\rm vac}^{\rm QCD}$ with $\rho_{\rm vac}^{\rm obs}$ indicates a discrepancy of 40 orders of magnitude.
The following table summarizes the vacuums of the electromagnetic, electroweak and quantum chromodynamic vacuums and illustrates their order of magnitude discrepency when compared to $\rho_{\rm vac}^{\rm obs}=10^{-47}({\rm GeV})^4$
\vspace{0.20in}
\vspace{0.20in}
\begin{table}[ht]
\captionstyle{\centering}
\caption{\textit{Quantum Vacuums and their contribution to the vacuum energy density together with their discrepancy from the experimentally measured value of $\Lambda$.}}
\vspace{0.20in}
\centering 
\begin{tabular}{c c c} 
\hline 
Vacuum & $({\rm GeV})^4$ & Magnitude Discrepancy \\ [0.5ex] 
\hline 
$\rho_{\rm EM}$     &   $10^8$        &       $55$       \\
$\rho_{\rm EW}$     &   $10^5$        &       $52$       \\
$\rho_{\rm QCD}$     &   $10^{-3}$        &       $40$       \\[1ex]
\hline 
\end{tabular}
\label{table:qvac} 
\end{table}
\vspace{0.20in}
\vspace{0.20in}
It is clear from Table 1.1 that the electromagnetic vacuum is several orders of magnitude higher than the next highest vacuum energy density. This is why the literature commonly discussed the \textit{electromagnetic} vacuum; this is because it represents the dominant contribution to the vacuum energy density.

\newpage


\chapter{Cosmology}
\label{chap:c}

One of the most profound results of modern cosmology is the evidence that 70\% of the energy density of the universe is in the form of an exotic energy with negative pressure that is currently driving an era of accelerated cosmological expansion \cite{wv, d, Carroll2004,Carroll1992,Rugh2002,weinberg1996,p}. This was first demonstrated by observations of Type Ia supernova made in 1998 by the Supernova Cosmology Project and the High-z Supernova Seach Team. The results have been well corroborated since then and have drawn huge interest as physicists attempt to explain the nature of this `dark' energy. For certain topologies, it has been shown that the Casimir energy leads to a non-singular de Sitter universe with accelerated expansion \cite{mt}, and possible links between Casimir energy and dark energy have been made in the literature \cite{gl}, \cite{milt}. The aim of this dissertation is to model the phenomenon of higher dimensional stability and the observed vacuum energy density as a consequence of Casimir energy. For this reason a section on cosmology is necessary.

In this chapter we review why Einstein was initially motivated to introduce $\Lambda$ into GR in the first place. We will appreciate the implications of $\Lambda$ in the context of negative pressure and accelerated expansion. We will also attempt to understand the fundamental nature of $\Lambda$ by attempting to explain its existence using the mathematics of QFT. We will also discover why the study of supernovae is of such importance when trying to obtain accurate measurements for $\Lambda$.

\section{Post Einsteinian Cosmology and the History of $\Lambda$}
\label{sec:pecathol}

The physical interpretation of GR \cite{Ell} is that spacetime is a Riemmanian manifold where $x^\mu$ are the coordinates on the manifold, and that there exists a metric $g_{\mu\nu}(x)$ which defines the line element,
\begin{equation}
ds^2=g_{\mu\nu}(x)dx^\mu dx^\nu.
\label{eq15a}
\end{equation}
Flat space corresponds to 
\begin{equation}
g_{\mu\nu}(x)=\eta_{\mu\nu}=diag(1,-1,-1,-1)
\label{eq15b}
\end{equation}
In general, the energy momentum tensor $T_{\mu\nu}$ is given by \cite{Peebles}
\begin{equation}
T_{\mu\nu}(x)=\frac{2}{\sqrt{-g}}\left( \frac{\partial(\sqrt{-g}\mathcal{L})}{\partial g^{\mu\nu}}-\frac{\partial}{\partial x^k}\frac{\partial(\sqrt{-g}\mathcal{L})}{\partial g^{\mu\nu}_{,k}}\right) \ ,
\label{eq15c}
\end{equation}
where $g$ is the determinant of the metric. One important feature of GR is that $g_{\mu\nu}(x)$ is considered to be a dynamical field from which one obtains Einstein's field equations, which are found by varying the Einstein Hilbert action,
\begin{equation}
S=-\frac{1}{2\kappa^2}\int d^4x \sqrt{g}R+\int d^4 x\mathcal{L}\sqrt{g},
\label{eq15d}
\end{equation}
where $\kappa^2 = 8\pi G/ c^4 $.
Upon completion of his General Theory of Relativity, Einstein applied his theory to the entire universe. He firmly believed in Machs principle, and the only way to satisy this was to assume that space is globally closed and that the metric tensor should be determined uniquely from the energy-momentum tensor. He also assumed that the universe was static, which was a reasonable assumption at the time because observational astronomy had not advanced to a level that contradicted this paradigm. In 1917, when a static solution to his equations could not be found, he introduced the cosmological constant \cite{E}, \cite{Einstein1}.
\begin{equation}
R_{\mu\nu}-\frac{1}{2}Rg_{\mu\nu}=8\pi GT_{\mu\nu}+\Lambda g_{\mu\nu} \ .
\label{eq16}
\end{equation}
For a static universe where $\dot{a}=0$, and using the energy-momentum law $\nabla_\nu T^{\mu\nu}=0$ derived from Einstein's equations, the following relation can be found.
\begin{equation}
8\pi G\rho=\frac{1}{a^2}=\Lambda \ ,
\label{eq17}
\end{equation}
where $a$ is the scale factor and $\rho$ is the energy density of the universe. Einstein found the connection between geometry and mass-energy extremely pleasing, as he believed this was a direct expression of Machian philosophy.

It is popularly believed that Hubble's discovery of galactic redshifting caused Einstein to retract his cosmological constant. The reality is more complicated, and of course more interesting. Einstein was almost immediately disappointed when his friend de Sitter proposed a static solution requiring no matter at all (and thus no cosmological constant). de Sitter began with the line element
\begin{equation}
ds^2=\frac{1}{{\rm cosh}^2Hr}\left[dt^2-dr^2-\frac{{\rm tanh}^2(Hr)}{H^2}(d\theta^2+{\rm sin}^2\theta d\phi^2)\right] \ ,
\label{eq18}
\end{equation}
where $H= \sqrt{ \frac{\Lambda}{3} }$ which will come to be known as Hubble's constant.
Of course, $\Lambda$ \textit{was} needed for a static universe, but over the period from about 1910 to the mid-1920's, Slipher observed that most galaxies have a redshift \cite{S}, indicating that indeed the universe might not be static. By 1922, Friedmann \cite{F} had described a class of cosmological solutions where the universe was free to expand or contract based on the `scale parameter' $a(t)$, with the metric 
\begin{equation}
ds^2=dt^2-a^2(t)\left(\frac{dr^2}{1-kr^2}+r^2(d\theta^2+{\rm sin}^2\theta d\phi^2)\right) \ .
\label{eq19}
\end{equation}
Even with the discovery and future confirmation of a nonstatic universe, dropping the cosmological constant was not something that came easily, because \textit{anything} that contributes to the energy density of the vacuum acts \textit{identically} like a cosmological constant. Solving Einsteins equation for eq.\ (\ref{eq19}) and setting $k=0$ we find
\begin{equation}
H^2=\left(\frac{\dot{a}}{a}\right)^2=\frac{8 \pi G}{3}\rho + \frac{\Lambda}{3} \ .
\label{eq20}
\end{equation}
From this perspective the cosmological constant contributes an extra term to the total vacuum energy
\begin{equation}
\rho_{\rm eff}=\left<\rho\right>+\frac{\Lambda}{8\pi G} ,
\label{eq21}
\end{equation}
and so we can rewrite Einstein's equation as 
\begin{equation}
H^2=\frac{8 \pi G}{3} \rho_{\rm eff} . 
\label{eq22}
\end{equation}
It is then straightforward to demonstrate an observational bound: 
\begin{equation}
\rho_{\rm eff}=\frac{3H^2}{8 \pi G}\approx10^{-47}({\rm GeV})^4 \ .
\label{eq23}
\end{equation}
The problem, however, is when we observe from eq.\ (\ref{eq21}) that
\begin{equation}
10^{-47} ({\rm GeV})^4=\frac{\Lambda}{8 \pi G}+\left<\rho \right> \ ,
\label{eq24}
\end{equation}
implies that the cosmological constant and $\left<\rho \right>$ almost perfectly cancel. The problems begin when we try to calculate a value for $\left<\rho \right>$.

\subsection{Cosmology and Quantum Field Theory} 
\label{sec:caqft}

As discussed earlier, calculations of $\left<\rho \right>$ utilize the formalisms of QFT in which one calculates the ground state of the vacuum energy by summing the zero-point energies of all normal modes of a field up to some cutoff. If we believe GR up to the Planck scale, then we can impose this as our cutoff. We see from eq.\ (\ref{eq9}) that the energy density of the vacuum is 
\begin{equation}
\left<\rho \right>_\Lambda\approx M^4_{\rm pl} \ .
\label{eq25}
\end{equation}
However, we know that $\left< \rho\right>+\frac{\Lambda}{8\pi G}$ is less than $10^{-47} ({\rm GeV})^4$, which implies that the two terms almost perfectly cancel. This `coincidence', is one of the deep mysteries of physics \cite{Rugh2002}, \cite{weinberg1996}. 

One (partial) fix to the vacuum energy calculations is the introduction of supersymmetry (SUSY). The basic idea is that all known particles have an associated superparticle whose spin differs by exactly one half (in $\hbar$ units). In the case of unbroken SUSY, i.e., when particle and supersymmetric partner have exactly equal masses, the superparticle additions to the vacuum energy perfectly cancel the particle contributions and the resulting vacuum energy is reduced to zero.\footnote{This is because the supersymmetric field contributions to the vacuum energy have an equal magnitude, but opposite sign to their non-SUSY counterparts.} However, when SUSY is broken and the difference of the particle mass and the supersymmetric partner mass is on a scale of $M_{\rm S} \approx 10^4 \ {\rm GeV}$, then the resulting vacuum energy density is on the order $(M_S)^4 \approx 10^{16} \ {\rm GeV}$ and the discrepancy is (only) off by a factor of $10^{60}$.

To make the situation even more puzzling, recent observations of distant supervova [9] indicate that the universe is not only expanding, but that it is \textit{accelerating}; that is, its rate of expansion $\frac{\dot{a}}{a}$ is increasing. The current problem is thus to to and explain why $\rho_{\rm eff}$ is greater than.

\subsection{Origins of Negative Pressure and Inflation}
\label{sec:oonpai}

The expansion of the universe is quantified by the scale factor a, which is a dimensionless parameter defined to have a value of 1 at the present epoch. The Friedmann equation governs the variation of a homogeneous, isotropic universe
\begin{equation}
H^2=\left(\frac{\dot{a}}{a}\right)^2=\frac{8\pi G\rho}{3}-\frac{k}{a^2}+\frac{\Lambda}{3},
\label{eq26}
\end{equation}
which is obtained after solving Einstein's equation for the $00$ components using the metric
\begin{equation}
ds^2=dt^2-a(t)d\vec{x}^2 \ .
\label{eq26a}
\end{equation}
From eq.\ (\ref{eq26}) we see that the expansion is determined by three factors: $\rho$ the (mass) energy density of the universe, k the curvature and $\Lambda$ the cosmological constant. It is $\Lambda$ which causes an \textit{acceleration} in the expansion. To understand this in more detail, we appeal to the time and space components of the zero-order Einstein equations where we set $k=0$,
\begin{equation}
\left(\frac{\dot{a}}{a}\right)^2=\frac{8\pi G}{3}\rho_{\rm eff},
\label{eq27}
\end{equation}
where $\rho_{{\rm eff}}=\rho+\frac{\Lambda}{8\pi G}$
and
\begin{equation}
\frac{\ddot{a}}{a}+\frac{1}{2}\left(\frac{\dot{a}}{a}\right)^2=-4\pi G p.
\label{eq28}
\end{equation}
Using eq.\ (\ref{eq27}) in eq.\ (\ref{eq28}) we see that
\begin{equation}
\frac{\ddot{a}}{a}=-\frac{4\pi G}{3}(\rho_{\rm eff}+3p)
\label{eq29}
\end{equation}
Observations indicate that $\ddot{a}/a>0$, and for this to happen the term in parenthesis on the right hand side must clearly be negative. Therefore, inflation requires
\begin{equation}
p<-\frac{\rho_{\rm eff}}{3}
\label{eq30}
\end{equation}
Because energy density is always positive it can be seen that the pressure must be negative. It is important to note that dark energy contributes a negative pressure and is unique to inflationary cosmology. Since matter and radiation both generate positive pressure, whatever drives inflation is neither of these \cite{Starobinksy,Guth2004}. 

\subsection{Supernova Cosmology and the Acceleration of the Universe}
\label{sec:sc}

Type 1a Supernovae (SNe Ia) are reliable tools for testing cosmological models due to their predictable luminosities and the fact that they are visible to a distance of about 500 Mpc \cite{Ba}. A supernova occurs when a star explodes, creating an extremely luminous object that can typically outshine its host galaxy for several weeks or even months. During this time, the supernova can radiate as much energy as our sun would over a period of ten billion years. A supernova will form if a white dwarf star accretes enough matter to reach the Chandrasekhar limit of about 1.38 solar masses. The Chandrasekhar limit is the maximum nonrotating mass that can be supported against gravitational collapse by electron degeneracy pressure. Above the Chandrasekhar limit, the star begins to collapse. Type Ia supernovae follow a characteristic light curve, allowing them to be used as a standard candle to measure the distance to their host galaxy. They can also be used to measure H, Hubble's constant, at relatively close distances. 

The Hubble diagram is a useful tool in cosmology and is often used in the study of the expansion of the universe. The redshift z is plotted against distance for some collection of objects. For low values of z we have
\begin{equation}
cz=Hd
\label{eq33}
\end{equation}
where $H$ is Hubble's constant and encodes the current rate of expansion of the universe. This quantity varies with comsological epoch, as does the redshift z of the objects being used to measure $H$; thus, $H \to H(z)$. Clearly $H(z)$ depends on both the geometry of the universe and its expansion history. So, in principle, the measure of $H(z)$ can teach us something about what the universe is made of.

In the late 1970's it was suggested that type Ia supernovae could be used to measure the deceleration parameter q at high redshift. Advancements in technology made this feasible and initially two major teams studied high-redshift supernova. These were the `Supernova Cosmology Project' (SCP) and the `High-Z Supernova Search Team' (HZT). Both teams discovered around 70 SNe and published similar results \cite{G,p}
which strongly favored a vacuum energy-dominated universe. Their results suggested an ``eternally expanding Universe which is accelerated by energy in the vacuum''. The type Ia supernovae were observed to be dimmer than would be expected in an empty universe $(\Omega_{\rm M})=0$ with $\Lambda=0$. An explanation for this observation is that a positive vacuum energy accelerates the universe. Clearly, for $\Omega_{\rm M}>0$, the problem is exacerbated since mass has the effect of decelerating the expansion.

Figure 2.1 shows a plot of magnitude against redshift for a range of high-z supernova \cite{Knop2003}.\footnote{Technically this is the `B-Magnitude' of the Johnson photometry system, a measure of blue light centered on a wavelength of 436 ${\rm nm}$.} Displayed three curves which depict the three cosmological possibilities: flat matter dominated, open, and flat with a cosmological constant. The high redshift data strongly favor a universe with about 75\% of the energy in the form of the cosmological constant, and suggest that we are currently experiencing an epoch of \textit{accelerated} expansion.

$\rho_\Lambda$ remains an unexplained parameter, and this vacuum energy is indistinguishable from dark energy. The precise cancellation of $\rho_\Lambda$ with $\rho$ is a deep mystery of physics. One interesting possibility is the modification of $\rho$ due to the non-trivial boundary conditions that higher dimensions impose. The calculations are similar in nature to Casimir energy calculations and so to understand the possible role of the quantum vacuum in the creation of $\rho_\Lambda$, we first review the Casimir effect.
\begin{singlespace}
\begin{figure}[H]
\begin{center}
\includegraphics[width=370pt]{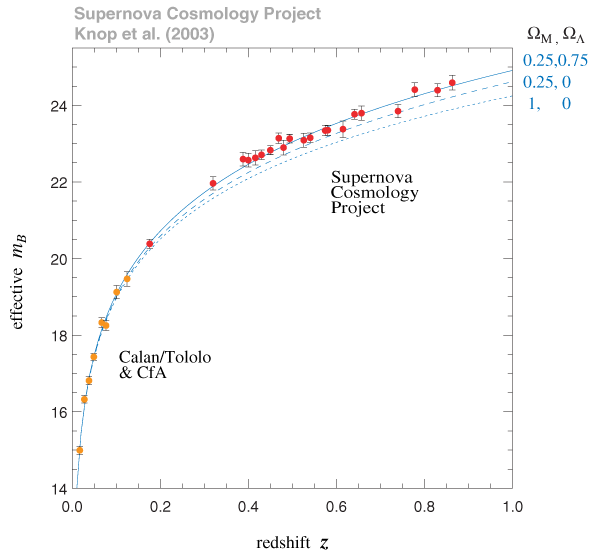}
\caption{\textit{Magnitude of high redshift supernova as a function of z. Three possible cosmological models are shown, $(\Omega_{\rm M},\Omega_\Lambda)=(1,0)$, $(\Omega_{\rm M},\Omega_\Lambda)=(0.25,0)$ and $(\Omega_{\rm M},\Omega_\Lambda)=(0.25,0.75)$. The data strongly supports a dark energy dominated universe.}}
\end{center}
\end{figure}
\end{singlespace}

\chapter{The Casimir Effect}
\label{chap:tce}

Arguably the most poignant demonstration of the reality of the quantum vacuum is the famous Casimir effect. In 1948, H. Casimir published a profound paper where he explained the van der Waals interaction in terms of the zero-point energy of a quantized field. The Casimir effect can be appreciated most simply in the interaction of a pair of neutral parallel plates. The presence of the plates modifies the quantum vacuum, and this modifcation causes the plates to be pulled toward each other with a force $F\propto\frac{1}{a^4}$ where $a$ is the plate separation. For many years, the paper remained unknown \cite{Bordag2001}, but from the 70's onward the Casimir effect received increasing attention, and over the last decade it has become very popular \cite{Balian2004}. 

The Casimir effect is a purely quantum effect. In classical electrodynamics the force between the plates is zero. The ideal scenario occurs at zero temperature when there are no real photons (only virtual photons) between the plates; thus, it is the ground state of the quantum electrodynamic vacuum which causes the attraction. The most important feature of the Casimir effect is that even though it is purely \textit{quantum} in nature, it manifests itself macroscopically. For example, for two parallel plates of area $A=1 \ {\rm cm}^2$ separated by a distance of $d=1{\rm \mu m}$ the force of attraction is $F \approx 1.3 \times 10^{-7} \ {\rm N}$. This force is certainly within the range of laboratory force-measuring techniques. Something that is unique to the Casimir force is that both the sign and the magnitude of the Casimir force is strongly dependent on the geometry of the plates \cite{Milton}. This makes the Casimir effect a good candidate for applications in nanotechnology.

Typically, the calculations of the expectation value of the vacuum is divergent, so some form of renormalization must be performed. For example, consider the calculation of the field vacuum expectation value (VEV) inside a metal cavity. Such a calculation will necessarily involve summing the energies of the standing waves in the cavity;
\begin{equation}
\left<E_{\rm vac}\right>=\frac{1}{2}\sum_{n=1}^\infty E_{\rm n},
\label{eq34}
\end{equation}
which is clearly divergent. The Casimir effect is studied in the context of a wide variety of field in physics, including gravitation and cosmology, condensed matter physics, atomic and molecular physics, quantum field theory and even nanotechnology \cite{mp}. Recently, high precision experiments have been performed that verify the theoretical predictions regarding the force. Some of the most important experiments will be discussed in Section 3.3

\section{Casimir's Original Calculations}
\label{sec:coc}

In this section, we review Casimir's original approach \cite{cas} first published in 1948 and the techniques taken to control the divergences associated with the calculations. We first consider a cubical cavity of volume $L^3$ bounded by perfectly conducting walls. A perfectly conducting square plate with side $L$ is placed in the cavity parallel to the x-y face. We consider the scenario where the plate is first close to the x-y face, and is then a distance $L/2$ away. In both cases the energy associated with the resonant frequencies in the cavity are divergent and therefore devoid of any physical meaning, but the difference between them will have a well defined value. The wave vectors in the cavity are given by
\begin{eqnarray}
k_{\rm x} &=& \frac{n_x \pi}{L} \nonumber \\
k_{\rm y} &=& \frac{n_y \pi}{L} \nonumber \\
k_{\rm z} &=& \frac{n_z \pi}{L}
\end{eqnarray}
For every k there exists two standing waves unless $n_i$ is zero. For large $L$ we can consider $k_{\rm x}$ and $k_{\rm y}$ a continuous variable over which we can integrate. Our expression for the ground state energy is,
\begin{equation}
E=\hbar c \frac{L^2}{\pi^2}\int_0^\infty \int_0^\infty \left( \frac{1}{2}\sqrt{k_{\rm x}^2+k_{\rm y}^2}+\sum_{n=1}^\infty \sqrt{\left(\frac{n\pi}{a}\right)^2+k_{\rm x}^2+k_{\rm y}^2} \right) dk_{\rm x} dk_{\rm y},
\end{equation}
where the first term in the parenthesis is simply the $n=0$ term.\footnote{We have retained $\hbar$ and $c$ in this calculation for clarity.} If we now make the substitution
\begin{equation}
k=\sqrt{k_{\rm x}^2+k_{\rm y}^2+k_{\rm z}^2}=\sqrt{x^2+k_{\rm z}^2},
\end{equation}
the integral becomes
\begin{equation}
E=\hbar c \frac{L^2}{\pi^2} \frac{\pi}{2}\sum_{(0)1}^\infty\int_0^\infty\sqrt{\left(\left(\frac{n\pi}{a}\right)^2+x^2\right)}xdx,
\end{equation}
where the notation $(0)1$ means that the term with $n=0$ has to be multiplied by $\frac{1}{2}$. For large $a$ the sum can be replaced by an integral and so the difference between the small and large $a$ cases is expressed by
\begin{eqnarray}
\delta E&=&\hbar c \frac{L^2}{\pi^2} \sum_{(0)1}^\infty\int_0^\infty\sqrt{\left(\left(\frac{n\pi}{a}\right)^2+x^2\right)}xdx  \nonumber \\
&-& \hbar c \frac{L^2}{\pi^2}  \int_0^\infty \int_0^\infty \sqrt{(k_{\rm z}^2+x^2)}xdx\left(\frac{a}{\pi}dk_{\rm z}    \right) \ .       
\end{eqnarray} 
Clearly the integrals are infinite and so we multiply them by some function $f(k/k_{\rm m})$ which tends to unity for $k<<k_{\rm m}$, but which tends to zero as $k/k_{\rm m} \rightarrow \infty$. The physical interpretation is that as the wavelength becomes shorter, the plates appear more transparent. If we now introduce the variable $u=a^2x^2/\pi^2$ our integral becomes
\begin{eqnarray}
\delta E = L^2\hbar c \frac{\pi^2}{4 a^3}  \sum_{(0)1}^\infty\int_0^\infty \sqrt{n^2+u} f(\pi\sqrt{n^2+u/ak_m})du \nonumber \\
- L^2\hbar c \frac{\pi^2}{4 a^3} \int_0^\infty \int_0^\infty\sqrt{n^2+u}f(\pi\sqrt{n^2+u/ak_{\rm m}})dudn \ .
\end{eqnarray}      
We now apply the Euler-Maclaurin formula:
\begin{equation}
\sum_{(0)1}^\infty F(n)-\int_0^\infty F(n)dn=-\frac{1}{12}F'(0)+\frac{1}{(24\times30)}F'''(0)+... \ .
\label{CasCalc1}
\end{equation}
Introducing $w=u+n^2$ we have
\begin{equation}
F(n)=\int_{n^2}^\infty \sqrt{w}f(w\pi/ak_{\rm m})dw.
\label{CasCalc2}
\end{equation}
Taking the relevant derivatives of eq.\ (\ref{CasCalc2}) in preperation for insertion into eq.\ (\ref{CasCalc1}),
\begin{eqnarray}
F'(n)&=&-2n^2f(n^2\pi/ak_{\rm m}) \nonumber \\
F'(0)&=&=0 \nonumber \\
F'''(0)&=&-4.
\end{eqnarray}
The higher derivatives will contain powers of $\pi/ak_{\rm m}$, and so we finally obtain
\begin{equation}
\delta E=-\hbar c \frac{\pi^2}{720}\frac{L^2}{a^3},
\end{equation}
valid for $ak_m>>1$. The force is obtained by taking the derivative,
\begin{equation}
F=\hbar c \frac{\pi^2}{240}\frac{L^2}{a^4},
\end{equation}
which is the famous Casimir force which `may be interpreted as a zero point pressure of the electromagnetic wave' \cite{cas}.

\section{Alternative Derivations}
\label{sec:ad}

A number of alternatives to Casimir's original calculation exist. The following is a brief review of some of the popular techniques.

\subsection{Green Function Approach for Parallel Plate Geometry}
\label{sec:gfafppg}

Derivation of the Casimir energy using a Green function approach is more physical and rigorous than the derivation already shown \cite{Milton}. We begin with the equation of motion for a massless scalar field $\phi$ for the geometry described in the previous section. The equation of motion produced by some source K is
\begin{equation}
-\partial_{\mu}\partial^{\mu}\phi=K \ .
\label{eq47}
\end{equation}
The corresponding Green's function satisfies
\begin{equation}
-\partial_{\mu}\partial^{\mu}G(x^\mu,x'^\mu)=\delta(x-x')\delta(t-t')\delta(z-z') \ ,
\label{eq48}
\end{equation}
where $\mu$ runs over x, z, and t. It is instructive to use separation of variables to express the Green's function as a product of reduced Green's function.
\begin{equation}
G(x^\mu,x'^\mu)=g(x,x')g(t,t')g(z,z') \ .
\label{eq49}
\end{equation}
However, because the physics is symmetric in x and t we can express $g(x,x')$ and $g(t,t')$ as delta functions, giving us
\begin{equation}
G(x,x')=\delta(x-x')\delta(t-t')g(z,z') \ .
\label{eq50}
\end{equation}
We express the delta functions in momentum space, giving us 
\begin{equation}
G(x,x')=\int\frac{dk}{2\pi}e^{i\vec{k}.(x-x')}\int\frac{d\omega}{2\pi}e^{-i\omega.(t-t')}g(z,z') \ ,
\label{eq51} 
\end{equation}
where we recall that the $\omega$ is really the $k_0$ integral in which $\hbar$ and c are set to 1. Substituting this Green's function back in to eq.\ (\ref{eq49}) we see
\begin{eqnarray}
-\partial_{\mu}\partial^{\mu}G(x^\mu,x'^\mu)&=& -(\partial_z^2+\partial_x^2+\partial_t^2) \label{eq51a} \\ 
&=&(-\partial_z^2-\lambda^2)G(x^\mu,x'^\mu) \ ,
\label{eq52} 
\end{eqnarray}
where $\lambda=\omega^2-k^2$. Recalling the separation-of-variables expression for the Green function we see
\begin{equation}
(-\partial_z^2-\lambda^2)\delta(x-x')\delta(t-t')g(z,z')=\delta(x-x')\delta(t-t')\delta(z-z') \ ,
\label{eq53}
\end{equation}
which implies
\begin{equation}
(-\partial_z^2-\lambda^2)g(z,z')=\delta(z-z').
\label{eq54}
\end{equation}
A general solution to eq.\ (\ref{eq54}) can be guessed easily.
\begin{equation}
g(z,z') = \begin{array}{cc}A\sin\lambda z,  &\mbox{ if }(0<z<z'<a)\\ B\sin\lambda(z-a), &\mbox{ if }(a>z>z'>1). \end{array} 
\label{eq62}
\end{equation}
Our reduced Green's function is continuous at $z=z'$; however, its derivative has a disconuity. These two conditions give us two equations,
\begin{eqnarray}
A\sin \lambda z'-B\sin\lambda(z-a)&=&0   \nonumber \\
A\lambda \cos\lambda z'-B\lambda \cos\lambda(z-a)&=&1.
\label{eq64}
\end{eqnarray}
The solution to this system of equations is
\begin{eqnarray}
A&=&\frac{1}{\lambda}\frac{\sin\lambda(z'-a)}{\sin\lambda a}   \nonumber \\
B&=&\frac{1}{\lambda}\frac{\sin\lambda(z')}{\sin\lambda a} \ , 
\label{eq66}
\end{eqnarray}
and so the final expression for our reduced Green's function is given by
\begin{equation}
g(z,z')=-\\frac{1}{\lambda \sin\lambda a}\sin\lambda z_<\sin\lambda(z_>-a)\ ,
\label{eq67}
\end{equation}
where we have used the notation $z_>(z_<)$as the greater (lesser) of z and z'.

Knowledge of the Green's function allows us to calculate the Casimir energy from the energy-momentum tensor, which is given by
\begin{equation}
T_{\mu\nu}=\partial_\mu\phi\partial_\nu\phi+g_{\mu\nu}\mathcal{L},
\label{eq68}
\end{equation}
where $\mathcal{L}$ is the Lagrange density given by
\begin{equation}
\mathcal{L}=-\frac{1}{2}\partial_\lambda\phi\partial^\lambda\phi\ .
\label{eq69}
\end{equation}
To extract the Casimir energy, the $T_{00}$ component is required, and the vacuum expectation value is
\begin{eqnarray}
\left<T_{00}\right>&=&-\frac{1}{2i\lambda}\frac{1}{\sin(\lambda a)}\left((\omega^2+k^2)\sin(\lambda z)\sin(\lambda(z-a))+\lambda^2\cos(\lambda z) \cos(\lambda(z-a))\right)  \nonumber \\
&=&-\frac{1}{2i\lambda \sin\lambda a}\left(\omega^2\cos\lambda-k^2\cos\lambda(2z-a)\right) \ .
\label{eq70}
\end{eqnarray}
We must now integrate over z to find the total energy per unit area. Integration of the second term gives a constant which is independent of a, and can thus be ignored. Integrating the first term we obtain
\begin{equation}
\int_0^a=dz\left<T_{00}\right>=\frac{\omega^2a}{2i\lambda}{\rm cot}\lambda a
\label{eq72}
\end{equation}
Our next task is to integrate the transverse momentum and the frequency to get the total energy per unit area. This is best done by performing a complex frequency rotation $\omega\rightarrow i\zeta$ and $\lambda \rightarrow i\sqrt{k^2+\zeta^2}=i\kappa$
\begin{equation}
E=-\frac{a}{2}\int\frac{d^dk}{(2\pi)^2}\int\frac{d\zeta}{2\pi}\frac{\zeta^2}{\kappa}{\rm coth}\kappa a
\label{eq73}
\end{equation}
Performing the integral, we obtain
\begin{equation}
E=-\frac{1}{2^{d+2}\pi^{d/2+1}}\frac{1}{a^{d+1}}\Gamma\left(1+\frac{d}{2}\right)\zeta(d+2)
\label{eq74}
\end{equation}

\subsection{Analytic Continuation}
\label{sec:ac}

Another technique that can be used to tame the divergences associated with calculation of the Casimir energy involves using analytic continuation, as described in \cite{HV} and \cite{aw}. For a very general proof, we work in d-dimensions and consider a scalar field that satisfies the free Klein Gordon equation in the absence of boundaries,
\begin{equation}
(\partial^2+m^2)\phi(x)=0 \ .
\label{eq75}
\end{equation}
Constraining the fields at $x=0$ and $x=a$, we impose Dirichlet boundary conditions, e.g.
\begin{equation}
\phi(0)=\phi(a)=0 \ .
\label{eq76}
\end{equation}
In the ground state, each mode contributes an energy
\begin{equation}
\omega_{\rm k}=\sqrt{\left(\frac{n\pi}{a}\right)^2+{\bf{k}}^2+m^2}
\label{eq78}
\end{equation} 
The total energy of the field between the plates is given by
\begin{equation}
E=\frac{L}{2\pi}^{d-1}\int d^{d-1}{\bf k}\sum_{n=1}^\infty \frac{1}{2}\omega_{\rm n} \ .
\label{eq79}
\end{equation}
This sum is clearly divergent, but can be regularized by using a process of analytic continuation. Using the formumla \cite{aw}
\begin{equation}
\int d^dk f(k)=\frac{2\pi^{d/2}}{\Gamma(\frac{d}{2})}\int k^{d-1}f(k)dk \ ,
\label{eq80}
\end{equation}
and substituting into eq.\ (\ref{eq79}), we obtain
\begin{equation}
E=\left(\frac{L}{2\pi}\right)^{d-1}\frac{2\pi^{(d-1)/2}}{\Gamma\frac{d-1}{2}}\sum_{n=1}^\infty\int_0^\infty \frac{1}{2}({\bf k}^2)^{(d-3)/2}d({\bf k}^2)\frac{1}{2}\sqrt{\left(\frac{n\pi}{a}\right)^2+{\bf k}^2+m^2  } \ .
\label{eq81}
\end{equation}
Using the well-known expression for the Beta function
\begin{equation}
\int_0^\infty t^r(1+t)^s dt=B(1+r,-s-r-1) \ ,
\label{eq82}
\end{equation}
and substituting in for the Beta function, we obtain
\begin{equation}
E=\frac{1}{2}\frac{\Gamma\left(\frac{-d}{2}\right)}{\Gamma\left(\frac{-1}{2}\right)}\pi^{(d+1)/2}\frac{(L/2)^{d-1}}{a^d}\sum_{n=1}^\infty\left[\left(\frac{am}{\pi}\right)^2+n^2\right]^{d/2}.
\end{equation}
Note that
\begin{equation}
B(M,N)=\frac{\Gamma(m)\Gamma(n)}{\Gamma(m+n)}.
\label{eq82a}
\end{equation}
At this stage, it is necessary to again introduce the Riemann zeta function, as the sum is formally divergent. We also take advantage of the reflection formula
\begin{equation}
\Gamma\frac{s}{2}\pi^{-s/2} \zeta(s)=\Gamma\left(\frac{1-s}{2}\right)\pi^{(s-1)/2}\zeta(1-s)\ ,
\label{eq83}
\end{equation}
and the reduplication \cite{aw} formula
\begin{equation}
\Gamma(s)\sqrt{\pi}=2^{s-1}\Gamma\left(\frac{s}{2}\right)\Gamma\left(\frac{1+s}{2}\right).
\label{eq84}
\end{equation}
This allows us to re-write the energy as
\begin{equation}
E=-\frac{L^{d-1}}{a^d}\Gamma\left(\frac{d+1}{2}\right)(4\pi)^{-(d+1)/2}\zeta(d+1)\ .
\label{eq85}
\end{equation}
The result is now finite for all d, and always negative. Again, the force is obtained by taking the derivative,$-\frac{\partial(E/L^{d-1})}{\partial a}$. For the case of d=1 we obtain
\begin{equation}
E=-\frac{\pi}{24L}
\label{eq86}
\end{equation} 
 which is consistent with the result derived from the Green's function approach.

\subsection{Riemann Zeta Function}
\label{sec:rzf}

The previous calculation of the Casimir energy was fairly involved; however, a far simpler procedure exists if we utilize the Riemman zeta function. First, recall the expression for the energy between the plates:
\begin{equation}
E_0(L)=\frac{\pi}{2L}\sum_{n=1}^\infty n \ .
\label{eq87}
\end{equation}
If we now use the definition of the Riemann zeta function,
\begin{equation}
\zeta(s)=\sum_{n=1}^\infty \frac{1}{n^s}
\label{eq88}
\end{equation}
we can rewrite eq.\ (\ref{eq87}) as:
\begin{equation}
E_0(L)=\frac{\pi}{2L}\sum_{n=1}^\infty \frac{1}{n^{-1}}=\frac{\pi}{2L}\zeta(-1) \ .
\label{eq89}
\end{equation}
However, $\zeta(-1)=-\frac{1}{12}$ (from analytic continuation), and so we quickly obtain the result:
\begin{equation}
E_0(L)=-\frac{\pi}{24L},
\label{eq90}
\end{equation}
which is the Casimir energy. Taking the derivative, we obtain the force of attraction between the plates:
\begin{equation}
\mathcal{F}(L)=\frac{\partial}{\partial L}E_0(L)=-\frac{\pi}{24L^2}
\label{eq91}
\end{equation}	
which is also in agreement with the previous results of this section. For the remainder of this dissertation, most calculations will be performed using $\zeta$-function techniques, which are largely preferred in the literature \cite{Kirsten2002}.

\section{Experimental Verification of the Casimir Effect}
\label{sec:ev}

Given the scenario of two parallel plates of area $A$ and infinite conductivity, separated by a distance $L$, the Casimir force is given by  
\begin{equation}
F(L)=-\frac{\pi^2}{240}\frac{\hbar c}{L^4}A \ ,
\label{eq91a}
\end{equation}
where we have included $\hbar$ and $c$ for the purpose of clarity. The force is a strong function of $L$. For a flat surface of $1 {\rm cm}^2$, and for a separation of $1  {\rm \mu m} $, the Casimir force of attraction is on the order of $10^{-7}{\rm N/cm^2}$. This is a small force, and the most severe limitation on the accuracy of various measurements is the sensitivity of the experimental techniques. The following sections review the developments in experimental techniques used to measure the Casimir force.

\subsection{Early Experiments by Spaarnay}
\label{sec:eebs}

M.J. Spaarnay performed the first measurements of the Casimir Force \cite{Sparnaay, Sparnaay1}. The experiment, which was based on a spring balance and parallel plates, set the benchmark for all future experiments. The work of Sparnaay also highlighted the problems associated with such high precision measurements and the fundamental requirements for an accurate experiment, as documented in \cite{Sparnaay, Sparnaay1}
\begin{itemize}
\vspace{-0.2in}
\begin{singlespace}
\item clean plates that are free of chemical impurities and dust particles;
\item precise and reproducible measurements of the separation between the two surfaces. Particularly important is a measurement of the average distance on contact of the two surfaces, which is nonzero due to surface roughness of the materials and the presence of dust;
\item low electrostatic charge on the surface and low potential difference between the surfaces.
\end{singlespace}
\end{itemize}

The sensitivity of the spring balance used in Sparnaay's measurement \cite{Sparnaay} was between $0.1 - 1$ Dynes. The extension of the spring was found by measuring the capacitance formed between the two plates. Care was also taken to isolate the experimental setup from vibrations. The springs that were used led to large hysteresis, which made determination of the separation difficult. The plates had to be electrically insulated from the apparatus, because even a potential difference as small as 17mV between the plates would overwhelm the Casimir Force.

Even though great care was taken with cleaning the equipment, dust particles larger than $2-3  {\rm \mu m}$ were observed on the plates. The chromium-steel and the chromium plates both generated an attractive force. Conversely, the aluminum plates created repulsive forces which were believed to be due to impurities on the aluminum surface. Given these adverse circumstances, only a general agreement with the Casimir force formula was achieved.

Since this pioneering work there have been a number of notable efforts in improving on the accuracy of Casimir force measurements.

\subsection{Improvements of the Casimir Force Measurement}
\label{sec:icfm}

Major improvements by Derjaguin's team \cite{Derjaguin} made force measurements which were in discrepancy with the theoretical prediction by 60\% \cite{Derjaguin1}. The experiment utilized curved plates, which avoided having to keep the plates parallel. This was accomplished by replacing one of the plates with a lens. The first reported use of this technique was in the measurement of the force between a silica lens and a plate \cite{Derjaguin1,Derjaguin3,Derjaguin4,Derjaguin5,Derjaguin6}. The Casimir force measurement was achieved by keeping one of the surfaces fixed, and attaching the other surface to the coil of a Galvanometer. The subsequent rotation of the coil led the the deflection of a beam of light, which was reflected off mirrors that were attached to the coil.

Mica cylinders were used by Tabor et al \cite{Tabor}, Israelachvili et al \cite{Israelachvili, Israelachvili1} and White et al \cite{White} to some success. Cleaved Muscovite mica greatly improved the surface smoothness, which introduced the possibility of bringing the two cylinders extremely close to one another, taking full advantage of the $1/L^4$ dependence of the Casimir force. Multiple beam interferometry was used for the separation measurement, with a reported resolution of $0.3 {\rm nm}$.

Major experimental improvements were made by P. van Blokland and J. Overbeek \cite{Blokland}. Their experiment was performed using a spring balance, and the force between a flat plate and a lens was measured. Both surfaces were coated with either $(100\ \pm{5})\ {\rm nm}$ or $(50\pm{5})\ {\rm nm}$ of chromium. Because the potential difference between the two plates leads to electrostatic forces which can complicate the experiment, the authors applied a compensating voltage at all times. 

The separation between the surfaces was determined by using a Schering bridge to measure the lens-plate capacitance. The authors estimated that the effects of surface roughness added contributions to the force of order 10\%. The relative uncertainty in the measurement of the Casimir force was reported to be $25\%$ near separations of $150 {\rm nm}$, but much larger around $500 {\rm nm}$. With all the uncertainties taken into account, Bordag \cite{Bordag2001} estimates that the accuracy of the experiment was of the order $50 \%$. This particular experiment was significant, because it was the first to address all the systematic errors and other factors (identified by Sparnaay) that are necessary to make a good Casimir force measurement.

\subsection{The Experiments of Lamoreaux}
\label{sec:tel}

A series of experiments performed by Lamoreaux \cite{Lamoreaux} marked the modern phase of Casimir force measurements. This experiment is also notable because it was contemporary with the development of theories involving compact dimensions, which utilized the Casimir force as a mechanism for moduli stability. This invigorated the theoretical and experimental community, and increased awareness as to the usefulness of the Casimir effect as a test for new forces in the submillimeter range.

The experiment used a balance based on the Torsion pendulum, which measured the Casimir force between a gold plated spherical lens and a flat plate. The lens was mounted on a piezo stack and the plate on one arm of the torsion balance. The remaining arm of the torsion balance formed the central electrode of a dual parallel plate capacitor. The Casimir force between the plate and the lens surface would result in a torque which would lead to a change in the torsion balance angle, which would then lead to a change in the capacitances, which would be detected through a phase sensitive circuit.

Initially claims of 5\% experimental agreement with the theoretical value of the Casimir force were made; however, it was later realized that finite temperature conductivity corrections could amount to as much as 20\% of the force attributed to the Casimir effect \cite{bgkm}. Lamoreaux subsequently highlighted two errors in his measurements \cite{Lamoreaux1} and calculated the finite conductivity corrections for gold to be 22\% and for copper 11\%. The measured value of the Casimir effect was thus adjusted by these values. Another error in the measurement was the radius of curvature of the lens, which corresponded to a 10.6\% increase in the theoretical value of the Casimir force.

In spite of these errors, the experiment by Lamoreaux is widely regarded as the introduction of a modern phase of high precision force detection, and stimulated a surge in theoretical activity \cite{Bordag2001}.

\subsection{Atomic Force Microscopes}
\label{sec:afm}

Atomic Force Microscopes (AFM) led to further progress in precision measurements of the Casimir force. Particularly notable is the experiment of Mohideen \cite{Mohideen}. A diagram of the experimental setup is shown in Figure 3.1. A Casimir force between the plate and the sphere causes the cantilever to flex, resulting in the deflection of a laser beam, which leads to a signal difference between the two photodiodes A and B. The plate was moved towards the sphere in $3.6 nm$ steps and the corresponding photodiode signal was measured. 

For separations on the order of $1 {\rm \mu m}$ between the interacting bodies, both the surface roughness and the finite conductivity of the boundary generate contributions to the Casimir force. An exact calculation of this force is impossible, but one can find approximate corrections \cite{Bordag2001}. 
\begin{singlespace}
\begin{figure}[H]
\begin{center}
\includegraphics[width=300pt]{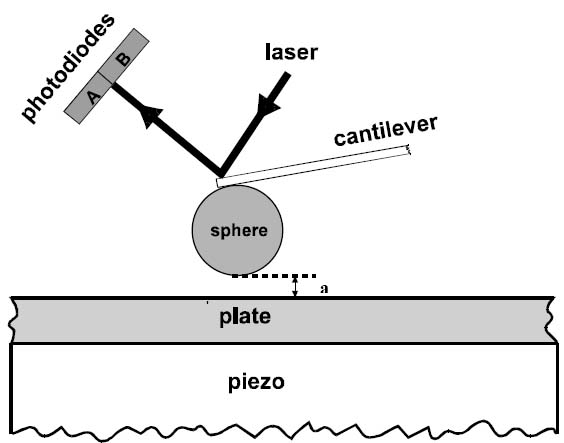}
\caption{\textit{Schematic of the experimental setup of AFM. When a voltage is applied to the piezoelectric element the plate is moved towards the sphere. }}
\end{center}
\end{figure}
\end{singlespace}
In the first published results of \cite{Mohideen} only the second order conductivity and roughness corrections were used to compare theory with experiment. Using these corrections, the rms deviation of the experiment from the theoretical force was determined to be $\sigma=1.6 {\rm pN}$ in the complete measurement range. This is on the order of 1\% accuracy. The experimental measurement plotted against the theoretical prediction is shown in Figure 3.2.

The close agreement between theory and experiment can be explained by the fact that the corrections due to surface roughness and finite conductivity are of a different sign, and in some ways, compensate for each other \cite{Mohideen}. A consequence of this is that the value of $\sigma$ was dependent on the separation interval. Hence, $\sigma$ was different if it was found at a small separation, than it was if found at large separation.
\begin{singlespace}
\begin{figure}[H]
\begin{center}
\includegraphics[width=300pt]{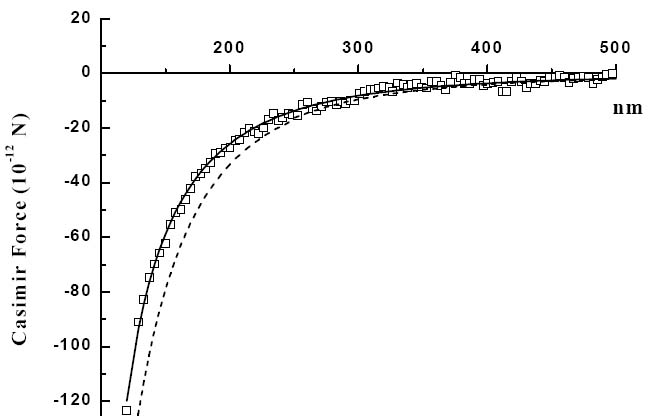}
\caption{\textit{Schematic of the experimental setup of AFM. When a voltage is applied to the piezoelectric element the plate is moved towards the sphere. The theoretical value of the Casimir force is shown as a dashed line, and then the Casimir force with corrections due to surface roughness and finite conductivity is shown as the solid line.}}
\end{center}
\end{figure}
\end{singlespace}
In \cite{MohideenandRoy}, improvements in this technique were reported, including:
\begin{itemize}
\vspace{-0.2in}
\begin{singlespace}
	\item  smoother metal coatings were used which reduced the surface roughness effects.
  \item  Reduction of total noise by vibrational isolation.
	\item  Independent electrostatic measurement of the separation of the surfaces.
	\item  Reduction in systematic errors due to the residual electrostatic force, scattered light, and instrument drift.
	\end{singlespace}
\end{itemize}
Surface roughness corrections were reported to be about 1.3 \% of the total measured force, which corresponded to a factor of 20 improvement of the previous experiment. The Casimir force as a function of plate separation is shown in Figure 3.3. The improvements to the previous measurements are demonstrated by a much closer agreement between experiment and theory.
\begin{singlespace}
\begin{figure}[H]
\begin{center}
\includegraphics[width=300pt]{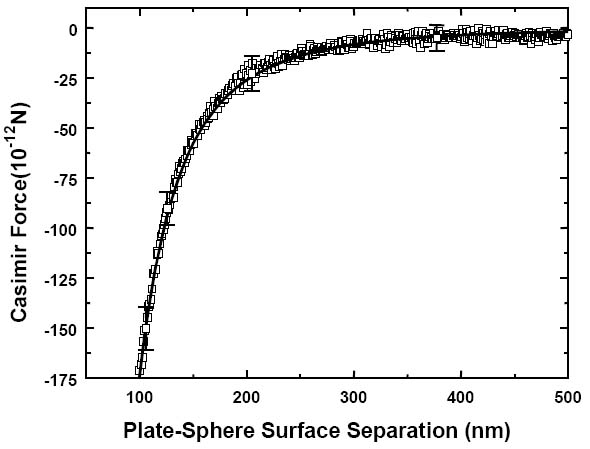}
\caption{\textit{Casimir force measurements as a function of distance [57]. The average measured force is shown as squares, and the error bars represent the standard deviation from 27 scans. The solid line is the theoretically predicted Casimir force, with the surface roughness and finite conductivity included.}}
\end{center}
\end{figure}
\end{singlespace}
The measurements of the Casimir force using an aluminum surface were conclusive to a precision of 1\%, demonstrating both the precision and accuracy of the AFM. Improvements on this experiment were performed by the same team \cite{MohideenandRoy}, with the main difference being that a gold surface was used for the AFM. Also, the aluminum was covered with a thin layer of Au/Pd, which reduced the effects of oxidation. The AFM cantilevers were also coated with $200 {\rm nm}$ of aluminum, which improved their thermal conductivity and decreased the thermally-induced noise that occurs when the AFM is operated in a vacuum. Full details of the experimental improvements can be found in \cite{MohideenandRoy}, which reports a standard deviation of $19 {\rm pN}$ and a precision of better than 1 \%. This experiment is among the most accurate measurements of the Casimir force to date.

\subsection{The Future of Casimir Force Measurement Experiments}
\label{sec:tfocfme}

Despite the conclusive results of Casimir force measurements using the AFM, the measurements are only sensitive to forces at surface separations $32 {\rm nm} < {d} < 1000 {\rm nm}$. Given the possibilities of force modifications due to submillimeter dimensions \cite{add,aadd}, there is a clear requirement for measurements of the Casimir force at smaller plate separations. At the other end of the spectrum, Casimir force measurements at separations above $1000 {\rm nm}$ are relevant to tests of supersymmetry breaking ($\approx 10 \ {\rm TeV}$)and string theory, as well as addressing long-range interactions. For example, in \cite{Mostepanenko1,Mostepanenko2,Mostepanenko3,Mostepanenko4,Mostepanenko5} it was shown that the Casimir effect leads to the strongest constraints on the Yukawa-type interactions, which means that the Casimir effect becomes a new non-accelerator test for massless elementary particles and new forces of nature. These sorts of tests become particularly significant with regard to the predictions of some models that the gravitational and gauge interactions may become unified at the weak scale \cite{add}.

To make the required experimental measurements for separations less than $60{\rm nm}$, smoother metal coatings are required \cite{Bordag2001}. Molecular Beam Epitaxy may be perfectly suited to this task because it produces atomic layer-by-layer growth. Even so, individual lattice steps of size $\pm 0.5{\rm nm}$ are unavoidable, and so the smallest separation possible for plates created by this method is on the order of $1 {\rm nm}$.

For the case of measurements greater than $1000 {\rm nm}$, measurements of the finite temperature corrections are necessary. These contributions to the Casimir force are significant for separations at or greater than this scale. Proposals to improve the AFM technique include: 

\begin{itemize}
\vspace{-0.2in}
\begin{singlespace}
	\item  lithographic fabrication of the cantilevers with large radius of curvature.
  \item  Interferometric detection of cantilever deflection.
	\item  Thermal noise reduction by reducing the temperature of the experimental setup.
	\item  Dynamic measurements.
	\item  Alternative boundaries including cubical and spherical boundaries.
\end{singlespace}
\end{itemize}

\chapter{Extra Dimensions}
\label{chap:ed}

$\indent$ In connection with the Casimir effect, extra dimensions provide a rich arena for us to generate models that explain the origin of $\left< \rho\right>$. Since their introduction into theoretical physics with Kaluza-Klein theory, extra dimensions have traditionally been considered to be small (probably Planck length). This solves the obvious conundrum that an extra dimension has never been seen. Extra dimensions are an integral component of string theory, a quantum theory of gravity which attempts to unify particle physics under a single mathematical structure. Other popular contemporary extra dimensional models include the Arkani-Hamed-Dimopoulos-Dvali (ADD) \cite{add} and the Randall-Sundrum (RS) models \cite{RS1,RS2}.\footnote{The entirety of the universe which includes our familiar 3 dimensions of space, 1 dimension of time and all additional spatial dimensions are referred to as the ``bulk''.} Both are attempts to explain what has become known as a heirachy problem in physics, which questions why the gravitational force is so much weaker than the other forces of nature. In the ADD model, it is proposed that the force carriers of the standard model (the photon, $W^+$, $W^-$, $Z^0$ and the gluons) are constrained to exist on the usual four dimensional spacetime, which we will call a 3-brane. Gravity, however, is free to move both on the 3-brane \textit{and} in the extra dimensions. These dimensions can be as large as a few  ${\rm \mu m}$ based on experimental upper limits. 

Because only gravity can propagate in the extra dimensions in the ADD models, we cannot `see' the extra dimensions. This is because the process of seeing is mediated through the photon which is restricted to our 3-spatial brane. Nor can we observe their effects through observation of weak or strong force interactions, which are also restricted to our 3-brane. It is the freedom of the graviton to propagate off of the brane that dilutes the field strength, accounting for the apparent weakness of gravity. In the RS models, it is the \textit{warping} of the extra dimensions that is the root cause of the weakness of gravity.

In this section we review the introduction of extra dimensions into theoretical physics, first looking at the Kaluza-Klein theory. We then review the ADD model of large extra dimensions and also the warped compactifications of the RS model.

\section{Historical Background Leading to Kaluza-Klein Theory}
\label{sec:hbltkkt}

It was Riemann, with his development of differential geometry in the nineteenth century, who gave us the tools necessary to study higher dimensional descriptions of the world. Riemann held the belief that three-dimensional space was not enough to provide an adequate description of nature \cite{k}. Improvements in physics led to Maxwell's unified theory of electricity and magnetism, and then GR which unified space and time with Special Relativity (SR). Inspired by these unifications, physicists of the early twentieth century wanted to unify gravity and electromagnetism. The first attempt was by Nordstrom in 1914, who used a scalar potential for the gravitational field. Later Weyl and Kaluza, using Einstein's tensor potential, followed two separate paths. Weyl's attempt involved an alteration of the geometry of spacetime in four dimensions. His early attempts had physical consequences which did not match experimental data. However, Weyl's work was extended by Einstein and Schrodinger independently in the Einstein-Schrodinger non-symmetric field theory, which is widely regarded as the most advanced unified field theory based on classical physics \cite{AE}.

\subsection{Kaluza's Idea} 
\label{sec:ki}

Kaluza initially postulated that a fifth spatial dimension could be introduced into Einstein's equations \cite{kaluza}, with the metric parameterized by
\begin{equation}
g_{ab}= \left( \begin{array}{ccc}
g_{\mu\nu}+\kappa^2\phi^2 A_\mu A_\nu & \kappa\phi^2 A_\mu  \\
\kappa\phi^2 A_\nu & \phi^2  \end{array} \right) \ ,
\label{eq92}
\end{equation}
where $\mu$ and $\nu$  run over 0,1,2,3 and $a$ and $b$ run over 0,1,2,3,4. $g_{\mu\nu}$ is the four-dimensional metric tensor, $A_\mu$ is the electromagnetic vector potential and $\phi$ is a scalar potential. $\kappa$ is a constant which can be scaled to ensure the correct multiplicative factors are included. The corresponding action of the system is
\begin{equation}
S^5=-\int d^5x \sqrt{\tilde{g}}R^{(5)} \ ,
\label{eq93}
\end{equation}
where $\tilde{g}$ is the five dimensional determinant of the metric and $R^{5}$ is the five dimensional Ricci scalar. To derive Kaluza's theory, we work with a model where we set $T_{ab}=0$. The motivation for this approach was Kaluza's belief that the higher dimensional universe is empty, and that matter in four dimensions is a manifestation of higher dimensional geometry. Thus, our five dimensional Einstein's equation reads
\begin{equation}
\tilde{R}_{ab}-\frac{1}{2}g_{ab}\tilde{R}=0
\label{eq94}
\end{equation}
Using the expression for the metric connection
\begin{equation}
\Gamma^a_{bc}=\frac{1}{2}g^{ad}(\partial_bg_{dc}+\partial_cg_{db}-\partial_dg_{bc}) \ ,
\label{eq95}
\end{equation}
and the Ricci tensor
\begin{equation}
R_{ab}=\partial_c\Gamma^c_{ab}-\partial_b\Gamma^c_{ac}+\Gamma^c_{ab}\Gamma^d_{cd}-\Gamma^c_{ad}\Gamma^d_{bc} \ ,
\label{eq96}
\end{equation}
we can solve eq.\ (\ref{eq94}) for the $\mu\nu$ components, the $\mu5$ and the $55$ components. We find
\begin{eqnarray}
G_{\mu\nu}&=&\frac{\kappa^2\phi^2}{2}T_{\mu\nu}-\frac{1}{\phi}\left[\nabla_\mu(\partial_\nu\phi)-g_{\mu\nu}\Box\phi\right] \ ,  \nonumber \\
\nabla^\mu F_{\mu\nu}&=&3\frac{\partial^\mu\phi}{\phi}F_{\mu\nu} \ ,  \nonumber \\ 
\Box\phi &=&\frac{\kappa^2 \phi^3}{4}F_{\mu\nu}F^{\mu\nu} \ ,
\label{eq99}
\end{eqnarray}
where 
\begin{equation}
F_{\mu\nu}=\partial_\mu A_\nu-\partial_\nu A_\mu
\label{eq100}
\end{equation}
is the Maxwell Tensor and
\begin{equation}
T_{\mu\nu}=\frac{1}{4}g_{\mu\nu}F_{\alpha\beta}-F_\mu ^\alpha F_\nu^\beta.
\label{eq101}
\end{equation}

The existence of the scalar field $\phi$ was an embarrassment for Kaluza, who originally set $\phi=1$ which, after setting $\kappa=4\sqrt{\pi G}$ reproduces Einstein's field equation
\begin{equation}
G_{\mu\nu}=8\pi G T_{\mu\nu} \ ,
\label{eq102}
\end{equation}
and Maxwell's equations
\begin{equation}
\nabla^\mu F_{\mu\nu}=0 \ .
\label{eq103}
\end{equation}
These remarkable results indicate, with the choice of metric parameterization given in  eq.\ (\ref{eq92}) and the higher dimensional Einstein tensor (\ref{eq94}), that four dimensional matter arises purely as an artifact of empty five-dimensional space-time. If, however, $\phi={\rm const}$ a Brans-Dicke type scalar tensor field theory is generated.

\subsection{The Klein Mechanism}
\label{sec:tkm}

Kaluza's idea suffered from a very obvious drawback. If there is a fifth dimension, where is it? In 1926, Oskar Klein suggested that the fifth dimension compactifies, so as to have the geometry of a circle of extremely small radius \cite{klein}. One way to envisage this spacetime is to imagine a hosepipe, Figure 4.1. It is only when we magnify the image, do we actually see the toroidal structure. From a long distance it looks like a one-dimensional line, but a closer inspection reveals that every point on the line is, in fact, a circle. Because the space has the topology $R^4\times S^1$, the higher dimensional periodicity allowed for a Fourier expansion of the field in the periodic dimension.
\begin{equation}
g_{\mu\nu}(\vec{x},y)=\sum_{n=-\infty}^\infty g^n_{\mu\nu}(x)e^{in\pi y/r},
\label{eq104}
\end{equation}
\begin{singlespace}
\begin{figure}[H]
\begin{center}
\includegraphics[width=200pt]{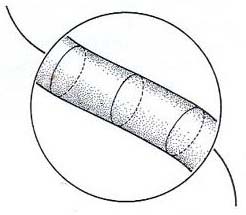}
\caption{\textit{Seen from a distance the thread appears one dimensional. Upon closer inspection we see a deeper structure.}}
\end{center}
\end{figure}
\end{singlespace}
\begin{equation}
A_{\mu}(\vec{x},y)=\sum_{n=-\infty}^\infty A^n_{\mu}(x)e^{in\pi y/r},
\label{eq105}
\end{equation}
\begin{equation}
\phi(\vec{x},y)=\sum_{n=-\infty}^\infty \phi(x)e^{in\pi y/r}.
\label{eq106}
\end{equation}
This form of the fields introduced the possibility of explaining the quantization of charge, even though the energy predictions were never verified. The `tower' of states also gave the mathematical potential to introduce the prediction of excited momentum states.

\subsection{Criticisms and Success of Kaluza-Klein Theory}
\label{sec:casokkt}

While KK theory is elegant in its simplicity, it is not without problems. One obvious criticism is that the theory is non-predictive, in that it does not extend Einstein's or Maxwell's theories, but merely synthesizes the formalism. A more serious criticism regards the introduction of the fifth dimension, which is seen as an artifical construct since our universe is clearly four dimensional. 

These criticisms aside, it is hard to not appreciate the elegance of KK theory and its potential for unifying gravity and electromagnetics.

\subsection{The Casimir Effect in Kaluza Klein Theory} 
\label{sec:tceikkt}

Some of the earliest work performed concerning the quantum dynamics of KK theories was carried out in the early 1980's by Appelquist and Chodos \cite{ac} \cite{ac1,Chodos1985}. Calculating the effective potential for the scalar field requires including the contributions from the massive $n\ne0$ modes. In five dimensions, the degrees of freedom described by eq.\ (\ref{eq104}) for the case of $n=0$ correspond to the graviton, the photon, and a scalar, which are all massless. In contrast, the five degrees of freedom corresponding to $g_{\mu\nu}^{(n)}(\vec{x})$ for $n\ne0$ are those of a massive spin 2 particle. Imposing the `cylindrical' gauge condition $g_{\mu5,5}(x)=0$ (all derivatives in the extra dimensions are zero) eliminates the unphysical massive modes associated with the vector and scalar fields.

Computation of the effective potential is analogous to the Casimir effect, where the `plates' correspond to the boundaries $y=0=2\pi R$, and the one loop effective potential is the zero-point energy associated with the excitations of the spin-two fields confined to the fifth dimension \cite{Brown1969}. Calculation of the Casimir energy begins with the path integral
\begin{equation}
Z=\int Dg_{\mu\nu}\delta{g_{\mu5,5}}e^{-S} \ ,
\label{eq106a}
\end{equation}
where $S$ is the action
\begin{equation}
S=-\frac{1}{16\pi G_D}\int d^Dx\sqrt{|g|}R^{(5)}
\label{eq106b}
\end{equation}
and $\delta{g_{\mu5,5}}$ is the Fadeev-Popov ghost determinant corresponding to the choice of the cylindrical gauge. Full details of the calculation are beyond the scope of this section, but can be found in \cite{ac}. After subtracting off the divergent component, we find
\begin{equation}
V_{\rm eff}(\phi)=-\frac{15}{4\pi^2}\zeta(5)\left(\frac{1}{2\pi R \phi^{1/3}}\right)^5 \ .
\label{eq106c}
\end{equation}
This term is interpreted to be the physical energy, which tends to contract the extra dimension. Eq.\ (\ref{eq106c}) contains no physical cutoff, and as such, there is no limit to the contraction. $V_{\rm eff}$ tends to $-\infty$ as $L_5 = R$ tends to zero. A similar lack of a cutoff arises in the Casimir effect; however, one expects that on physical grounds, a natural cutoff will appear. This cutoff is the Planck length $L_{\rm p}$, and Appelquist and Chodos demonstrate that their calculation of the one loop effective potential is only reliable if $L_5>>L_{\rm p}$.

There are many remarkable aspects to this result. Firstly, the fact that the spin-two field generates a potential in the extra dimension opens up the possibility of adding additional fields to investigate the possibility that the Casimir energy may, in fact, stabilize the extra dimension. Another tantalizing possibility is that the cosmological constant can be explained as an artifact of higher dimensional vacuum energy. This seminal work has opened up a rich field of reserarch in regarding the utility of the Casimir effect in cosmology and higher dimensional theories  \cite{Danos2008,Bailin1987,Duff1994}.

\section{String Theory}
\label{sec:st}

KK theory remained largely ignored and was considered somewhat obscure for the first half of the twentieth century, as were the speculations regarding additional spatial dimensions. However, the birth of string theory generated a renewed interest in the idea, largely due to string theory's promise of being a quantum theory of gravity. 

\subsection{History of String Theory}
\label{sec:host}

Unification is one of the main themes in the history of science and is a guiding principle in theoretical physics. Countless examples exist where diverse and seemingly unrelated phenomena have been understood in terms of a small number of underlying principles. In the 1940s, it was demonstrated that quantum mechanics and electromagnetism could be accurately described by quantum field theory, and by the 1970s the weak and strong nuclear forces could also be described using QFT.

The full theory, called the standard model of particle physics, is arguably the most succesful physical theory to date. The SM is defined by the gauge group $SU(3)_c\otimes SU(2)_L\otimes U(1_Y)$ and the left handed matter representations $L_{i} = \left( \begin{array}{c} \nu \\ e \end{array}\right)_{i}$, $\bar{e}_{i}$, $Q_{i} = \left( \begin{array}{c} u \\ d \end{array}\right)_{i}$, $\bar{u}_{i}$, $d_{i}$ (where $i$ runs from $1$ to $3$). This, together with the complex scalar Higgs particle $h = \left( \begin{array}{c} h^{+} \\ h^{0} \end{array} \right)$, successfully describes most phenomena associated with electromagnetic, weak, and strong interactions consistent with current experimental limits of several hundred GeV, corresponding to distances on the order of $10^{-18}{\rm m}$.  

Incorporating gravitational interactions into the SM using the methods of quantum field theory has proven to be one of the most challenging problems facing theoretical physics today. Other salient problems include explaining why the 48 spin-$\frac{1}{2}$ matter fields are organized into three distinct generations,\footnote{The left-handed anti-neutrino $N_i$ is now sometimes included as part of the SM. The count of 48 includes $N_i$ for each generation.} why there are 21 free parameters in the theory that are used to determine the particle dynamics appearing in the Lagrangian, and what generates the fermionic mass heirachy. 

In the late 80's, it was speculated that a more pleasing form for the SM would incorporate symmetry relating bosons and fermions. This supersymmetry, as it is called, requires that the masses of the fermion and boson pairs should be equal; however, if this were the case, particles and their associated super-particles should be produced with equal probability during particle accelerator experiments. As this is certainly not the case, supersymmetry must be \textit{broken} at the energies we have so far probed (up to 1 TeV).

String theory is currently the best candidate for a quantum theory of gravity. The divergences associated with point particle gravitational interactions are removed in string theory via the extended nature of the string. In fact, string theory has a number of compelling features, and upon construction of a relativistic quantum theory of one dimensional objects, one discovers:

\begin{itemize}
\vspace{-0.2in}
\begin{singlespace}
	\item \textit{Gravity.} From its birth string theory includes a prediction of closed spin 2 particles associated with the graviton. 
	\item \textit{Grand Unification.} String theories lead to gauge groups large enough to include the standard model, $SU(3)_c\otimes SU(2)_L\otimes U(1_Y)$.
	\item \textit{Extra Dimensions.} A fascinating feature of string theory is the prediction that additional dimensions of space exist.
	\item \textit{Chiral Gauge Couplings.} String theory allows chiral gauge couplings.
	\item \textit{Supersymmetry.} Ten dimensional string theory requires spacetime supersymmetry.
	\item \textit{A single adjustable parameter.} String theory contains only one free parameter in the string Lagrangian: namely, the string tension.
\end{singlespace}
\end{itemize}

One of the central themes of this dissertation is the issue of higher dimensional stability. Another later theme investigates string model building. For this reason, string theory plays a supportive role in this dissertation, and is worthy of deeper discussion. In the next section, we review the simplest of string theories and also the origin of the extra dimensions.

\subsection{Fundamentals of Bosonic String Theory}
\label{sec:fobst}

The original version of string theory is bosonic string theory, which is by far the simplest of all the string theories. Although it does have many attractive qualities, it also contains in particular two features which render it unphysical. Firstly, it contains a state which has a negative mass-squared, which we call a tachyon. Tachyons travel faster than light, which conflicts with much of what we understand about physics. The second unattractive feature is that it cannot describe fermions. Even with these features considered, bosonic string theory is extremly useful, as it helps us to develop some of the mathematical formalism required to understand theories which are more physical. The remainder of this section is devoted to the development of some of the formalisms of the theory, and also an explanation of where the addional spatial dimensions arise. 

We first consider a string propagating in flat space, parameterized by $X^{\mu}$.  The index $\mu$ runs over the dimensions of space-time. The string sweeps out a \textit{world-sheet}, defined by a space-like coordinate $\sigma$, and a time-like coordinate $\tau$. These coordinates do not represent physical coordinates in spacetime, but locate points on the string itself. It is convenient to map the world-sheet coordinates to the upper-half complex plane which we call the Teichm\"uller space: $z=\exp\left[\sigma + i \tau\right]$ and $\bar{z}=\exp\left[\sigma - i \tau\right]$. By convention, derivatives with respect to world-sheet coordinates are denoted $\partial \equiv \frac{\partial}{\partial z}$ and $\bar{\partial} \equiv \frac{\partial}{\partial \bar{z}}$. In the case of a closed string we have a right-moving coordinate $z$, and a left-moving coordinate $\bar{z}$.

To construct the action for the string, we proceed in a similar way to a classical point particle which is dependent only on the area of the world sheet. This analogy allows us to construct the Nambu-Goto action \cite{Gasperini},
\begin{equation}
	S_{NG} = -\frac{1}{2 \pi \alpha'} \int_{W} d \tau d \sigma \left(-\det h_{ab}\right)^{\frac{1}{2}},
\label{eq106d}
\end{equation}
where $h_{ab}$ is the induced metric of the world-sheet and is given by
\begin{equation}
	h_{ab} = \partial_a X^{\mu} \partial_b X^{\nu} g_{\mu \nu} = \partial_a X^{\mu} \partial_b X_{\mu} \ ,
\end{equation}
where $a,b$ run over the world-sheet coordinates $\sigma, \tau$. $\alpha'$ is the string coupling, related to the tension in the string by $T=\frac{1}{2 \pi \alpha'}$.

This action has two symmetries from which we can infer a new metric which will avoid the mathematically dangerous square root present in the NG action. These symmetries are:
\begin{itemize}
\vspace{-0.2in}
\begin{singlespace}
	\item the isometry group of flat space-time, the Poincar\'e group in $D$-dimensions, corresponding to translations and Lorentz transformations, and
	\item two-dimensional (world-sheet) diffeomorphism invariance, which tells us that the action does not depend on the manner in which we choose our coordinates.
\end{singlespace}
\end{itemize}
Both of these will ensure that we can always give our world-sheet a Lorentz metric, called $\gamma$:
\begin{equation}
	\gamma_{ab} = \rm{diag}\left(- +\right).
	\label{eq106f}
\end{equation}
One can rewrite eq.\ (\ref{eq106d}) in terms of this Lorentzian metric:
\begin{equation}
	S_P = -\frac{1}{4 \pi \alpha'}\int_W d \tau d \sigma \sqrt{ \gamma}\gamma^{ab} \partial_a X^{\mu} \partial_b X_{\mu}.
	\label{eq106g}
\end{equation}
This action is called the Polyakov action, in which a new symmetry called Weyl invariance has become apparent.  This is a scale-invariance of the world-sheet metric, $\gamma_{ab} \rightarrow \Lambda(\sigma)\gamma_{ab}$. The action is also invariant under translations and infinitesimal Lorentz transformations. The Polyakov action is the action of $D$ bosonic fields living in two dimensions \cite{Gasperini}. The importance of these world `sheet bosons' will become apparent shortly. 

The string equations of motion can be found by varying the action with respect to $X^\mu$ and by imposing Neumann boundary conditions.\footnote{The open strings can also have Dirichlet boundary conditions, which lead to an important development in string theory: namely, the discovery of D(irichlet)-branes \cite{polchinski}.} We first define 
\begin{equation}
	S=\int_{\tau_1}^{\tau_2}d\tau\int_0^\pi d\sigma L(\dot{X},X'),
	\label{eq106h}
\end{equation}
and
\begin{equation}
	L(\dot{X},X')=-\frac{1}{4 \pi \alpha'} \sqrt{ \gamma} \gamma^{ab} \partial_a X^{\mu} \partial_b X_{\mu}.
\label{eq106i}
\end{equation}
Integrating by parts we obtain
\begin{eqnarray}
\delta S &=&\int_{\tau_1}^{\tau_2}d\tau \int_0^\pi d\sigma\frac{\partial L}{\partial(\partial_a X^\mu)}\delta\partial_aX^\mu \nonumber \\
&=&\int_{\tau_1}^{\tau_2}d\tau\int_0^\pi d\sigma(\partial_a\frac{\partial L}{\partial(\partial_aX^\mu)})\delta X^\mu +
\int_0^\pi d\sigma\left[\frac{\partial L}{\partial \dot{X^\mu}}\delta X^\mu\right]_{\tau_1}^{\tau_2} \nonumber \\
&+&\int_{\tau_1}^{\tau_2}(\frac{\partial L}{\partial {X'^\mu}}\delta X^\mu)_0^\pi \ .
\label{eq106k}          
\end{eqnarray}
The first term corresponds to the Euler-Lagrange equations, the second does not contribute because $\delta X=0$ at $\tau_1$ and $\tau_2$, and the third term represents the variational contribution at the spatial ends of the action integral, which is fixed by the boundary conditions. The action is stationary provided the string satisfies
\begin{equation}
\partial_a(\sqrt{-\gamma}\gamma^{ab}\partial_bX_\mu)=0, 
\label{eq106l}
\end{equation}
The equations of motion can be simplified using the three symmetries of the Polyakov action: 
\begin{itemize}
\vspace{-0.2in}
\begin{singlespace}
\item global invariance under Poincare transformations $X^\mu(\xi)\rightarrow \Lambda^\mu_\nu X^\nu(\xi)+a^\mu$;
\item local invariance under reparametrization of the world-sheet manifold, $\xi^a\rightarrow {\tilde{\xi}}^a(\xi)$;
\item conformal invariance under local transformations of the world-sheet metric $\gamma_{ab}\rightarrow \gamma_{ab}e^{(2\omega\xi)}$.
\end{singlespace}
\end{itemize}
Thanks to these symmetries, it is always possible to choose a conformal gauge where the world-sheet metric is characterized by a flat geometry $\gamma_{ab}=\eta_{ab}$. In this gauge, eq.\ (\ref{eq106l}) simplifies to the wave equation \cite{Gasperini}
\begin{equation}
\ddot{X}_\mu-X''_\mu=0,
\label{eq106n}
\end{equation}
Varying the NG action with respect to the metric gives the additional constraint
\begin{equation}
T_{ab}=\partial_aX^\mu\partial_bX^\nu\eta_{\mu\nu}-\frac{1}{2}\gamma_{ab}\gamma^{ij}\partial_iX^\mu\partial_jX^\nu\eta_{\mu\nu}=0.
\label{eq106o}
\end{equation}
We can now form the ``Virasoro constraints'': 
\begin{eqnarray}
\frac{1}{2}(T_{00}+T_{10})&=\frac{1}{4}\eta_{\mu\nu}(\dot{X^\mu}+X'^\mu)(\dot{X^\nu}+X'^\nu) \nonumber \\
\frac{1}{2}(T_{00}-T_{10})&=\frac{1}{4}\eta_{\mu\nu}(\dot{X^\mu}+X'^\mu)(\dot{X^\nu}-X'^\nu)
\label{eq106p}
\end{eqnarray}
For solving the equations of motion, it is convenient to introduce ``light-cone'' coordinates $\xi^{\pm}$ which are defined as
\begin{eqnarray}
\xi^{\pm} &= \tau\pm\sigma  \ ; \nonumber \newline
\partial_{\pm} &= \frac{1}{2}(\partial_\tau\pm\partial_\sigma) \ ; \nonumber \newline
\end{eqnarray}
\begin{eqnarray}
\tau &= \frac{1}{2}(\xi^++\xi^-) \ ; \nonumber \\
\sigma &= \frac{1}{2}(\xi^+-\xi^-) \ .
\end{eqnarray}
Using these coordinates, the wave equation can now be written simply as
\begin{equation}
\partial_+\partial_-X^\mu=0,
\label{eq106r}
\end{equation}
and is solved by a linear combination of left- and right-moving waves,
\begin{equation}
X^\mu(\xi)=X^\mu_L(\xi^+)+X^\mu_R(\xi^-).
\label{eq106s}
\end{equation}
In these coordinates the Virasoro constraints take the form
\begin{eqnarray}
T_{++}&=\partial_+X^\mu\partial_+X_\mu=0 \nonumber \\
T_{--}&=\partial_-X^\mu\partial_-X_\mu=0.
\label{eq106t}
\end{eqnarray}
We can now investigate the classical solutions of the bosonic open string in flat space.

\subsection{The Classical Open String}
\label{sec:tcos}

When the string does not have coincident ends, we must impose boundary conditions at each end of the string. One possibility is to impose $\frac{\partial L}{\partial X'^\mu}=0$ at both boundaries where 
\begin{equation}
L=\frac{1}{4\pi\alpha '}(\dot{X}^\mu \dot{X}_\mu-X'^\mu X'^\mu).
\label{eq106u}
\end{equation}
From this one obtains the Neumann boundary conditions
\begin{eqnarray}
X'^\mu|_{\sigma=0}   \nonumber \\
X'^\mu|_{\sigma=\pi}.
\label{eq106v}
\end{eqnarray}
These equations guarantee that no momentum flows off the string \cite{Gasperini}. The solutions to the open string equations of motion can be separated into left and right moving modes and expanded as a Fourier series.
\begin{eqnarray}
X^{\mu}_R\left(\xi^-\right)     &=     \frac{1}{2}x^{\mu}_0 +\alpha'p^\mu(\tau-\sigma)+ i\sqrt{\frac{\alpha'}{2}} \sum_{n\neq0}\frac{\alpha_n^\mu}{n}e^{-in(\tau-\sigma)}   \nonumber \\
X^{\mu}_L\left(\xi^+\right)     &=     \frac{1}{2}x^{\mu}_0 +\alpha'p^\mu(\tau+\sigma)+ i\sqrt{\frac{\alpha'}{2}} \sum_{n\neq0}\frac{\tilde{\alpha}_n^\mu}{n}e^{-in(\tau+\sigma)}.
\label{eq106x}
\end{eqnarray}
The open string solution which satisfies Dirichlet boundary conditions can be written in the form
\begin{equation}
X^\mu(\sigma,\tau)=x^\mu_1+(x^\mu_2-x^\mu_1)\frac{\sigma}{\pi}+i\sqrt{2\alpha'}\sum_{n\neq0}\frac{\alpha_n^\mu}{n}e^{i(n\tau)}sin(n\sigma),
\label{eq106y}
\end{equation}
where $x_1$ and $x_2$ are the positions of the ends of the strings. To impose the Virasoro constraints, we define the operator $\alpha_0^\mu=p^\mu\sqrt{2\alpha'}$ which allows us to include the $n=0$ mode in the Fourier series. The light cone gradients then become
\begin{equation}
\partial_{\pm}X^\mu=\sqrt{\frac{\alpha'}{2}}\sum_{n=-\infty}^{\infty}\alpha_n^\mu e^{in(\tau\pm\sigma)}
\label{eq106z}
\end{equation}
Imposing the boundary conditions $X'^\mu(-\sigma)=-X'^\mu(\sigma)$ and $\dot{X}^\mu(-\sigma)=\dot{X}^\mu(\sigma)$ the solution becomes periodic with a period of $2\pi$. The Virasoro functional can be defined on the extended interval $[-\pi,\pi]$ and gives the constraint
\begin{eqnarray}
L_m &=&\frac{1}{2\pi\alpha '}\int_\pi^\pi d\sigma T_{++}e^{im(\tau+\sigma)} \nonumber \\
 &=&\frac{1}{2\pi\alpha'}\int_\pi^\pi d\sigma\partial_+X^\mu\partial_+X_\mu e^{im(\tau+\sigma)} \nonumber \\
 &=&\frac{1}{4\pi}\int_\pi^\pi d\sigma\sum_n \sum_k \alpha_n^\mu \alpha_{k\mu}e^{-i(\tau+\sigma)(n+k-m)} \ , 
\end{eqnarray}
which can finally be expressed in terms of raising and lowering operators
\begin{eqnarray}
L_m&=\frac{1}{2}\dsum_{n=\infty}^\infty \alpha_{m-n}^\mu\alpha_{n\alpha} \nonumber \\
   &=0 \ .
\label{eq106z1}
\end{eqnarray}

\subsection{Quantization and the Open String Spectrum}
\label{sec:qatoss}

We may quantize the bosonic string by promoting the classical variables to operators and by replacing the Poisson brackets with commutators according to $\{A,B\}\rightarrow i[A,B]$, and thus we obtain for the Fourier coefficients \cite{polchinski},
\begin{eqnarray}
	\left[\alpha_m^{\mu},\alpha_n^{\nu}\right]  &=  \left[\bar{\alpha}_m^{\mu},\bar{\alpha}_n^{\nu}\right]=m \delta_{m+n,0}\eta^{\mu \nu} \nonumber \\
	\left[x^{\mu},p^{\nu}\right]                &=  i \eta^{\mu \nu};\;\left[\alpha_m^{\mu},\bar{\alpha}_n^{\nu}\right]=0.
\end{eqnarray}
In the light cone gauge, the Virasoro functional can be written as:
\begin{eqnarray}
L_0&=\frac{1}{2}\sum_{n=-\infty}^{\infty}\alpha_{-n}^\mu \alpha_{n\mu}=\frac{1}{2}\alpha_0^\mu\alpha_{0\mu}+\frac{1}{2}\sum_{n=\ne 0}\alpha_{-n}^\mu \alpha_{n\mu} \nonumber \\ 
&=\alpha'p^\mu p_\mu+\frac{1}{2}\sum_{n\ne 0}(2\alpha_{-n}^+\alpha_n^--\alpha_{-n}^i\alpha_n^i) \nonumber \\
&=\alpha'p^\mu p_\mu-\frac{1}{2}\sum_{n=1}^\infty\left(\alpha_{-n}^i\alpha_n^i+\alpha_{n}^i\alpha_{-n}^i\right) \ .
\label{eq106z2}
\end{eqnarray}
By applying the commutation rule $[\alpha_n^i,\alpha_{-n}^j]=n\delta^{ij}$, for $n>0$ we obtain
\begin{equation}
L_0=\alpha'p^\mu p_\mu-\frac{1}{2}\sum_{n=1}^\infty\left(2\alpha_{-n}^i\alpha_n^i+(D-2)n\right)
\label{eq106z3}
\end{equation}
We can see that the last term in equation eq.\ (\ref{eq106z3}) is divergent; however, we can use the definition of the Riemann $\zeta$-function;
\begin{equation}
\zeta(x)=\sum_{n=1}^\infty\frac{1}{n^x},
\label{eq106z4}
\end{equation}
and so this term becomes $\zeta(-1)=-\frac{1}{12}$ which gives us
\begin{equation}
L_0=\alpha'p^2-\sum_{n=1}^\infty N_n+\frac{D-2}{24}
\label{eq106z5}
\end{equation}
where the number operator $N_n=\alpha_{-n}^i\alpha_{n}^i$. The classical constraint $L_0=0$ for an open string corresponds to the mass-shell condition H=0 from which we determine the mass spectrum of the open string states as
\begin{equation}
\alpha'M^2=\sum_{n=1}^\infty N_n -\frac{D-2}{24}
\label{eq106z6}
\end{equation}
This condition is only consistent in a fixed number of dimensions $D$. The ground state of this theory corresponding to $N_n=0$ for all n is associated with the mass
\begin{equation}
M^2=-\frac{D-2}{24\alpha'}
\label{eq106z7}
\end{equation}
If $D>2$ one discovers that $M^2<0$; this level describes a tachyon, which indicates a pathology of the theory.

The first excited level is associated with the mass
\begin{equation}
\alpha'M^2=\frac{1}{\alpha'}\left(1-\frac{D-2}{24}\right)
\label{eq106z8}
\end{equation}
The spectrum of the open bosonic string is only compatible with a Lorentz invariant description of physical states if the vector at level $N=1$ is characterized by the condition $M^2=0$, which implies that the theory \textit{must} be formulated in a space-time with the critical number of dimensions:
\begin{equation}
D=26
\label{eq106z9}
\end{equation}
This remarkable result demonstrates why bosonic string theory is formulated in 26 dimensions. A parallel argument proves that D$=10$ is required for supersymmetric strings.

\subsection{The Five String Theories and M-Theory}
\label{sec:tfstamt}

D$=10$ supersymmetric string theory can be classified into one of five theories: Type I, Type IIA, Type IIB, SO(32) and $E_8 \times E_8$, each of which has its own descriptive territory. Type I string theory is a theory of unoriented open and closed strings which has been found to be particularly useful in M-theory extensions of string model building. Type II string theories model oriented closed strings, which can be classified into chiral and non-chiral. The non-chiral Type IIA can be distinguished from the chiral Type IIB theories by the massless Ramond-Ramond sector states.  Bosons present in Type IIA include a Maxwell field $A_{\mu}$ and a three-index antisymmetric gauge field $A_{\mu \nu \rho}$.  In contrast, Type IIB theories contain a scalar field A, a Kalb-Ramond field $A_{\mu \nu}$, and a totally antisymmetric gauge field $A_{\mu \nu \rho\sigma}$. The final two theories are the heterotic string theories, SO(32) and $E_8 \times E_8$. The term heterotic refers to the combination of a left-moving open bosonic string with a right-moving open superstring. A consequence of this feature is that the theory is effectively set in a ten-dimensional space-time.

It was discovered in the late 80's by Witten that the five theories could be transformed into each other by using what are called dualities \cite{Horava1996}. T-duality relates models with different radii of compactified dimensions, and S-duality relates models with inverse coupling strengths. The resulting theory connecting all 5 string theories is called M-theory and is believed to be the underlying fundamental theory whose low energy limits are ten-dimensional string theory and eleven-dimensional supergravity. 

\section{Large Extra Dimensions}
\label{sec:led}

Models with large extra dimensions have enjoyed a revived interest in physics. This began with the Arkani-Hamed, Dimopoulos and Dvali (ADD) proposal to lower the quantum gravity scale to the 10 to 100 TeV scale (accessible to the next generation of particle accelerators) by embedding the standard model fields in a 3+1 dimensional brane existing in a higher dimensional bulk spacetime \cite{add,add1}. Gravity is free to propagate in the bulk, which effectively dilutes the strength of gravity. This idea was inspired by M-theory, where it was recognized that the scale of quantum gravity could be lowered from the Planck scale to the GUT scale \cite{W1,hw,l,aadd}.

\noindent The assumptions underlying the ADD model are:

\begin{itemize}
\vspace{-0.2in}
\begin{singlespace}
\item n-extra dimensions compactified on a torus with volume $V_n=(2\pi r)^n$.
\item Standard Model fields are localized to the brane.
\item Gravity can propagate in the bulk.
\item There is no cosmological constant in the bulk or on the boundary
\item The brane is stiff.
\end{singlespace}
\end{itemize}
The action for this model can be broken into
\begin{equation}
S_{\rm tot}=S_{\rm bulk}+S_{\rm brane}
\label{eq106d1}
\end{equation}
Focusing on the bulk action, we have the higher dimensional Einstein-Hilbert action, as in the KK model studied earler.
\begin{equation}
S_{\rm bulk}=-\frac{1}{2}\int d^{4+n}x\sqrt{-g^{4+n}}\tilde{M}^{n+2}\tilde{R} \ ,
\label{eq106d2}
\end{equation}
where $\tilde{M}$ is the n-dimensional Planck mass and $\tilde{R}$ is the $4+n$ dimensional Ricci scalar. We now integrate out the extra dimensions of the action.
\begin{eqnarray}
S_{\rm bulk}  &=   -\frac{1}{2}\tilde{M}^{n+2}\int d^{4}x\int d\Omega_nr^n\sqrt{-g^{(4)}}R^{(4)} \nonumber \\
          &=   -\frac{1}{2}\tilde{M}^{n+2}(2\pi r)^n\int d^{4}x\sqrt{-g^{(4)}}R^{(4)}.
\label{eq106g1}
\end{eqnarray}
We can see from this equation that what we perceive as the Planck scale is, in fact a quantity that is $\textit{derived}$ from a more fundamental quantum gravity scale and the volume of the extra dimensions, 
\begin{equation}
M_{\rm Pl}^2=(2\pi r)^n\tilde{M}^{n+2} \ .
\label{eq106h2}
\end{equation}
This remarkable result is one possible solution to the heirachy problem regarding the apparent weakness of gravity. Physically, the graviton is diluted across the bulk with a diminished intersection with the SM brane. At large distances gravity behaves according to the familiar $\frac{1}{r^2}$ relation; however, close to length scales on the order of the extra dimensions, one can expect to see deviations from this.

\subsection{Constraints on Deviations From Newtonian Gravity}
\label{sec:codfng}

Any deviations from Newtonian gravity at short distances will provide compelling evidence for the existence of extra dimensions. Gravity, however, is the least accurately measured of all the forces due to its weakness. There have been a number of experiments testing for deviations from Newtonian gravity. For a good review of these, see \cite{ahn}. The strongest constraints come from the Eot-Wash experiment \cite{hkhagss}, which shows no deviations from Newtonian gravity down to 200 microns.

A fairly straightforward calculation allows us to estimate the size of the largest extra dimensions based on the formula
\begin{equation}
r=\left(\frac{1}{2\pi}\frac{M_{\rm Pl}^2}{\tilde{M}^{n+2}}\right)^{1/n}.
\label{eq106i1}
\end{equation}
Clearly, the size of the extra dimensions (assuming they are all equivalent) depends on the fundamental Planck scale $\tilde{M}$. The lowest value this could take is $\tilde{M} \approx 1 \ TeV$ which is a scale up to which we `trust' the SM. Using this, we obtain estimations on the size of the extra dimensions (Table 4.1).

It is clear that, according to the constraints discussed above, $n=1$ is ruled out by solar system tests of Newtonian gravity. From these naive calculations and experimental validation we can see that $n=3$ is the minimum number of extra dimensions.\footnote{It is interesting to point out that, on galactic scales, possible evidence for deviations from Newtonian gravity may be seen in the form of anomalous galactic rotation curves. This is the main motivation for models which introduce dark matter; for good reviews see \cite{jkg} \cite{o} \cite{bhs}.}
\vspace{0.20in}
\vspace{0.20in}
\begin{table}[ht]
\caption{\textit{Table illustrating the predicted size of the extra dimensions.}}
\vspace{0.2in}
\centering 
\begin{tabular}{c c} 
\hline 
Number of extra dimension & r (m) \\ [0.5ex] 
\hline 
$n=1$         &                   $\approx 10^{12}$         \\
$n=2$         &                   $\approx 10^{-3}$         \\
$n=3$         &                   $\approx 10^{-8}$          \\
$\dot{\dot{\dot{}}}$   &          $ $                       \\
$n=6$         &                   $\approx 10^{-11}$        \\
\hline 
\end{tabular}
\label{table:xdr} 
\end{table}
\vspace{0.10in}

\section{Warped Extra Dimensions}
\label{sec:wed}

The idea that our universe can be modelled as a (mem)brane existing in a higher dimensional bulk spacetime has received a huge amount of attention in recent years \cite{russell2000, Maartens2001, Langlois2002, Rasanen2002, Deruelle2002, Toporensky2006, Binetruy1999, Khoury2001, Tziolas2007, GDvali, Binetruy2001}. It is possible that the brane energy density affects the spacetime curvature, and an approximation can be achieved by first considering a model where branes are located at the two ends of a periodic fifth dimension. To ensure stability of the model \cite{pl}, two branes are required to balance the bulk energy. To get a stable metric the effects of the brane on the spacetime must be compensated by a negative cosmological constant in the bulk. Thus, the fifth dimension can be considered a slice of AdS space bounded by flat branes, and the price of keeping the branes \textit{flat} is to introduce curvature into the fifth dimension. Such models are termed warped extra dimensions, and the first such mention of the idea was by Rubakov and Shaposhnikov \cite{rs} who came up with an innovative solution to the heirachy problem. They suggested that the vacuum energy of the matter fields on the brane could be almost cancelled by the bulk vacuum, which would leave a small but non-zero $\Lambda$ for the brane observer. Gogberashvili gave the first exact solution for a warped metric \cite{Gogberashvili}; however, the models of Randall and Sundrum are the best known of the warped extra dimensional models.

\subsection{Randall-Sundrum Model}
\label{sec:trsm}

The Randall-Sundrum (RS1) model \cite{RS1} proposes a novel geometrical solution to the hierachy problem. The hierachy problem questions why gravity is so much weaker than the weak force (which is $10^{32}$ stronger), and why the Higgs boson is so much lighter than the Planck mass. The Higgs boson is one component of the SU(2) doublet Higgs field, which is a scalar field that permeates all space. The non-zero vacuum expectation value (VEV) of the Higgs field gives mass to all the elementary particles. The non-zero VEV spontaneously breaks the electroweak gauge symmetry, which is referred to as the \textit{Higgs Mechanism}. This is arguably the simplest mechanism that is capable of giving mass to the gauge bosons. Although the SM does not predict the mass of the Higgs boson, if it has a mass between 115 and 180 GeV then the standard model is valid at scales all the way up to the Planck mass \cite{Cheng}.

One would expect that the square of the Higgs mass would make the mass of the Higgs enormous unless there is a fine-tuning cancellation between the quadratic radiative corrections and the bare mass. Supersymmetry is one possible solution to this problem, whereby the quantum corrections arising from the supersymmetric partner to the Higgs provides equal magnitude but opposite sign contributions to the mass. The RS1 model also provides a nice solution to this problem by proposing the existence of a \textit{warp factor} which appears in the metric and has the effect of diluting the strength of the Higgs, if its field is localized near the visible brane.

In the RS1 setup, the Standard model fields are not confined to one of two 3-branes which lie at the endpoints (i.e., fixed points) of an $S^1/\IZ_2$  orbifold, except for the Higgs field. One of the branes physically corresponds to `our' universe and is sometimes referred to as the IR or `visible' brane. The closer a SM field is to the visible brane, the greater its coupling to the Higgs and therefore the greater the mass.

The second brane is the UV or `hidden' brane. 
The line element in RS1 is described by the metric
\begin{equation}
ds^2=e^{-2kr_c|\varphi|}\eta_{\mu\nu}dx^\mu dx^\nu - r_c^2 d\varphi^2,
\label{eq108}
\end{equation}
where the points $(x^\mu,\varphi)$ and $(x^\mu,-\varphi)$ are identified with each other, $x^\mu$ are the standard four dimensional coordinates and $|\varphi|\leq\pi$. The exponential factor is referred to as the warp factor and is an appealing feature in the RS1 model, as it can generate a TeV mass scale from the Planck scale in the higher dimensional theory, and reduce the effective gravitational strength on the visible brane through the supression factor $e^{2kr_c|\varphi|}$, while retaining a bulk width that is only a couple of orders of magnitude above the Planck scale. 

\subsection{Theory of Warped Extra Dimensions}
\label{sec:rss}

The classical action for the RS1 setup is 
\begin{equation}
S=S_{\rm gravity}+S_{\rm vis}+S_{\rm hid}
\label{eq111a}
\end{equation}
where
\begin{equation}
S_{\rm gravity}=\int d^4x\int_{-\pi}^\pi d\phi\sqrt{-G}(2M^3R-\Lambda)
\label{eq111b}
\end{equation}
where G is the five dimensional metric:
\begin{eqnarray}
S_{\rm vis}  &=  \int d^4x \sqrt{g_{\rm vis}}(\mathcal{L}_{\rm vis}-V_{\rm vis}) \nonumber \\
S_{\rm hid}  &=  \int d^4x \sqrt{g_{\rm hid}}(\mathcal{L}_{\rm hid}-V_{\rm hid})
\label{eq111d}
\end{eqnarray}
where $\mathcal{L}$ is the lagrangian for any matter fields on the hidden or visible brane, and $V$ is the constant vacuum energy which has been separated out. The five-dimensional Einstein equation for this action is
\begin{eqnarray}
\sqrt{-G}(R_{MN}-\frac{1}{2}G_{MN}R)&=-\frac{1}{4M}  [ \Lambda\sqrt{-G}G_{MN} \nonumber \\
&+ V_{\rm vis}\sqrt{-g_{\rm vis}}g^{\rm vis}_{\mu\nu} \delta^\mu_M\delta^\nu_N\delta(\phi-\pi) \nonumber \\
&+ V_{\rm hid}\sqrt{-g_{\rm hid}}g^{\rm hid}_{\mu\nu}\delta^\mu_M\delta^\nu_N\delta(\phi) ] . 
\label{eq111e}
\end{eqnarray}
The three terms on the right side represent a bulk cosmological constant, a brane tension localized to the visible brane, and a brane tension localized to the hidden brane. The solution satisfying the ansatz that four dimensional Poincare invariance is preserved is of the form
\begin{equation}
ds^2=e^{-\sigma(\phi)}\eta_{\mu\nu}dx^\mu dx^\nu+r^2d\phi^2 \ ,
\label{eq111e1}
\end{equation}
where r is the radius of the compactified fifth dimension before any orbifolding of the extra dimension. Using a code constructed in Mathematica, we solved Einstein's equations to obtain for the Einstein tensor
\begin{equation}
G_{11}=\frac{e^{-2\sigma}}{r^2}(-6\sigma'^2+3\sigma'') \ ,
\label{eq111g}
\end{equation}
and
\begin{equation}
G_{55}=6\sigma'^2 \ .
\label{eq111g1}
\end{equation}
Using the 55 component and inserting the appropriate value for the energy momentum tensor using eq.\ (\ref{eq111e}) we obtain 
\begin{equation}
6\sigma'^2=\frac{1}{4M^3}\Lambda r_c^2.
\label{eq111g2}
\end{equation}
Rearranging we find
\begin{equation}
\frac{6\sigma'^2}{r_c^2}=\frac{-\Lambda}{4M^3} \ ,
\label{eq111f}
\end{equation}
where 
\begin{equation}
\sigma'=\frac{d\sigma}{d\phi} \ .
\label{eq111g3}
\end{equation}
Integrating with respect to $\phi$ we obtain
\begin{equation}
\sigma=r_c|\phi|\sqrt{-\Lambda}{24M^3} \ ,
\label{eq111h}
\end{equation}
where the integration constant is omitted because it just amounts to a rescaling of $x^\mu$. For this solution to be real, we must have $\Lambda<0$ indicating that the spacetime between the two branes is a slice of Anti-deSitter space. 

Now solving for the $G_{11}$ component in a similar fashion we obtain
\begin{equation}
\frac{3\sigma''}{r_c^2}=\frac{V_{\rm hid}}{4M^3r_c}\delta(\phi)+\frac{V_{\rm vis}}{4M^3r_c}\delta(\phi-\pi).
\label{eq111i}
\end{equation}
If we now use our expression for $\sigma$ in eq.\ (\ref{eq111h}) and insert it into eq.\ (\ref{eq111i}), we only find a solution if $V_{hid}$ and $V_{vis}$ are related through a scale $k$,
\begin{equation}
V_{\rm hid}=-V_{\rm vis}=24M^3k \ ,
\label{eq111j}
\end{equation}
where
\begin{equation}
\Lambda=-24M^3k^2.
\label{eq111k}
\end{equation}
From this we can see that the brane tensions of the visible and hidden branes are equal in magnitude, but opposite in sign, indicating a perfect balancing. From this, we obtain the final solution to the bulk metric.
\begin{equation}
ds^2=e^{-2kr_c|\phi|}\eta_{\mu\nu}dx^\mu dx^\nu+r^2d\phi^2.
\label{eq111l}
\end{equation}
The physical implications for this metric are profound. Consider, for example, the action of a Higgs field \cite{RS1}, 
\begin{equation}
S_{\rm vis}=\int d^4x\sqrt{-g_{\rm vis}}\left(g^{\mu\nu}_{\rm vis}D_\mu H^\dagger D_\nu H-\lambda(|H|^2-v_0^2)^2\right),
\label{eq109}
\end{equation}
where the $\lambda$ is a lagrange multiplier which ensures $|H|^2=v_0^2$, fixing the mass parameter of the Higgs. If we now substitute eq.\ (\ref{eq111l}) into this action, we find
\begin{equation}
S_{\rm vis}=\int d^4x\sqrt{-g_{\rm hid}}e^{-4kr_c\pi}\left(g^{\mu\nu}_{\rm hid}e^{2kr_c\pi}D_\mu H^\dagger D_\nu H-\lambda(|H|^2-v_0^2)^2\right),
\label{eq110}
\end{equation}
If we now redefine $H\rightarrow e^{kr_c\pi}H$ we see, 
\begin{equation}
S_{\rm eff}=\int d^4x\sqrt{-g_{\rm hid}}\left(g^{\mu\nu}_{\rm hid}D_\mu H^\dagger D_\nu H-\lambda(|H|^2-e^{2kr_c\pi}v_0^2)^2\right),
\label{eq111}
\end{equation}
The remarkable feature of the RS1 model is that a field with mass $m_0$ on the $\varphi=0$ (hidden) brane will have a reduced physical mass of $m\approx {e}^{-2\pi k r_c} m_0$ on the $\varphi=\pi$ (hidden) brane. Typically $2\pi kr_c\approx 12$. In this model, the branes themselves remain static and flat.
\begin{singlespace}
\begin{figure}[H]
\begin{center}
\includegraphics[width=220pt]{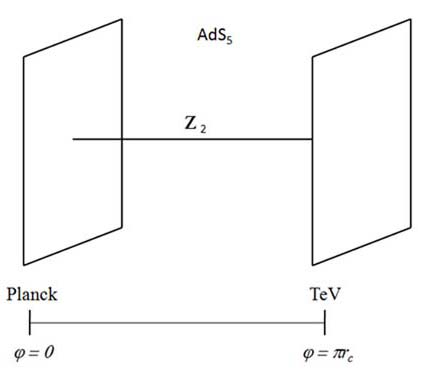}
\caption{\textit{In the Randall Sundrum model the two branes are located at the fixed ends of an $S^1/Z_2$ orbifold and the choice of metric offers a unique solution to the Heirachy problem.}}
\end{center}
\end{figure}
\end{singlespace}
In the RS1 model the radius $r_c$ is associated with the VEV of a massless four dimensional scalar field. This modulus field has zero potential and $r_c$ is thus not determined in the model. For the scenario to be relevant, a mechanism is required to stabilize the $r_c$. In Section 6 we propose a new method to stabilize $r_c$ using the Casimir effect \cite{oc3}.


\section{The Casimir Effect in Cosmology and Theories with Extra Dimensions}
\label{sec:tceicatwed}

As we have discovered in Chapters 1 and 2, the quantum vacuum energy and $\Lambda$ cancel to a physically unrealistic high degree creating an extremely small, but non-zero $\rho_{\rm eff}$. Compactified extra dimensions introduce non-trivial boundary conditions to the quantum vacuum and Casimir-type calculations become important when finding the resulting vacuum energy. Any significant reduction in $\rho$ lessens the impact of this `cancellation coincidence' and, as we shall see in the remaining chapters, the additional freedom the extra spatial dimensions afford us provide us with this opportunity.

Although tantalizing, the study of extra dimensions opens up a whole new set of questions. For example, if extra dimensions exist and are hidden from us due to their compact nature, then what \textit{keeps} them small? Why do they not expand to be large like the three macroscopic dimensions we are familiar with, or conversely, why not perpetually shrink? 

By experimenting with different quantum fields within the compact extra dimension, it is possible to find configurations that stabilize them. We will see that Casimir energy ostensibly holds the promise of explaining both the smallness of $\rho_{\rm eff}$ \textit{and} the mystery of higher dimensional stability. Of course, there are many details that remain to be understood, but higher dimensional quantum vacuum energy is a compelling candidate in explaining some of these problems in physics. 

In the next chapter we will discuss a novel modification of the higher dimensional quantum vacuum energy that is introduced when we allow Lorentz invariance to be broken in \textit{only} the hidden dimension. We will see that the additional properties that this proposal introduces will provide us with added freedom in stabilizing the extra dimensions and creating a smaller $\rho$, which will be investigated fully in Chapter 6.

\chapter{Lorentz Violating Fields}
\label{chap:lvf}

In this section we derive an original solution to the Casimir energy in the scenario of one additional spatial dimension in the presence of a field with a non-zero VEV in \textit{only} the fifth dimension. We find that we can generate an enhancement of the Casimir energy that is tuned by a single parameter. This will give us additional freedom in later chapters with our moduli stabilization models.

Recently, Carroll \cite{ct} investigated the role of Lorentz violating fields in hiding extra dimensions. A simple mechanism to implement local Lorentz violation is to postulate the existence of a tensor field with a non-zero expectation value which couples to standard model fields. The most elementary realization of this is to consider a single spacelike vector field with a fixed norm. This field selects a `preferred' frame at each point in spacetime, and any fields that couple to it will experience a local violation of Lorentz invariance. One novel feature of this research was the demonstration that it allowed different spacings in the KK towers \cite{ct}. The model worked in a five dimensional flat spacetime, and the Lorentz violating field is a spacelike five-vector with VEV $u^a=(0,0,0,0,v)$.

Although the scenario clearly violates Lorentz invariance, it has been well-explored in the literature \cite{k0,kp}, and tests of Lorentz invariance violations have recently received a lot of attention for a possible role in cosmology \cite{mnpss,fhl,kdm,acnp,b,fnpa}. It was demonstrated by Kostelecky that spontaneous Lorentz breaking may occur in the context of some string theories \cite{ks}. In the standard model, spontaneous symmetry breakdown occurs when symmetries of the Lagrangian are not obeyed by the ground state of the theory. This occurs when the perturbative vacuum is unstable. The same ideas apply in covariant string theory which, unlike the standard model, typically involves interactions that could destabilize the vacuum and generate nonzero expectation values for Lorentz tensors (including vectors).\cite{ck}

Recently, the authors calculated the Casimir energy for the case of a scalar field coupled to a field localized only to the fifth dimension \cite{oc}. One novel feature of the setup was the demonstration that it allowed different spacings in the Kaluza Klein towers \cite{ct}, and consequentially the generation of an $\textit{enhanced}$ Casimir energy. By including this feature in our stabilization scenario, we will have increased freedom to generate the minimum of potential, which will be controlled by a single parameter which encodes the ratio of the 5th dimensional field VEV to the mass parameter. In this section, we discuss motivations for this study, and show our calculations and results.

\section{Motivations}
\label{sec:mot}

Four-dimensional Lorentz invariance is a basic ingredient in the SM (and all local relativistic quantum field theories); this has been verified by numerous experiments \cite{k0}. However, motivation does exist for deeper study into possible Lorentz violation along the extra compact dimensions, or in the 3+1 spacetime at sufficiently small distance scales. One reason is that quantitive statements regarding the degree with which nature preserves Lorentz symmetry are expressed within a framework which \textit{allows} for violations \cite{kp}. Another compelling reason is that the sensitivity of current tests implies that highly supressed Lorentz violations might arise at scales well beyond SM physics.

It has been shown that spontaneous Lorentz breaking may occur in the context of some string theories \cite{ks}. In the SM, spontaneous symmetry breakdown occurs when symmetries of the Lagrangian are not obeyed by the ground state of the theory. This occurs when the perturbative vacuum is unstable. The same ideas apply in covariant string theory which, unlike the SM, typically involve interactions that could destabilize the vacuum and generate nonzero expectation values for Lorentz tensors (including vectors) \cite{ck}.

In \cite{ct}, spacetime is modelled as flat and five dimensional, with the Lorentz violating field taking the form of a spacelike five-vector $u^a=(0,0,0,0,v)$. The fifth dimension is compactified on a circle. We first define an antisymmetric `Lorentz Violating Tensor' $\xi^{ab}$  in terms of $u^a$
\begin{equation}
\xi_{ab}=(\nabla_a u_b -\nabla_b u_a)
\label{eq113},
\end{equation}
we can form the following action: 
\begin{equation}
S=M_*\int d^5x \sqrt{g} \left[-\frac{1}{4}\xi_{ab}\xi^{ab}-\lambda(u_au^a-v^2)+\sum_{i=1} \mathcal{L}_i\right].
\label{eq114}
\end{equation}
Here the indices $a, b$ run from 0 to 4. $\lambda$ is a Lagrange multiplier which ensures $u^au_a=v^2$, and we take $v^2>0$. The $\mathcal{L}_i$ can represent various interaction terms. Here we only investigate interactions with a scalar field. This form of the Lagrangian ensures the theory remains stable and propagates one massless scalar and one massless pseudoscalar \cite{dgb}. Of interest is the KK tower generated by the Lorentz violating field in the compact dimension in the context of moduli stabilization. 

Clearly, all fields which propagate in the bulk will give Casimir contributions to the vacuum energy, and a natural extension of the study of Lorentz violating fields is whether these could provide an energy spectrum which stabilized the extra dimension. In the next section, we calculate the effective vacuum potential due to a Lorentz violating tensor field coupling with a scalar field with periodic boundary conditions. We will focus on the background geometry of the RS1 model, although the techniques employed here may be used in alternative geometries. 

\subsection{KK Spectrum with Lorentz Violating Vectors}
\label{sec:kksfpsiwlvv}

We consider a real scalar field $\phi$ coupled to a Lorentz violating spacelike five vector $u^a$ with a vacuum expection value in the compact extra dimension. The Lagrangian is \cite{ct}
\begin{equation}
\mathcal{L}_{\phi}=  \frac{1}{2} (\partial\phi)^2 -\frac{1}{2} m^2\phi^2-\frac{1}{2\mu^2_\phi}u^a u^b\partial_a\phi\partial_b\phi.
\label{eq116}
\end{equation}
The indices a and b run from 0 to 4. The mass scale $\mu_\phi$ is added for dimensional consistency. The background solution has the form $u^a=(0,0,0,0,v)$, which ensures four dimensional Lorentz invariance is preserved. 

Using the five-dimensional Euler-Lagrange equation
\begin{equation}
\partial_a \left(\frac{\partial \mathcal{L}}{\partial(\partial_a\phi)}\right)-\frac{\partial \mathcal{L}}{\partial \phi}=0 \ ,
\label{eq117}
\end{equation}
and plugging in for the Lagrangian we obtain
\begin{equation}
\partial_a\partial^a\phi-m^2\phi=\mu^{-2}_\phi\partial_a(u^au^b\partial_b\phi).
\label{eq118}
\end{equation}
The scalar can be expressed in momentum space as
\begin{equation}
\phi \propto e^{ik_ax^a}=e^{ik_\mu x^\mu}e^{ik_5y}
\label{eq119}
\end{equation}
where $\mu=0,1,2,3$. Calculating each term in the Euler Lagrange eq. (\ref{eq117}),
\begin{eqnarray}
\partial_a\partial^a\phi &=&\partial_\mu\partial^\mu\phi+\partial_y\partial^y\phi
\label{eq120}\\ 
                         &=&  -k_\mu k^\mu\phi-k_5 k^5\phi .
\label{eq121} 
\end{eqnarray}
For the term involving the VEV of the Lorentz violating field, it is clear that the only nonzero index values are $a=b=5$; thus, we quickly obtain
\begin{eqnarray}
\mu^{-2}_\phi\partial_a(u^au^b\partial_b\phi)&=&\mu^{-2}_\phi\partial_5(u^5u^5\partial_5\phi) \nonumber \\
&=&  \frac{v^2k_5^2}{\mu_\phi^2}\phi
\label{eq123}. 
\end{eqnarray}
where we have used the fixed norm constraint $u^au_a=v^2$ obtained from the equation of motion for $\lambda$. Choosing $v^2>0$ ensures that the vector will be timelike. We now compactify the fifth dimension on a circle of radius R ($k_5=\frac{n\pi}{R}$), with $\IZ_2$ symmetry which identifies $u^a\rightarrow -u^a$. 

The orbifolding will not effect the coupling of the \textit{scalar} field to the Lorentz violating field, but would significantly effect the couplings for more complex fields (fermions for example), and so we include this procedure for completeness and for its relevance in the RS paradigm. The effect of the orbifolding for the scalar is essentially to remove all odd (even) scalar modes under $y\rightarrow -y$ for even (odd) periodicity scalar fields. In each case, this amounts to eliminating half of the modes in summation over $n$ in eq.\ (\ref{eq7}) for example \cite{pp}. Thus, for both periodicities, the net result is a reduction of the Casimir energy be a factor of $\half$.    

If we now impose periodic boundary conditions on the wave vector in the fifth dimension;
\begin{equation}
k_5=\frac{n\pi}{R} \ ,
\end{equation}
and substitute eqs.\ (\ref{eq121}) and \ (\ref{eq123}) into eq.\ (\ref{eq117}), we obtain
\begin{equation}
-k_\mu k^\mu=m^2+(1+\alpha_\phi^2)\left(\frac{n\pi}{R}\right)^2 \ ,
\label{eq124}
\end{equation}
where $\alpha_\phi=\frac{v}{\mu_\phi}$ is the ratio of the Lorentz violating VEV to the mass parameter.

We thus see that, as shown in \cite{ct}, with the addition of a Lorentz violating field the mass spectrum of the extra dimensional KK tower is modified by non-zero $\alpha_\phi$: 
\begin{equation}
m^2_{\rm KK}=k^2+(1+\alpha_\phi^2) \left( \frac{n\pi}{R} \right) ^2
\label{eq125}
\end{equation}
The value of $\alpha_\phi$ depends on the choice of the mass scale, which should be on the order of the Planck scale.

\subsection{Scalar Field Coupled to a Lorentz Violating Vector}
\label{sec:ceoasfctalvv}

We now apply the results of the previous section to calculate the higher dimensional Casimir energy \cite{oc}. 

The Casimir energy due to the KK modes of a scalar field, obeying periodic boundary conditions compactified on $S^1$ and interacting with a Lorentz violating vector field, is  
\begin{equation}
E=\frac{1}{2}{\sum_{n=-\infty}^{\infty}}'\int\frac{d^4k}{(2\pi)^4}log\left( k^2+(1+\alpha_\phi^2)\left(\frac{n\pi}{R} \right)^2         \right), 
\label{eq126}
\end{equation}
where the prime on the summation indicates that the $m=0$ term is omitted. We can rewrite the log as a derivative, and then, after a Mellin transformation, perform a dimensional regularization on the integral and the summation
\begin{eqnarray}
E &=&\frac{1}{2}\frac{\partial}{\partial s}|_{s=0}{\sum_{n=-\infty}^{\infty}}'\int\frac{d^4k}{(2\pi)^4}\left( k^2+\xi n^2 \right)^{-s} \nonumber \\
  &=&\frac{1}{2}\frac{\partial}{\partial s}\zeta^+(s) |_{s=0}
  \label{eq127},
\end{eqnarray}
where the periodic scalar function is defined as
\begin{equation}
\zeta^+(s)={\sum_{n=-\infty}^{\infty}}'\int\frac{d^4k}{(2\pi)^4}\frac{1}{\Gamma(s)}\int_0^\infty dte^{(k^2+\xi n^2)t}t^{s-1}. 
\label{eq129}
\end{equation}
Here we have made the substitution $\xi=\frac{\pi^2(1+\alpha^2_\phi)}{R^2}$ and used the identity
\begin{equation}
z^{-s}=\frac{1}{\Gamma(s)}\int_0^\infty dt e^{-zt}t^{s-1}.
\label{eq130}
\end{equation}
We first perform the k integral, 
\begin{equation}
\int d^4k e^{-k^2t}=\frac{\pi^2}{16t^2},
\label{eq131}
\end{equation}
and now calculate
\begin{equation}
\zeta^+(s)=\frac{\pi^2}{(2\pi)^4}\frac{1}{\Gamma(s)}{\sum_{n=-\infty}^{\infty}}'\int_0^\infty dt e^{-\xi n^2t}t^{s-3}.
\label{eq132}
\end{equation}
Making the substitution $x=\xi n^2 t$ gives us
\begin{eqnarray}
t=\frac{x}{\xi n^2}\quad{\rm and}\quad dt=\frac{dx}{\xi n^2}.
\label{eq133}\end{eqnarray}
Now substituting back into eq.\ (\ref{eq132}),
\begin{equation}
\zeta^+(s)=\frac{\pi^2}{(2\pi)^4}\frac{1}{\Gamma(s)}{\sum_{n=-\infty}^{\infty}}'\int_0^\infty \frac{dx}{\xi n^2} e^{-x}\left(\frac{x}{\xi n^2}\right)^{s-3}
\label{eq134}.
\end{equation}
We can express the $x$ integral in terms of the Gamma function
\begin{equation}
\zeta^+(s)=\frac{\xi^{2-s} \pi^2}{(2\pi)^4}\frac{\Gamma(s-2)}{\Gamma(s)}   {\sum_{n=-\infty}^{\infty}}'\frac{1}{n^{2s-4}}.
\label{eq135}
\end{equation}
We immediately recognise the infinite sum as the Riemann Zeta function, so we finally obtain
\begin{equation}
\zeta^+(s)=\frac{\xi^{2-s} \pi^2}{(2\pi)^4}\frac{\Gamma(s-2)}{\Gamma(s)}   \zeta(2s-4).
\label{eq135a}
\end{equation}
After expressing the Gamma functions as 
\begin{equation}
\frac{\Gamma(s-2)}{\Gamma(s)}=\frac{\Gamma(s-2)}{(s-2)(s-1)\Gamma(s-2)}, 
\label{eq136}
\end{equation}
plugging back into eq.\ (\ref{eq127}), and performing the derivative with respect to s evaluated at $s=0$ we obtain
\begin{equation}
E=-\frac{\pi^2}{2\pi^4} \left(    \frac{(1+\alpha^2_\phi)^2\pi^2}{R^2}\right)^2\zeta'(-4).
\label{eq137}
\end{equation}
However, the derivative of the zeta function is known to be
\begin{equation}
\zeta'(-4)=\frac{3}{4\pi^4}\zeta(5),
\label{eq138}
\end{equation}    
and so we find our final expression for the Casmir energy in an $S^1/\IZ_2$ orbifold for a scalar field with periodic boundary conditions coupled to a Lorentz violating vector field to be
\begin{equation}
E=-\frac{3(1+\alpha^2_\phi)^2}{64\pi^2}\frac{1}{R^4}\zeta(5).
\label{eq139}
\end{equation}
We see that the Casimir energy is proportional to $1/R^4$, where $R$ is the size of the extra dimension. We also see that the term $\alpha$ serves as a 
tuning parameter, whose value will adjust the Casimir energy density. We will see in Section 6 that this tuning gives us additional freedom in our models, and that the adjustment will play an important mechanism in moduli stabilization.

\subsection{Discussion of Results}
\label{sec:dor}

Here we have derived an \textit{original} expression for the Casimir energy for a scalar field which couples to a higher dimensional vector. We thus find that the Casimir energy contribution of a scalar field with periodic boundary conditons interacting with a Lorentz violating vector field remains attractive and tends to shrink the extra dimension. Thus, stabilization is not achieved with only scalars interacting with a Lorentz violating vector field. Note, however, that the expression for the effective potential takes into account the Casimir energy contribution from the bulk, but is incomplete because there can be additional contributions from the branes and other possible fields.

In the context of radius stabilization, we have calculated the one loop corrections arising from a scalar field with periodic boundary conditions interacting with a Lorentz violating vector field in the compactified extra dimension of the RS spacetime. The compactification scheme appears with enhanced sensitivity to the presence of periodic scalars interacting with Lorentz violating vectors (and tensors in general). In particular, the contributions are attractive, inducing the extra dimension to shrink in size. Thus, a net positive contribution to the Casimir force from additional fields is required for stabilization.

We will see in the next section that the parameter $\alpha_\phi$ gives us additional freedom that will help us generate stable minima of the potential when we also include phenomenologically viable fields such as the Higgs field and SM particles. Thus, $\alpha_\phi$ will serve as a fine-tuning parameter to aid us with the construction of viable stabilization models.



\chapter{Moduli Stability}
\label{chap:ms}

In this section, we investigate the role of Casimir energy as a mechanism for brane stability in five-dimensional models with the fifth dimension compactified on an $S^1/\IZ_2$ orbifold, which includes the Randall-Sundrum two brane model. We employ a $\zeta$-function regularization technique utilizing the Schwinger proper time method and the Jacobi's theta function identity to perform an \textit{original derivation} of the one-loop effective potential. 

We show that the combination of the Casimir energies of a scalar Higgs field, the three generations of SM fermions and one additional massive non-SM scalar in the bulk produce a non-trivial minimum of the potential. In particular, we consider a scalar field with a coupling in the bulk to a Lorentz violating vector particle localized to the compactified dimension. Such a scalar may provide a natural means of the fine-tuning needed for stabilization of the brane separation. Lastly, we briefly discuss the possibility that Casimir energy plays a role in generating the currently observed epoch of cosmological inflation by examining a simple higher-dimensional anisotropic metric.

\subsection{Background}
\label{chap:b}

Naively, one might expect an extra dimension to either contract to the Planck length or to inflate to macroscopic scales, and so the question of stabilization becomes important. Negative energy is a vital component to all realistic stabilization schemes \cite{adnv}. Negative-tension orientifold planes are the source of this negative energy in string theory, whereas in the SM, this source is the Casimir energy.\newpage

One attractive feature of the Casimir energy in stabilization schemes is that it is an inherent property of the quantum vacuum, and does not need to be added `by hand'. Additionally, the Casimir effect can easily be extended to regions of non-trivial topology \cite{aw,ke}, adding to its theoretical attractiveness. For example, on $S^1$, a circular manifold, one can associate $0$ and $2\pi$ with the location of the plates, and the Casimir energy can be calculated. This becomes relevant when we consider models with additional spatial dimensions \cite{gl}.

Since the pioneering work of Appelquist and Chodos \cite{ac,ac1}, it has been known that the Casmir effect due to quantum \textit{gravitational} fluctuations can generate a minimum of the vacuum potential in KK models. This minimum prevents the extra dimension from continuing to either shrink or expand. Extensions of this work include demonstrating that quantized fermionic and bosonic fields, as well as massive twisted bosons, could stabilize the fifth dimension against collapse \cite{rr}. 

KK setups in which the extra dimension is an $S^1$ or $S^1/\IZ_2$ topology are not the only extra dimensional scenarios, and the utility of the Casimir energy as a stabilization mechanism has proven to be a rich field of research. For example, toroidal topologies have been examined \cite{Ito:2003,bcp}, as have more sophisticated surfaces. More exotic spacetimes have also been studied. For example, Anti-de Sitter(AdS) space and brane world scenarios have all been investigated \cite{fh,gpt,enoo,gpt1,nos,Mukohyama,fmt,hkp,kt,nos1,ns,ss,Goldberger2000,Brevik2000,Rubakov2001,Gabadadze2003,Csaki2005,Perez-Lorenzana2006,Csaki2004,Norman2004}.

\textit{Classical} stabilization forces have also been investigated; for example Gell-Mann and Zweibach \cite{gmz} examined the stabilization effect due to a scalar field along an extra dimension. Their work was expanded upon by the famous work of Goldberger and Wise \cite{gw}, who analyzed the classical stabilization forces in the context of brane worlds. However, it was later discovered that this is not useful for the stabilization of two positive tension branes \cite{kop,bhlm}. As well as possibly playing a role in the stabilization of higher dimensions, Casimir energy can also be investigated in the context of cosmology.

In Section 6.1 we begin by reviewing the derivation of the 4-dimensional effective theory with a discrete KK tower for the scalar field. In Section 6.2 we derive in detail the Casimir energy for a periodic and anti-periodic massive scalar field using $\zeta$-function techniques. Although it is quite possible that this specific derivation has appeared in the literature, we have not come across it and so we show the calculation in its entirety. In Section 6.3 we review the derivation for the Casimir energy in the case of an exotic coupling of a massive scalar field to an antisymmetric Lorentz violating tensor. The resulting expression for an enhanced Casimir energy was first derived in \cite{oc}, and will be investigated as a component in the stabilization scenarios we study. In Section 6.4 we explore which field combinations generate a minimum of the Casimir energy. Finally, in Section 6.5 we discuss the role of the Casimir energy in the dynamics of cosmological evolution.

\section{Scalar Field in Randall-Sundrum Background}
\label{sec:sfirsb}

In this section we review the equation of motion of a scalar field in the RS1 setup. In this scenario, the heirachy between the electroweak scale and the Planck scale is generated by introducing a fifth dimension compactified on an $S^1/Z^2$ orbifold with large curvature. At low energies, a negative bulk cosmological constant prevents gravity from propagating in the extra dimensions. Two 3-branes with opposite tension are located at the orbifold fixed points. The line element in RS is described by the metric eq.\ (\ref{eq108}). This review follows closely \cite{gw}.

Consider a free scalar field in the bulk
\begin{equation}
\mathcal{L}=  \frac{1}{2}G_{AB}\partial_A\Phi \partial_B\Phi-\frac{1}{2} m^2\Phi^2.
\label{eq140}
\end{equation}
Solving the equation of motion, we obtain
\begin{equation}
e^{-2kR|\varphi|}\eta_{\mu\nu}\partial_\mu\Phi \partial_\nu\Phi+\frac{1}{R^2}\Phi\partial_\phi(e^{-4kR|\varphi|}\partial_\phi)-m^2e^{-4kR|\varphi|}\Phi^2=0.
\label{eq141}
\end{equation}
To separate out the extra dimensional contributions, we first use separation of variables and express the field as
\begin{equation}
\Phi(x,\phi)=\sum_n\psi_n(x)\frac{y_n(\phi)}{\sqrt{R}}
\label{eq142}
\end{equation}
and find the equation for y to be
\begin{equation}
-\frac{1}{R^2}\frac{d}{d\phi}\left(e^{-4kR|\varphi|}\frac{dy_n}{d\phi}\right) +m^2e^{-4kR|\varphi|}y_n=m_n^2e^{-2kR|\varphi|}y_n.
\label{eq143}
\end{equation}
The bulk scalar manifests itself in four dimensions as tower of scalars with mass $m_n$. To solve equation eq.\ (\ref{eq5}) it is useful to perform a change of variable: $z_n=m_n e^{kR|\varphi|}/k$ and $f_n=e^{-2kR|\varphi|}/y_n$. We can now write eq.\ (\ref{eq143}) as 
\begin{equation}
z_n^2\frac{d^2f_n}{dz_n^2}+z_n\frac{df_n}{dz_n}+\left[z_n^2-\left(4+\frac{m^2}{k^2} \right)\right]f_n=0.
\label{eq144}
\end{equation}
The solutions to this equation are Bessel functions:
\begin{equation}
y_n(\phi)=\frac{e^{2kR|\varphi|}} {N_n}\left[ J_\nu\left(\frac{M_ne^{kR|\varphi|}}{k}\right)+b_{n\nu}Y_\nu\left(\frac{M_ne^{kR|\varphi|}}{k}\right)\right]
\label{eq145}
\end{equation}
To satisfy the boundary conditions at $y=0$ and $y=\pi R$ the argument of the Bessel function has to satisfy
\begin{equation}
\frac{M_ne^{kR}}{k}\approx\pi(N+\frac{1}{4})\ \ \ \ N\geq1
\label{eq146}
\end{equation}
and we have a 4-dimensional effective theory with a discrete KK spectrum for the scalar field with exponentially suppressed masses.

\subsection{Higher Dimensional Casimir Energy Calculations}
\label{chap:hdcec}

The Casimir energy generated from the quantum fluctuations in the large dimensions are insignificant when they are compared to the contributions arising from the compact dimensions, because the energy is inversely proportional to volume of the space. Therefore, our first Casimir energy calculation focuses on the Casimir energy for a field with boundary conditions on the $S^1$ compactification. We use $\zeta$-function techniques inspired by those discussed in the literature \cite{e1,e3,e4,ekz,e5,Inami,Elizalde2006,Kirsten2002,Kirsten1991,Bordag2002,Elizalde1994,Elizalde1994a,Kirsten1994}.

For a massive field, we can express the modes of the vacuum in RS1 as \cite{ftz}
\begin{equation}
E_n=\sqrt{{\bf k}^2+\left(\frac{\pi n}{r_c}\right)^2+M_n^2},
\label{eq147}
\end{equation}
with $M_n$ set by eq.\ (\ref{eq146}), and $r_c$ the radius of the compact extra dimension. As usual, we have used natural units. The Casimir energy is given by
\begin{equation}
V^+=\frac{1}{2}{\sum_{n=-\infty}^{\infty}}' \int \frac{d^4k}{(2\pi)^4}{\rm log}({\bf k}^2+\left(\frac{n\pi}{r_c}\right)^2+M_n^2),
\label{eq148}
\end{equation}
where the prime on the summation indicates that the $n=0$ term is excluded. For purposes of regularization, we will write this as
\begin{equation}
V^+=\frac{1}{2}{\sum_{n=-\infty}^{\infty}}'\int\frac{d^4k}{(2\pi)^4}\int_0^\infty\frac{ds}{s}e^{-({\bf k}^2+\left(\frac{n\pi}{r_c}\right)^2+M_n^2)s}.
\label{eq149}
\end{equation}
We first perform the Gaussian integration (the k-integral)
\begin{equation}
\int_0^\infty d^4ke^{-{\bf k}^2s}=\frac{\pi^2}{s^2},
\label{eq150}
\end{equation}
and are left with the remaining calculation;
\begin{equation}
V^+=\frac{1}{2}\frac{\pi^2}{(2\pi)^4} {\sum_{n=-\infty}^{\infty}}'\int_0^\infty ds \ \frac{1}{s^3}e^{-\left((\frac{n\pi}{r_c})^2+M_n^2\right)s}.
\label{eq151}
\end{equation}
To help us solve this equation we will use the Poisson Resummation formula:\footnote{Sometimes called Jacobi's theta function identity.}
\begin{equation}
{\sum_{n=-\infty}^\infty}' e^{-(n+z)^2t}=\sqrt{\frac{\pi}{t}}\sum_{n=1}^\infty e^{-\pi^2n^2/t}cos(2\pi nz),
\label{eq152}
\end{equation}
to rewrite the summation of eq.\ (\ref{eq151}). Setting $z=0$ we obtain,
\begin{equation}
\sum_{n=-\infty}^\infty e^{-(\frac{n\pi}{r_c})^2}=\sqrt{\frac{1}{\pi s}}\sum_{n=1}^\infty e^{r_c^2n^2/s}.
\label{eq153}
\end{equation}
Inserting this back into eq.\ (\ref{eq151}) we see that our exponential term can now be expressed as
\begin{equation}
\frac{1}{2}\frac{\pi^2}{(2\pi)^4} \sum_{n=1}^\infty \sqrt{\frac{1}{\pi s}}e^{-(r_c^2n^2/s+M_n^2s)},
\label{eq154}
\end{equation} 
and inserting back into eq.\ (\ref{eq151}) our expression for the Casimir energy density now becomes
\begin{equation}
V^+=\frac{1}{2}\frac{\pi^2}{(2\pi)^4} \sqrt{\frac{1}{\pi}}\sum_{n=1}^\infty \int_0^\infty ds \ . \frac{1}{s^{7/2}}e^{-(M_nr_cn(\frac{M_ns}{r_cn}+\frac{r_cn}{M_ns}))}.
\label{eq155}
\end{equation} 
If we now set $x=\frac{M_ns}{r_cn}$ we can write eq.\ (\ref{eq17}) as
\begin{equation}
V^+=\frac{1}{2}\frac{\pi^2}{(2\pi)^4} r_c^{-5/2}M_n^{5/2}\sum_{n=1}^\infty \frac{1}{n^{5/2}}\int_0^\infty dx x^{-7/2}e^{-M_nr_cn(x+\frac{1}{x})}.
\label{eq156}
\end{equation} 
The integral is easily solved using the following expression for the Modified Bessel function of the Second kind:
\begin{equation}
K_\nu(z)=\frac{1}{2}\int_0^\infty dx x^{\nu-1}e^{-z/2(x+\frac{1}{x})}.
\label{eq157}
\end{equation} 
Using eq.\ (\ref{eq157}) in eq.\ (\ref{eq156}), and recognizing the infinite sum as the Riemann zeta function, we obtain our final expression for the Casimir energy density of a massive scalar field in the five dimensional setup.
\begin{equation}
V^+=-\frac{\zeta(5/2)}{32\pi^2}\frac{M_n^{5/2}}{r_c^{5/2}}\sum_{n=1}^\infty K_{5/2}(2M_nr_cn).
\label{eq158}
\end{equation} 
It is straightforward to extend this expression to include antiperiodic fields. Recalling eq.\ (\ref{eq152}), we see that for antiperiodic fields we can make the substition $n \rightarrow n+1/2$ which ensures the summation is over integer multiples of 1/2. This implies our z term in the Poisson Resummation fomula is now non-zero $(z=1/2)$, so we simply have to include the cosine term in our final Casimir energy expression. Thus, the Casimir energy for anti-periodic fields in our five dimensional setup becomes
\begin{equation}
V^-=-\frac{\zeta(5/2)}{32\pi^2}\frac{M_n^{5/2}}{r_c^{5/2}}\sum_{n=1}^\infty K_{5/2}(2M_nr_cn){\rm cos}(n \pi).
\label{eq159}
\end{equation}
We now wish to find an expression of the Casimir energy due to a \textit{massless} scalar, which will also be used as a component in the stabilization investigation. The necessary calculation is
\begin{equation}
V^+_{\rm massless}=\frac{1}{2}{\sum_{n=-\infty}^{\infty}}'\int\frac{d^4k}{(2\pi)^4}log(k^2+\left(\frac{n\pi}{r_c}\right)^2)
\label{eq160}
\end{equation}
This calculation is well-known in the literature, and so we simply quote the result
\begin{equation}
V^+_{\rm massless}=-\frac{3\zeta(5)}{64\pi^2}\frac{1}{r_c^4}.
\label{eq161}
\end{equation}
The $\IZ^2$ constraint requires that we identify points on a circle related by the reflection $y=-y$.
Neglecting any brane contributions, the $S^1/\IZ^2$ orbifolding simply forces us to ignore all modes odd for $V^+$ and even for $V^-$ under this reflection, which means we discard half of the modes in the summation  eq.\ (\ref{eq148}) \cite{pp}. Our final expressions for the Casimir energy are thus simply multiplied by a factor of $\frac{1}{2}$.

From the expression for the Casimir contribution for a periodic massive scalar field it is straightforward to enumerate the Casimir contributions of all other massive and massless fields by using knowledge of five-dimensional supersymmetry multiplets \cite{pp},
\begin{equation}
V^+_{\rm fermion}(r)=-4V^{+}(r),
\label{eq23a}
\end{equation}
\begin{equation}
V^-_{\rm fermion}(r)=\frac{15}{4}V^{+}(r),
\label{eq23b}
\end{equation}
\begin{equation}
V^+_{\rm higgs}(r)=2V^{+}(r),
\label{eq23c}
\end{equation}
where the positive sign on the potential indicated a periodic field and a negative sign indicates an antiperiodic field.

\subsection{Investigating Higher Dimensional Stability}
\label{chap:ihds}

We now explore the possibility of stabilization scenarios which involve the fields we have discussed. The basic ingredients will be the Casimir energy density of periodic massive scalar fields $V^{+}$ (e.g., the Higgs), an exotic periodic scalar field with a coupling to a Lorentz violating vector in the extra dimension $V^{+}_{lv}$, and massive periodic fermionic fields $\tilde{V}^{+}_i$. 
Because we are phenomenologically motivated, we choose the fermion field masses to be those of the three generation of the SM. Once we add the Casimir energy density contributions from the SM fields, we investigate which additional fields are necessary to generate a stable minimum of the potential. The masses of the SM fields can all be found in Appendix A.

\subsection{Standard Model Fields}
\label{chap:smf}

The first scenario we investigate involves populating the extra dimension with the SM fermionic fields
\begin{equation}
\tilde{V}^{+}_{\rm ferm}\equiv\sum_{i=1} \tilde{V}^{+}_{i},
\label{eq168}
\end{equation} 
where the index i runs over all of the SM fermionic fields, apart from the left-handed antineutrino, and for which the masses are given in Appendix A. We also include the contribution from a bosonic Higgs-like field $V^{+}_{higgs}$. For computation of $V^{\rm tot}(r)$, we have normalized the masses of the SM particles in terms of the Z-boson mass. The top quark provides the majority of the contribution to the total mass of the fields. The potential is plotted in Figure 6.1 as a function of the radius of the fifth dimension. We investigate three possible Higgs masses: 115, 150 and 200 GeV. Here, the lower limit is based on accelerator evidence (or lack there of), and the upper limit is based on theoretical predictions. 

Our expression for the total Casimir energy density is given by
\begin{equation}
V^{\rm tot}(r)=\tilde{V}^{+}_{\rm ferm} + V^{+}_{\rm higgs},    
\label{eq169}
\end{equation} 
where our energy densities are calculated using eq.\ (\ref{eq158}) and eq.\ (\ref{eq168}). For computation of $V^{\rm tot}(r)$, we have normalized the masses of the SM particles in terms of the Z-boson mass. The potential is plotted in Figure 6.1 as a function of the radius of the fifth dimension. We find that that no stable minimum develops for this specific combination of fields, and that the range of Higgs values has negligible bearing on the overall shape of the Casimir potential. We conclude that additional field contributions are necessary for the generation of a stable minimum.
\begin{singlespace}
\begin{figure}[H]
\begin{center}
\includegraphics[width=430pt]{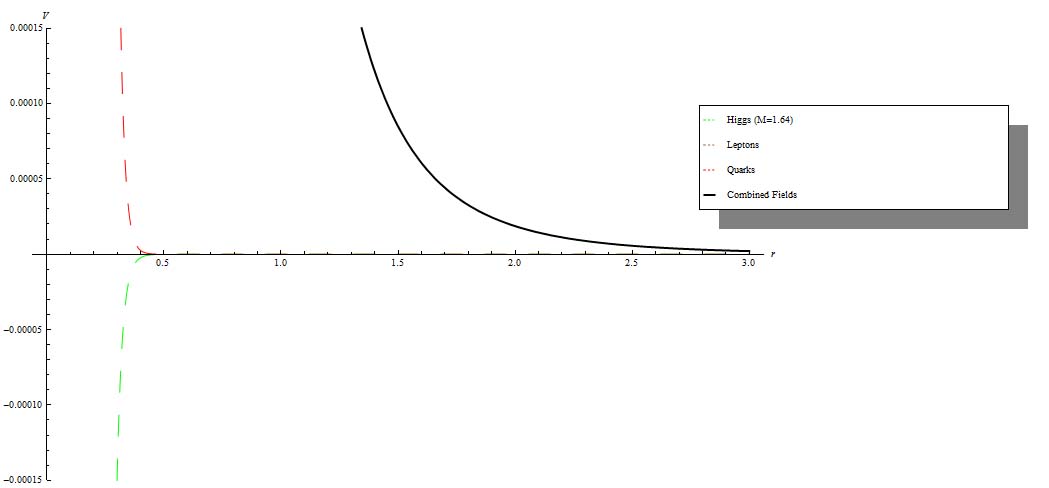}
\caption{\textit{The total contribution to the Casimir energy due to the standard model fermions and the Higgs field. The variation in Casimir energy density for the three values of the Higgs mass is shown; however, the change is so minute that it cannot be discerned from the single solid black line, which also hides the contribution from the leptons also behind the black line. No stable minimum of the energy density is found with this field configuration.}}
\end{center}
\end{figure}
\end{singlespace}

\subsection{SM Fields, a Higgs Field and an Exotic Massive Fermion}
\label{chap:smfahfaaaemf}

Because these field contributions alone are not adequate to generate a minimum of the potential, we add a contribution $\tilde{V}^{-}_E$ from some exotic antiperiodic massive fermionic field for which the mass is selected `by hand' to ensure a stable minimum.\footnote{This has some phenomenological motivations. For example, work by Mohapatra and others have motivated the possibility of a `light sterile bulk neutrino' as an explanation for solar and atmospheric neutrino oscillations \cite{cmy,cmy1,ddg,addm} The premise here is to postulate the existence of a gauge singlet neutrino in the bulk which can couple to leptons in the brane. This coupling leads to a suppression of the Dirac neutrino masses and is largely due to the large bulk volume that suppresses the effective Yukawa couplings of the KK modes of the bulk neutrino to the fields in the brane.}. Our expression for the total Casimir energy density is given by
\begin{equation}
V^{\rm tot}(r)=\tilde{V}^{+}_{\rm ferm} + V^{+}_{\rm higgs}+\tilde{V}^-_{\rm E},    
\label{eq170}
\end{equation} 
where our energy densities are calculated using eqs.\ (\ref{eq158}) and \ (\ref{eq159}). 
\begin{singlespace}
\begin{figure}[H]
\begin{center}
\includegraphics[width=430pt]{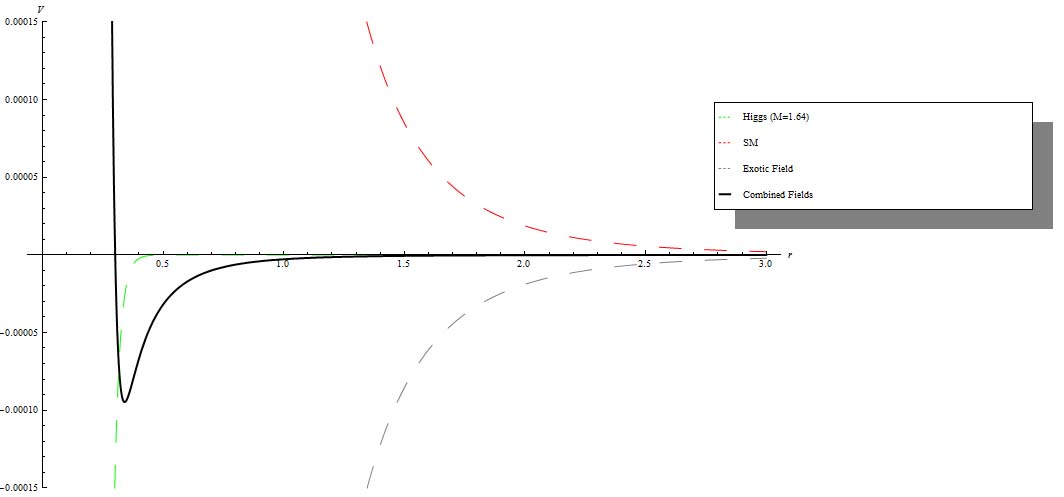}
\caption{\textit{With the addition of an antiperiodic massive fermionic field we see that a stable minimum of the potential develops.}}
\end{center}
\end{figure}
\end{singlespace}
We find that when along with the SM fermions (including anti-particles except the left handed antineutrino) and the Higgs field, that when an additional contribution from a massive \textit{antiperiodic} fermionic field having a mass of $m=0.02$ is added, a stable minimum develops. The stable minimum in this scenario is negative and therefore corresponds to an AdS solution. However, an additional positive contribution from the brane tension can easily be added to raise the overall potential above zero so that the minimum sits in a region of positive potential, thus generating a deSitter space.

Analysis of Figure 6.2 demonstrates that if the radius r is less than the critical value of $r=0.4$, the extra dimension tends to grow. However, as $r \rightarrow 0.4$, this growth is supressed and the extra dimension is stabilized. Conversely, if we start with a radius higher than the critical value, the extra dimension tends to shrink until the minimum is reached. Once the size of the fifth dimension is stabilized, the large dimensions experience increasingly more Casimir energy as they continue to expand, which is a salient feature of dark energy.

\begin{singlespace}
\subsection{Higgs Field and a Massless Scalar Field Coupled to a Lorentz Violating Vector}
\label{chap:hfaamsfctalvv}
\end{singlespace}

For our next study we consider the case of the Higgs field and a single massless scalar field with coupling to a Lorentz violating vector field of the type discussed in Chapter 5.  We explore the case of the massless scalar being both periodic and antiperiodic. With this choice of fields, our expression for the Casimir energy in the compact fifth dimension becomes:
\begin{equation}
V^{\rm tot}(r)=V^{+}_{\rm higgs}+(1+\alpha_\phi^2)^2V^{\pm}_{\rm massless},          
\label{eq171}
\end{equation} 
which we plot in Figure 6.3 for the case of a periodic field and Figure 4 for the antiperiodic fields. We also include the contributions for a range of coupling parameters $\alpha_\phi$.
\begin{singlespace}
\begin{figure}[H]
\begin{center}
\includegraphics[width=430pt]{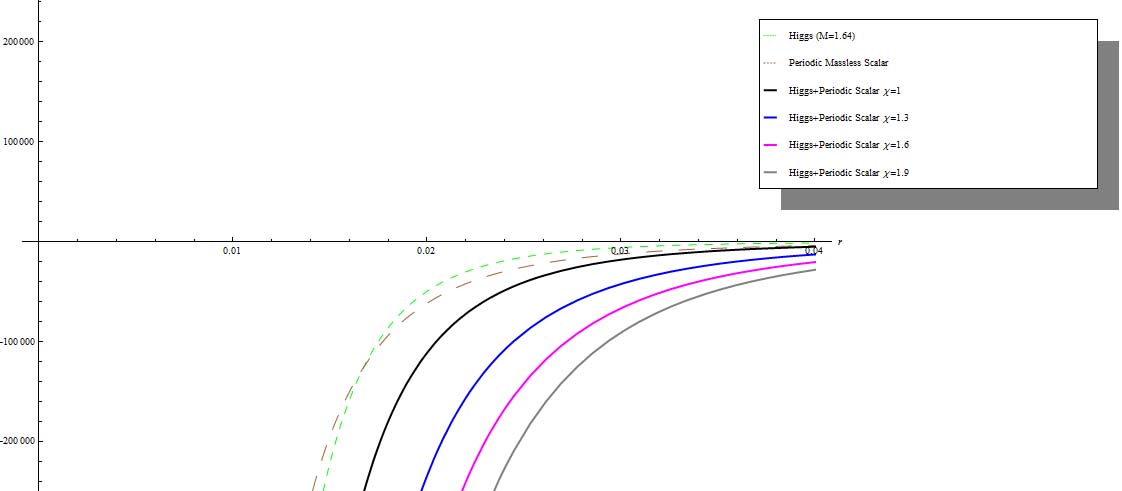}
\caption{\textit{The image illustrates a scenario with a Higgs field and a periodic massless scalar coupled to a Lorentz violating vector. It is clear that no stable minimum occurs for this choice of fields. The coupling parameter is encoded via $\chi=(1+\alpha_\phi^2)^2$.}}
\end{center}
\end{figure}
\end{singlespace}
It is clear from Figure 6.3 that no stable minimum is obtained for the case of a Higgs field and a periodic massless scalar enhanced by $\chi$. This is because all the fields in this scenario contribute a negative Casimir energy, and therefore, there are no compensating positive contributions which would allow for the creation of the stable minimum. This vacuum is pathological and has no finite minimum at finite r. The energy density drops off aymptotically for all values of $\alpha_\phi^2$.
\begin{singlespace}
\begin{figure}[H]
\begin{center}
\includegraphics[width=430pt]{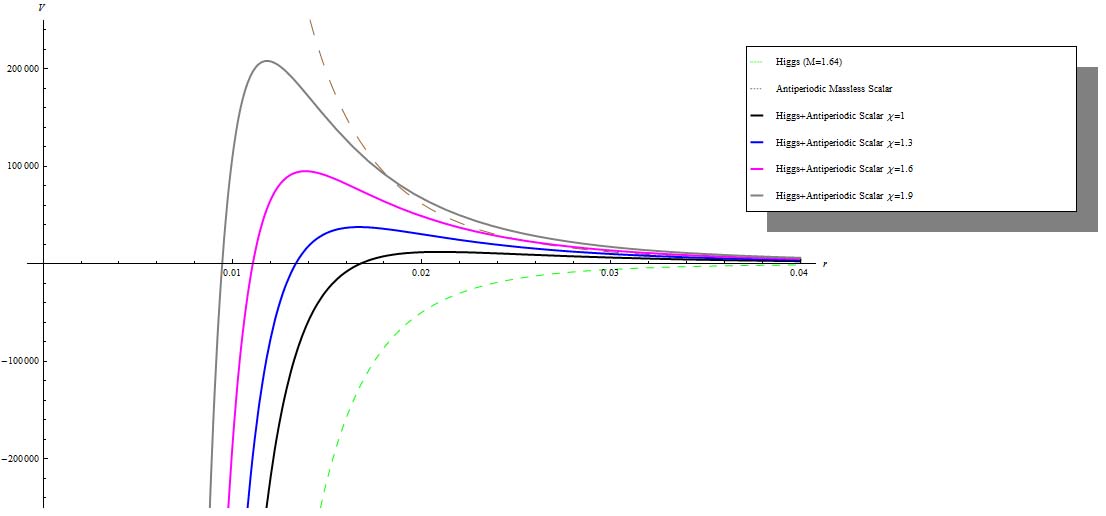}
\caption{\textit{This image illustrates a scenario with a Higgs field and an antiperiodic massless scalar coupled to a Lorentz violating vector. Again, no stable minimum occurs for this choice of fields. The coupling parameter is encoded via $\chi=(1+\alpha_\phi^2)^2$. }}
\end{center}
\end{figure}
\end{singlespace}
However, for the antiperiodic scalar shown in Figure 6.4, we also find an unstable vacuum. If the radius begins at a distance less than the critical point, then from the perspective of an observer located in the bulk, a vacuum in the spacetime manifold would first be nucleated and then expand close to the speed of light. See, for example, discussions on false vacuum decay by Fabinger and Horava \cite{fh}.

For the universe to be deSitter in this scenario, the branes would have to start out separated by a distance above the critical value of around 0.015. The branes would steadily roll down the potential and the brane separation would grow larger as the Casimir energy density decreased. In this setup, the possibility of vacuum tunnelling through the maximum exists, and this situation would represent a catastrophic fall into an ADS space. Such instabilites were studied in a KK scenario by Witten \cite{Witten:1982}.

\begin{singlespace}
\subsection{Higgs Field, Standard Model Fermions and a Massless Scalar Field Coupled to a Lorentz Violating Vector}
\label{chap:hfsmfaamsfctalvv}
\end{singlespace}

In this study we analyze the case of a Higgs field, the SM fermions and a single massless (anti)periodic scalar field with coupling to a Lorentz violating vector field. With these fields our Casimir energy in the compact fifth dimension becomes:
\begin{equation}
V^{\rm tot}(r)=\tilde{V}^{+}_{ferm} + V^{+}_{higgs}+(1+\alpha_\phi^2)^2V^{\pm}_{massless},       
\label{eq172}
\end{equation} 
which we plot in Figure 5 for periodic and in Figure 6 for antiperiodic massless scalar fields. 

We see that in the case of a periodic massless scalar field (Figure 6.5), as $\alpha_\phi^4$ is increased the Casimir energy is enhanced, and consequentially the depth of the minimum increases while the stable minimum is located at progressively smaller radii. One nice feature of this field contribution is that the coupling parameter $\alpha_\phi$ is proportional to the VEV of the Lorentz-violating field. Thus, stabilization could correspond to minimization of a potential by this VEV. Hence, the apparant fine-tuning becomes a natural outcome. This is an advantage of having a massless scalar field that is coupled to a Lorentz-violating field. The case of an antiperiodic field is illustrated in Figure 6.6. We can see that even with a range of values of $\chi$, no stable minimum is created.

\subsection{DeSitter Minimum}
\label{chap:dm}

The minimums so far discussed have all been located at a negative Casimir potential, indicating an AdS space. Because we know that our universe is expanding, we now discuss a scenario which returns a positive minimum. For this we need at least two additional exotic fields, and our expression for the Casimir energy in the compact fifth dimension becomes:
\begin{equation}
V^{\rm tot}(r)=\tilde{V}^{+}_{\rm ferm} + V^{+}_{\rm higgs}+\tilde{V}^{-}_{\rm E}+V^+_{\rm E}. 
\label{eq173}
\end{equation}

\begin{singlespace}
\begin{figure}[H]
\begin{center}
\includegraphics[width=430pt]{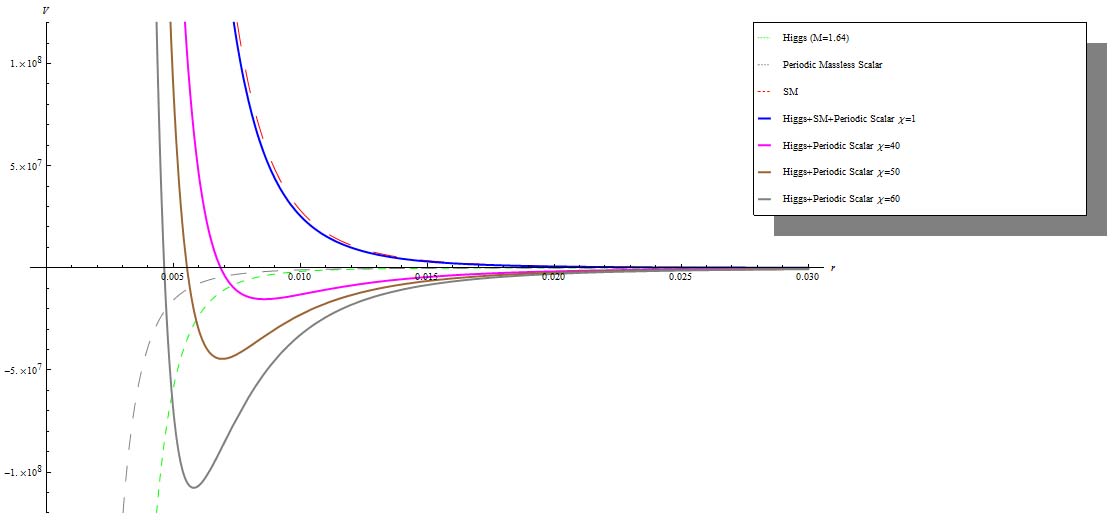}
\caption{\textit{In this image we show the contributions to the Casimir energy density from a Higgs field, the Standard Model fields and a periodic scalar field coupled to a Lorentz violating vector. As the parameter $\chi$ is increased the minimum of the potential is also decreased as is the radius of dimensional stabilization. Again here $\chi=(1+\alpha_\phi^2)^2$.}}
\end{center}
\end{figure}
\end{singlespace}
\begin{singlespace}
\begin{figure}[H]
\begin{center}
\includegraphics[width=430pt]{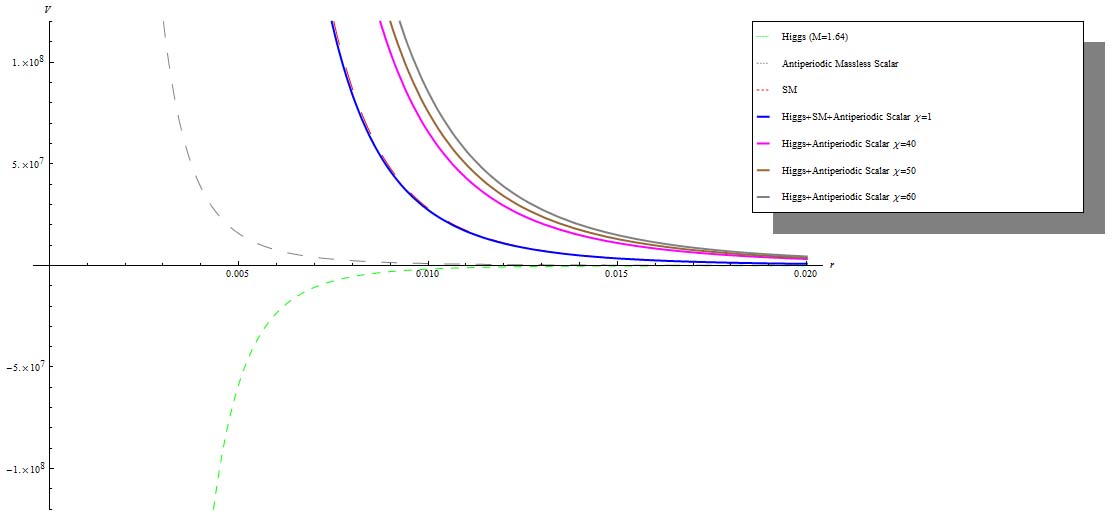}
\caption{\textit{Here we show the Casimir energy density with a Higgs field, the Standard Model fields and an antiperiodic scalar field coupled to a Lorentz violating vector. A stable minimum of the potential is not achieved for any value of $\chi$.}}
\end{center}
\end{figure}
\end{singlespace}
\newpage

The first exotic field, $\tilde{V}^{-}_E$ is an antiperiodic fermion and the second exotic field $V^+_E$ is simply a massive periodic scalar field. In this example, the mass of $\tilde{V}^{+}$ is chosen to be 1.1 and the mass of $V^+_E$ is selected to be 1.8. Decreasing the mass of either of the exotic fields results in the minimum becoming deeper. It is clear from Figure 6.7 that a stable minimum of the Casimir energy density is created and that the minimum is located at $V(r)>0$, demonstrating a dS minima. The nice feature of this model is that the mass of the mass of the exotic particles can be tuned to create a minimum that lies extremely close to the zero potential, which gives us a way recreate the experimentally determined value of $\rho_{\rm eff}$.
\begin{singlespace}
\begin{figure}[H]
\begin{center}
\includegraphics[width=430pt]{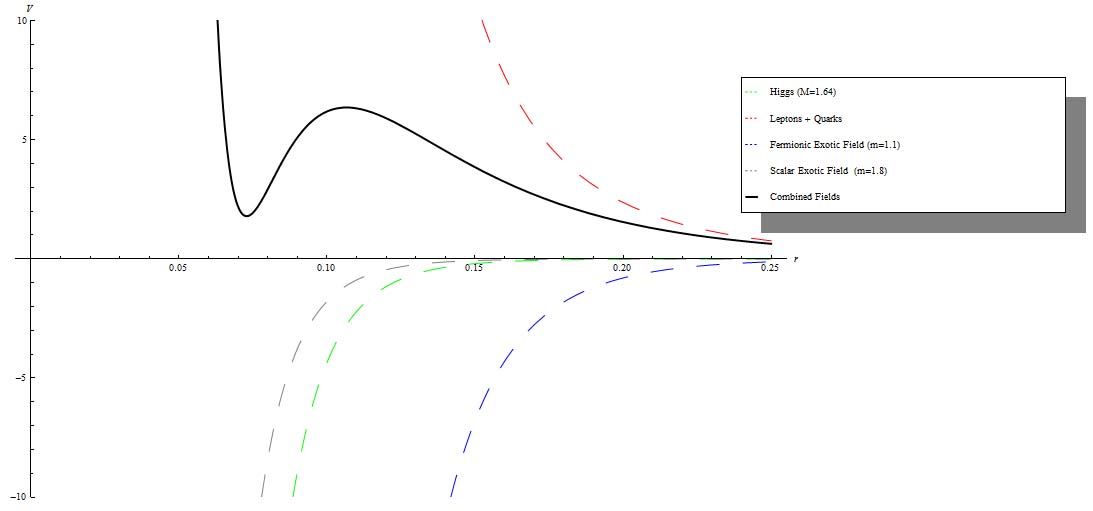}
\caption{\textit{ This field configuration consists of a Higgs field, the SM fields, an antiperiodic exotic massive fermionic field and a massive periodic scalar field. The minimum of the Casimir energy sits at a positive energy density, indicating a deSitter spacetime.}}
\end{center}
\end{figure}
\end{singlespace}
This field configuration allows the possibility of vacuum tunnelling out of the minimum ($r>0.11$), leading to an eternally inflating extra dimension.   
\section{Periodic $T^2$ Spacetimes}
\label{chap:PT2s}

The investigation performed above focussed on the case of a single extra dimension orbifolded via an $S^1/\IZ_2$ scheme. Arguably, this is one of the simplest of higher dimensional scenarios. We now illustrate the calculations of the Casimir energy density in six dimensions compactified on a Torus, giving us a higher dimensional $T^2$ spacetime. This type of compactification was employed in the composite-Higgs scheme and is phenomenologically interesting \cite{ACDH:2000,ahb}. 

\section{$T^2$ Calculations}
\label{chap:T2calcs}

The Casimir energy for a scalar field with KK modes is expressed as \cite{i,pp}
\begin{equation}
V^{(+,+)}=\frac{1}{2}{\sum_{m,n=-\infty}^{\infty}} \int \frac{d^4k}{(2\pi)^4}{\rm log}\left(  \vec{k}^2+\left(\frac{n}{r_1}\right)^2+\left(\frac{m}{r_2}\right)^2\right),
\label{eq173a}
\end{equation}
where $r_1$ and $r_2$ are the radii of the two compact higher dimensions and the $+$ signs after the $V$ indicate that the KK modes are periodic. The sums run from $-\infty$ to $+\infty$. As before we can express this as:
\begin{equation}
V^{(+,+)}=-\frac{1}{2}{\sum_{m,n=-\infty}^{\infty}}\frac{\partial}{\partial s} \mid_{s=0} \int\frac{d^4 k}{(2\pi)^4}\left(\vec{k}^2+\left(\frac{n}{r_1}\right)^2+\left(\frac{m}{r_2}\right)^2\right)
\label{eq173b}
\end{equation}
We first perform the k-integral noting that
\begin{equation}
\int d^4 k (k^2+\alpha^2)^{-s}=-\frac{\pi^2}{16}\frac{\partial}{\partial s}\mid_{s=-2}\frac{\alpha^{-s}}{s(s+1)}.
\label{eq173c}
\end{equation}
The problem that remains is to calculate the double sum:
\begin{equation}
I=\sum_{m,n}\left(\frac{n^2}{r_1^2}+\frac{m^2}{r_2^2}\right)^{-s}.
\label{eq173c1}
\end{equation}
We can immediately express this as 
\begin{equation}
I=\sum_{m,n}r_1^{2s}\left(n^2+a^2 m^2\right)^{-s},
\end{equation}
where $a=r_1^2/r_2^2$. We first decompose the double sum as follows
\begin{equation}
I=\sum_{n=-\infty}^\infty n^{-2s}+{\sum_{m}}'\sum_{n}\left(n^2+a^2 m^2\right)^{-s}.
\label{eq173d}
\end{equation}
We recognise the first term as the Riemann $\zeta$-function and we can express the second term using the following definition of the gamma function:
\begin{equation}
\Gamma(s)=z^{s}\int_0^\infty dt \ t^{s-1}e^{-zt}.
\label{eq173e}
\end{equation}
Our expression for the double sum now read:
\begin{equation}
I=-2\zeta{(2s)}+{\sum_{m}}'\sum_{n} \frac{1}{\Gamma(s)}\int_0^\infty dt \ t^{s-1}e^{-(n^2+a^2m^2)t}.
\label{eq173f}
\end{equation}
The factor of two in front of the Zeta function is a consequence of the summation running from $n=-\infty$ instead of $n=1$. We now use Poisson resummation on the sum in $n$:
\begin{equation}
{\sum_{n=-\infty}^\infty}' e^{-n^2 t}=\sqrt{\frac{\pi}{t}}\sum_{n=0}^\infty e^{-\pi^2 n^2/t}.
\label{eq173g}
\end{equation}
Our double summation calculation now reads
\begin{equation}
I={\sum_m}'\frac{\sqrt{\pi}}{\Gamma(s)}\int_0^\infty dt \ t^{s-3/2}e^{-a^2m^2t}\left(1+2\sum_{n=1}^\infty e^{-\pi^2 n^2/t}\right) \ .
\label{eq173h}
\end{equation}
We first solve the m-summation. Defining $x\equiv a^2m^2t$ and $dx\equiv a^2m^2dt$ we can rewrite this as
\begin{eqnarray}
\frac{\sqrt{\pi}}{\Gamma(s)} {\sum_m}'\int_0^\infty dt \ t^{s-3/2}e^{-a^2m^2t} &=& \frac{\sqrt{\pi}}{\Gamma(s)} {\sum_m}'\int_0^\infty dx \frac{1}{a^2m^2}  \frac{x}{a^2m^2}^{s-3/2} x^{s-3/2}e^{-ax} \nonumber \\
&=&\frac{\sqrt{\pi}}{\Gamma(s)} \frac{1}{a^2}^{s-1/2} {\sum_m}' (\frac{1}{m^2})^{s-1/2}\int_0^\infty x^{s-3/2}e^{-ax}dx \nonumber \\
&=& 2 \sqrt{\pi}a^{1/2-s} \zeta{(2s-1)}\frac{\Gamma(s-1/2)}{\Gamma(s)},
\end{eqnarray}
where we have used the integral representation of the gamma function eq.\ (\ref{eq173e}) and the definition of the Riemann $\zeta$-function.

We now turn to the summation over both n and m;
\begin{eqnarray}
\frac{\sqrt{\pi}}{\Gamma(s)}{\sum_{m=-\infty}^{\infty}}'\sum_{n=1}\int_0^\infty dt \ t^{s-3/2} e^{-(a^2m^2t+\pi^2n^2/t)} \nonumber \\
= \frac{\sqrt{\pi}}{\Gamma(s)}{\sum_{m=-\infty}^{\infty}}'\sum_{n=1}\int_0^\infty dt \ t^{s-3/2} e^{-am\pi n(x+1/x)},
\label{eq173i}
\end{eqnarray}
where we have defined $x=amt /  n \pi$. Using this expression we can express the integral in terms of the modifed Bessel function of the second kind using 
\begin{equation}
K_\nu(z)=\frac{1}{2}\int_0^\infty dx x^{\nu-1}e^{-z/2(x+\frac{1}{x})}.
\nonumber
\end{equation} 
Our final expression for the double summation reads:
\begin{eqnarray}
I&=&-2\zeta(2s)+2\sqrt{\pi}\frac{\Gamma(s-1/2)}{\Gamma(s)}a^{(1-2s)}\zeta(2s-1) \nonumber \\
&+&\frac{8\pi^s}{\Gamma(s)}\sum_{m=1}^\infty \sum_{n=1}^\infty \left(\frac{n}{m}\right)^{(s-1/2)}K_{s-\frac{1}{2}}(2\pi a m).
\end{eqnarray} 
We can now insert this into eq.\ (\ref{eq173c}) to obtain the expression for the Casimir energy density with two compactified toroidal spacetime dimensions where the quantum fields obey periodic boundary conditions;
In obtaining the above equation we have used the the following \cite{i};
\begin{eqnarray}
\zeta'(-4)&=\frac{3}{4\pi^4}\zeta(5), \ \ \ \ \ \ \    \frac{\Gamma'(-2)}{\Gamma(-2)^2}=-2, \nonumber \\  
\Gamma(-\frac{5}{2})&=-\frac{8}{15}\sqrt{\pi}, \ \ \ \ \ \ \  \zeta(-5)=-\frac{15}{4\pi^6}, \nonumber \\ 
\zeta(6) &= -\frac{1}{252}.
\end{eqnarray} 
For the case of antiperiodic fields we follow a similar procedure, however the double summation in eq.\ (\ref{eq173c1}) must be modified:
\begin{equation}
I=\sum_{m,n}r_1^{2s}\left( (n+\frac{1}{2})^2+a^2 (m+\frac{1}{2})^2\right)^{-s},
\end{equation}
The Poisson resummation formula reads;
\begin{equation}
{\sum_{n=-\infty}^\infty} e^{-(n+z)^2t}=\sqrt{\frac{\pi}{t}}\sum_{n=1}^\infty e^{-\pi^2n^2/t}{\rm cos}(2\pi nz).
\end{equation}
For the case of the periodic fields we chose $z=0$, however for the antiperiodic fields we must use $z=\frac{1}{2}$. Also, the infinite sums we will encounter now take the form:
\begin{equation}
{\sum_{n=-\infty}^\infty} \frac{1}{(n+\frac{1}{2})^s}
\end{equation}
and so we must use the less well known \textit{Hurwitz} zeta function:
\begin{equation}
\zeta(s,\nu)={\sum_{n=0}^\infty} \frac{1}{(n+\nu)^s}.
\end{equation}

Using the formulae derived in this section we now have expressions to calculate the Casimir energy density for the following boundary conditions: $V^{(+,+)}$, $V^{(-,-)}$ , $V^{(-,+)}$ and $V^{(+,-)}$. It would be straightforward to perform a study similar to that performed in Sections 6.1.4 thru 6.1.8, however we leave the details for future studies.

\section{Cosmological Dynamics}
\label{chap:cd}

We now briefly change directions in our discussion to examine of some of the implications of generic higher dimensions in a cosmological context. We will focus on a $d+n+1=D$ dimensional anisotropic metric for simplicity. We consider a toy universe in which all the energy density content is due to Casimir energy contributions from the higher dimensions. We will see that in this setup, the Casimir energy density can, under certain conditions, lead to an accelerated expansion scenario in the three large spatial dimensions. Following as in \cite{gl}, we start by considering a homogeneous and anisotropic metric 
\begin{equation}
ds^2=-dt^2+a(t)^2d\vec{x}^2+b^2(t)d\vec{y}^2 \ ,
\label{eq174}
\end{equation} 
where a(t) and b(t) are the scale factors in the three large dimensions and the compact dimensions. We obtain the equations of motion by varying the d+1 dimensional Einstein-Hilbert action, 
\begin{equation}
S=\int d^4xdy\sqrt{g}\left(\frac{M^{3}}{16\pi}R_{D}-\rho_{D}\right) \ ,
\label{eq175}
\end{equation} 
with respect to the five-dimensional metric $g_{ab}$, from which we obtain the Einstein equations
\begin{equation}
3H_a^2+3nH_aH_b+\frac{n}{2}(n-1)H_b^2=8\pi G \rho_{D} 
\label{eq176}
\end{equation} 
\begin{equation}
\dot{H}_a+3H_a^2+nH_aH_b=\frac{8\pi G }{2+n}\left( \rho_{D}+(n-1)p_a-np_b\right) 
\label{eq177}
\end{equation} 
\begin{equation}
\dot{H}_b+nH_b^2+3H_aH_b=\frac{8\pi G }{2+n}\left( \rho_{D}+2p_b-3p_a\right) 
\label{eq178}
\end{equation} 
Let us first consider eq.\ (\ref{eq176}) and analyze the simple scenario where the extra dimension has already found its minimum of potential, implying $H_b=0$. Using the relation
\begin{equation}
\dot{H}_a=\frac{\ddot{a}}{a}+\left(\frac{\dot{a}}{a}\right)^2 \ ,
\label{eq179}
\end{equation} 
we find that eq.\ (\ref{eq176}) becomes
\begin{equation}
\frac{\ddot{a}}{a}+4H_a^2=\frac{8\pi G }{2+n}\left( \rho_{6D}+(n-1)p_a-np_b\right) \ ,
\label{eq180}
\end{equation}
and then using eq.\ (\ref{eq179}) we obtain
\begin{equation}
\frac{\ddot{a}}{a}=-\frac{8\pi G }{2+n}\left[ \left(\frac{5+4n}{3}\right)\rho_{6D}  +(n-1)p_a-np_b\right] \ .
\label{eq181}
\end{equation}
For the case of n=2 we see that this equation simplifies to
\begin{equation}
\frac{\ddot{a}}{a}=-\frac{8\pi G}{4} \left[ \frac{13}{3}\rho_{6D} + p_a-2p_b\right] \ ,
\label{eq182}
\end{equation}
which implies that our current epoch of cosmological acceleration requires
\begin{equation}
\rho_{\rm 6D}\geq\frac{1}{13}\left(3p_a-6p_b\right)
\label{eq183}
\end{equation}

\section{Discussion}
\label{chap:d}

We have investigated the possibility of moduli stability using the Casimir effect in RS1. We have calculated the one loop corrections arising from a massive scalar field with periodic boundary conditions in the compactified extra dimension by applying the Schwinger proper time technique and exploiting the Jacobi Theta function. We have populated the bulk with numerous fields in an attempt to uncover stabilization scenarios with an emphasis on phenomenologically motivated field content. Extending on our work in Chapter 5, we have explored the implications of the existence of a five-dimensional vector field with a VEV in the compact dimension coupling to one of the scalar fields, and noted its relevance as a tuning parameter. 

We have demonstrated that the Casimir energy of the SM fields, in conjunction with the Higgs field, cannot provide the necessary potential to stabilize the extra dimension. The fermionic nature of the SM fields contribute a \textit{positive} Casimir energy which is not balanced by the Higgs field when we include the three lepton generations (and their anti-particles, excepting the left-handed antineutrinos) and the six quarks with three color degrees of freedom (and their anti-particles). We have also investigated the possibility of stability for a range of Higgs masses based on experimental lower limits and theoretical upper limits, and found the same result.

We have investigated the possibility of adding an exotic massive anti-periodic fermionic field to this field setup, and discovered that a light field (m=0.02 in normalized units) is sufficient to generate a stable minimum of the potential. The minimum is located at a negative energy density, which corresponds to an AdS spacetime. Reduction of the mass of the exotic field causes the minimum to become deeper. Our justification for the addition of this field is the possibility of the existence of a light sterile bulk neutrino.

Next, we built on previous work by considering the effects on the Casimir energy of a scalar field coupled to a Lorentz violating vector field. This field, which is completely charactered by the parameter $\chi$, allows for fine tuning of the Casimir energy and the stabilization radius. Fine tuning is a generic feature of stabilization schemes, and in this model the simple addition of a vector field in \textit{only} the fifth dimension creates additional freedom for the stabilization schemes. We discovered that no stable minimum of the potential can be found with either a single periodic or antiperiodic massless scalar field coupled to a Lorentz violating vector in the case of a Higgs vacuum. However, when the SM fields are included a periodic massless scalar field coupled to a Lorentz violating vector \textit{can} lead to a stable minimum, but this is dependent on the parameter $\chi$. Our motivations for studying the phenomenology of Lorentz violating fields stem from the recent surge of activity regarding the possibility of Lorentz invariance violations and the potential role of Lorentz violating fields in cosmology \cite{Ackerman,Dulaney:2008}.

We also outlined a higher dimensional field configuration, which creates a positive energy density minimum of the Casimir energy. We find that at least two additional exotic fields are required. Our example highlights a possible connection between dark energy, the heirachy problem, and additional bulk fields. The capability of this model to explain so many apparantly unrelated phenomenon under the common framework of extra dimensional boundary conditions makes this model particularly appealing. 

Since releasing our results, we have received interest from a group at Fermilab performing similar research. They have discovered that if standard model fields (without the Higgs boson) populate the bulk, the fermion condensation can also stabilize the two branes in the RS model \cite{Bai2008}. From recent correspondence we have learned that they felt that our work was a more \textit{natural} way to stabilize the extra dimension, and has given them new thoughts with regards to the issue.

There has also been recent interest in the literature relating dark energy to Casimir energy, and for this reason we have briefly reviewed cosmological aspects of extra dimensions by considering an anisotropic cosmology. Using simple arguments, we have found the relation between the Casimir energy density and the pressure in both large and compact dimensions necessary for accelerated cosmological expansion.



\chapter{Emerging Possibilities in Spacecraft Propulsion}
\label{chap:episp}

In this chapter, we take an excursion from pure research and explore some of the exciting possibilities that are opened up when we introduce higher dimensions into physics. Specifically, we discuss the concept of \textit{field propulsion}, which is a hypothetical system of propelling spacecraft beyond the conventional rocket technology. Recently we published a paper outlining the physics of such a device, and performed calculations regarding the energy requirements to create a warping of spacetime, and also calculations defining maximum obtainable speeds \cite{oc1, oc2}. We also published a \textit{layman's} version of this idea, which subsequently received an reasonable amount of attention \cite{oc2}; this reflects a public interest in this area. We feel that pursuing such exciting areas is an excellent way to attract fresh young minds to the field of physics, and also to explore how physics places ultimate limitations on technological developments.

\subsection{Motivations for Studing New Forms of Propulsion}
\label{chap:mfsnfp}

The universe is truly vast, and current propulsion technology severely restricts us to the exploration of our own solar system. If we wanted to visit even the nearest star systems, we would be faced with transit times of many tens of thousands of years at best. A compelling reason for why we might actually want to visit other stars is the recent evidence of `extrasolar planets,' which are planets that orbit stars other than our sun. To date, we know of at least 250 extrasolar planets. Even more exciting is the possibility that some of these planets may be `Earth-like', with the theoretical capability to support life. \newpage

Recently, a Swiss team discovered a planet designated Gliese 581c; this planet orbits the star Gliese, 20.4 light years away from earth. The planet is remarkable in that it is the only known extrasolar planet so far that exists in the area known as the `habitable zone' of a star (Fig 7.1). This is the area surrounding a star where surface temperatures could maintain water in a liquid state. Gliese 581c is believed to be roughly 5 times the mass of Earth, and to have a similar surface temperature to that of Earth.
\begin{singlespace}
\begin{figure}[H]
\begin{center}
\includegraphics[width=400pt]{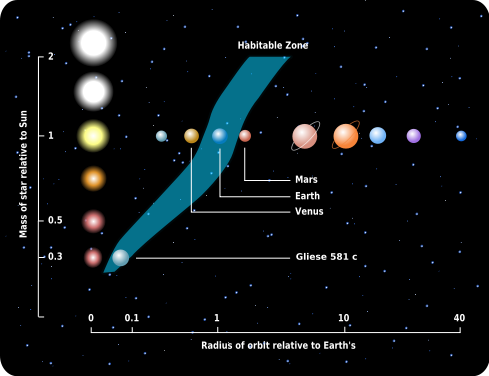}
\caption{\textit{Gliese lies within the 'Habitable Zone' of its host star, meaning that liquid water could exist on its surface.}}
\end{center}
\end{figure}
\end{singlespace}
Given these striking similarities, one naturally wonders if Gliese 581c might harbor life. Although it may be indirectly possible to verify the existence of life using observational techniques here on Earth, the depth of analysis would be particularly limited. On the other hand, we could obtain data of much more scientific value if we could actually visit these planets with probes, or even with humans, to better understand the origin and development of life, and to see if it exists only in the form of elementary organisms or if indeed intelligence has evolved. The discovery of life outside of Earth would be of huge scientific significance, and for this reason we feel justified in exploring novel possibilities regarding spaceraft propulsion.

\section{Current and Future Propulsion Technology}
\label{chap:cafpt}

In this section, we take a look at the limitations of current propulsion technology and review some emerging possibilities in spacecraft propulsion technology.

\subsection{Rocket Propulsion}
\label{chap:rp}

Conventional space exploration technology utilizes rocket propulsion, which obtains thrust by the reaction of the rocket to the ejection of a fast moving fluid from a rocket engine. Combustion of propellent against the inside of combustion chambers generates huge forces, which accelerate the gas to extremely high speeds, which in turn exerts a large thrust on the rocket. Rocket technology can be traced back to the 13th century, where it was used by the Chinese for fireworks and warfare. Rocket technology was used throughout the 20th century to generate the speeds necessary to reach Earth orbit, and enabled human spaceflight to the moon. To date, the fastest rocket propelled spacecraft is the Voyager 1 probe, which utilzed a gravitational assist maneuvre to drive it to its current speed of $38,600 mph$ or about $0.006\%$ the speed of light. Travelling at this speed, it would take the Voyager probe about $70,000$ years to reach the closest star. Clearly, rocket technolgy is vastly inadequate when considered in the context of interstellar exploration.

\subsection{Solar Sail}
\label{chap:ss}

A solar sail is a device which accelerates using photon radiation pressure from the sun. The physics of solar sails is well known, but the technology required to manage large sails has not yet been developed. The concept requires a spacecraft to be attached to a large membrane which reflects light from the sun; the resulting radiation pressure would theoretically generate a small amount of thrust. Solar sail propulsion would allow a spacecraft to escape the solar system with cruise speeds much higher than conventional rockets. NASA has successfully tested solar sail technology on small scales in vacuum chambers, but such technology has yet to be used in space. Because the momentum transfer from the sunlight is very low, solar sails have very low thrusts. Even so-called `supersails,' which are sails manufactured using thin film of aluminum $50 {\rm nm}$ thick, would only accelerate at $60 {\rm mm/s}^2$

It has been proposed that Earth-based lasers could push solar sails. Given a sufficiently powerful laser and a mirror that would be large enough to keep the laser focused on the sail, it has been suggested that a solar sail \textit{could} accelerate to a significant fraction of the speed of light. However, the precisely shaped optical lenses would have to be wider than Earth, and the lasers would have to be far more powerful than anything humanity has created to date.

\subsection{Nuclear Propulsion}
\label{chap:np}

A spacecraft powered by nuclear propulsion derives its thrust from nuclear fission or nuclear fusion. This technology was first seriously studied by Stanislaw Ulam and Frederick de Hoffman in 1944 as a spinoff of their work in the Manhattan Project.
\looseness=1
One possible way to achieve nuclear propulsion is to heat a fluid by pumping it through a nuclear reactor, and to then let the fluid expand through a nozzle. The fuel used for nuclear fission contains over a million times as much energy per unit mass as chemical fuel does, and so nuclear propulsion appears to be a promising alternative propulsion system. However, the approach is limited by the temperature at which a reactor can operate.
Because hydrogen is the lightest substance and therefore consists of the fastest-moving particles at any temperature, it is the best working fluid to use.
\looseness=0
Arguably, the most serious attempt to realize a working nuclear rocket was project Orion, which was an advanced rocket design studied in the 1960s. The project explored the feasibility of building nuclear-pulse rockets powered by nuclear fission. The suggestion by Stanislaw Ulam and Cornelius Everett was to release atomic bombs behind a spacecraft, followed by disks made of solid propellant. After the bombs exploded, the material of the disks would vaporize and convert to hot plasma, some of which would strike the pusher plate, thus driving the craft forwards.

A variety of mission profiles was considered, and the most ambitious was an interstellar version. The spacecraft would have a mass of 40 million tons powered by the sequential release of ten million atomic bombs, which would explode 60m to the rear of the vehicle. A maximum speed of $10\%$ the speed of light was suggested.

Due to the ethical, legal and most importantly the safety issues associated with transporting nuclear devices into space and subsequently detonating them to propel a spacecraft, the Orion project was ultimately forgotten. 

\subsection{Wormholes}
\label{chap:w}

One of the major limiting features of interstellar exploration is the ultimate speed limit imposed on spacecraft by Special Relativity (SR). In SR, an object's mass increases as its velocity increases and asymptotically approaches infinity as the velocity approaches the speed of light ($3\times10^8 {\rm m/s}$), according to the equation:
\begin{equation}
m'=\gamma m
\label{eq184}
\end{equation} 
where c is the speed of light and $\gamma$ is the Lorentz factor, 
\begin{equation}
\gamma=\frac{1}{\sqrt{1-\frac{v^2}{c^2}}} \ .
\label{eq185}
\end{equation} 
Given that the closest star is slightly over $4$ light years away, and that the galaxy has a diameter of around $100,000$ light years, it appears that interstellar exploration would be an extremely time-consuming process. However, there are two loop-holes to the relativistic speed of light limit. The first is a \textit{wormole}, which will be discussed in this section, and the second is the \textit{warp drive}, which will be discussed in the following section.

In essence, a wormhole is a shortcut through spacetime. A wormhole has two mouths which are connected by the wormhole throat, and if the wormhole is traversable, matter can travel from one mouth two the other. 
\begin{singlespace}
\begin{figure}[H]
\begin{center}
\includegraphics[width=200pt]{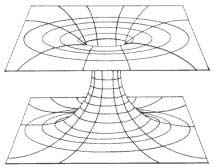}
\end{center}
\caption{\textit{A wormhole is a shortcut through spacetime.}}
\end{figure}
\end{singlespace}
Lorentzian wormholes are not excluded from the framework of GR. However, the plausibility of wormhole solutions to Einstein's field equations is uncertain. Also, it is unknown whether quantum theories of gravity allow them. One feature of wormholes is the requirement that negative energy is required to hold open the throat. This is in violation of the Weak Energy Principle (WEP), but certain quantum phenomenon are also known to violate this principle. The Casimir effect is a good example, furthermore string/M-theory allows negative energy configurations.

Because SR only applies locally, wormholes allow faster-than-light travel. For any two points connected by a wormhole, the time taken to traverse it would be less than the time it would take a beam of light to make the journey if it took a path outside the wormhole. The possibility of using wormholes to explore the universe was popularized by a 1988 paper \cite{mty} by Morris, Thorne and Yurtsever, who explored the technological constraints imposed on an arbitrarily advanced civilization. 

\section{Warp Drive}
\label{chap:wd}

An alternative to the wormhole idea is the warp drive, which involves creating an asymmetric bubble of locally contracting/expanding spacetime around a spacecraft, effectively stretching and compressing space itself. Over the last decade, there has been a respectable level of scientific interest regarding the concept of the `warp drive' \cite{Ford:1994bj,Ford:1996er,Pfenning:1997wh,Pfenning:1997rg,Everett:1997hb,Everett:1995nn,Visser:1998ua,VanDenBroeck:1999sn,Lobo:2004wq,Hiscock:1997ya,GonzalezDiaz:1999db,VanDenBroeck:1999xs,Puthoff:1996my,Natario:2004zr,Lobo:2004an,Bennett:1995gp,GonzalezDiaz:2007zzb}. Recently, we explored a novel approach to generate the asymmetric bubble of spacetime required to create such a warp drive \cite{oc1,oc2,Obousy:2005ug}, which we will summarize in this section.

As discussed throughout this dissertation, certain classes of higher dimensional models suggest that the Casimir effect is a candidate for the cosmological constant. We demonstrate that a sufficiently advanced civilization could, in principal, manipulate the radius of the extra dimension to locally adjust the value of the cosmological constant. This adjustment could be tuned to generate an expansion/contraction of spacetime around a spacecraft, creating an exotic form of field-propulsion.  Due to the fact that spacetime expansion itself is not restricted by relativity, a faster-than-light `warp drive' could be created. Calculations of the energy requirements of such a drive are performed and an `ultimate' speed limit, based on the Planckian limits on the size of the extra dimensions, is found.

\section{Warp Drive Background}
\label{sec:wbd}

The term `warp drive' originated in science fiction. A 1994 paper by theoretical physicist Miguel Alcubierre placed the concept on a more scientific foundation \cite{alc}. Alcubierre's paper demonstrated that a solution to Einstein's field equations could `stretch' space in a way such that space itself would expand behind a hypothetical spacecraft, while contracting in front of the craft, creating the effect of motion (Fig. 7.3). In contrast to the conventional technology that results in movement of the craft through space, in this theory space itself moves around the spacecraft. This is a radical departure from the traditional concept of motion, because the spacecraft is, in a classical sense, motionless within a hypothetical bubble of transient spacetime. In a manner identical to the inflationary stage of the universe, the spacecraft would have a relative speed, defined as change of proper spatial distance over proper spatial time, \textit{faster than the speed of light}.
\begin{singlespace}
\begin{figure}[H]
\begin{center}
\includegraphics[width=350pt]{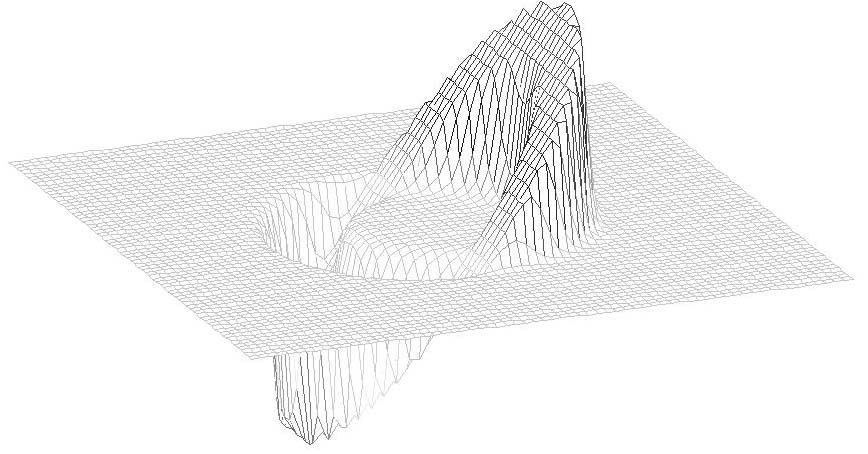}
\caption{\textit{The Alcubierre `top-hat' metric. A bubble of assymetric spacetime curvature surrounds a spacecraft which would sit in the center of the bubble. The disturbance in the positve z direction represents positive dark energy and the disturbace below represents negative dark energy. The space immediately surrounding the spacecraft would be expanding/contracting behind/in front of the craft. In this image the ship would `move' from right to left.}}
\end{center}
\end{figure}
\end{singlespace}
What is particularly appealing about this approach to propulsion is that the spacecraft could effectively travel faster than the speed of light. SR forbids objects from moving through space at or above the speed of light, but the fabric of space itself is not restricted in any way. Thus, even though the spacecraft cannot travel faster than light in a local sense, it could make a round trip between two points in an arbitrarily short period of time as measured by an observer on board the spacecraft. 

In a manner similar to \cite{mty}, warp drives provide a unique and inspiring opportunity to ask the question `what constraints do the laws of physics place on the abilities of an arbitrarily advanced civilization.' Warp drives and space exploration in general, also serves as an excellent way to 
attract young minds into the field. We feel that this work, which we hope will be perceived as exciting by the younger generation, will help to inspire the next generation of scientists. Indeed, the media success of our work \cite{ocMedia1, ocMedia2, ocMedia3, ocMedia5, ocMedia6, ocMedia7, ocMedia8} demonstrates a huge public appeal to this sort ot work.

Here we discuss an original mechanism to generate the necessary `warp bubble'. The main focus of the section is to demonstrate that the manipulation of the radius of one, or more, of the extra dimensions found in higher dimensional quantum gravity theories, especially those that are based on or inspired by string/M-theory, creates a $\textit{local}$ asymmetry in $\rho_{\rm eff}$ which could be used to propel a space vehicle.  

Warp drives have not been the sole interest of theoretical physicists, as was demonstrated by the formation of the NASA Breakthrough Propulsion Program and the British Aerospace Project Greenglow, both of whose purpose was to investigate and expand on these ideas regarding exotic field propulsion.

At such an early stage in the theoretical development of the ideas presented in this paper, it is challenging to make predictions on how this `warp drive' might function. Naively, one could envision a spacecraft with an exotic power generator that could create the necessary energies to locally manipulate the extra dimension(s), and the technology to perform this manipulation. In this way, an advanced spacecraft would expand/contract the compactified spacetime around it, thereby creating the propulsion effect.

\subsection{The Physics of Warp Drives}
\label{sec:tpowd}

Numerous papers discussing the idea of warp drives have emerged in the literature in recent years \cite{Ford:1994bj,Ford:1996er,Pfenning:1997wh,Pfenning:1997rg,Everett:1997hb,Everett:1995nn,Visser:1998ua,VanDenBroeck:1999sn,Lobo:2004wq,Hiscock:1997ya,GonzalezDiaz:1999db,VanDenBroeck:1999xs,Puthoff:1996my,Natario:2004zr,Lobo:2004an,Bennett:1995gp,GonzalezDiaz:2007zzb}. The basic idea is to formulate a solution to Einstein's equations whereby a warp bubble is driven by a local expansion of spacetime behind the bubble and a contraction ahead of the bubble. One common feature of these papers is that their physical foundation is the GR. An element missing from all the papers is that there is little or no suggestion as to $\textit{how}$ such a warp bubble may be created.

The aim of this section is $\textbf{not}$ to discuss the plausibility of warp drive, the questions associated with violation of the null energy condition, or issues regarding causality. The aim of this paper is to suggest that a warp bubble could be generated using ideas and mathematics from $\textit{quantum field theory}$, and to hypothesize $\textit{how}$ such a bubble could be created by a sufficiently advanced technology.

$\indent$By associating the cosmological constant with the Casimir Energy due to the KK modes of vacuum fluctuations in higher dimensions, especially in the context of M-theory derived or inspired models, it is possible to form a relationship between $\Lambda$ and the radius of the compact extra dimension. We know from Section 5 that
\begin{equation}
V^+_{\rm massless}=\frac{1}{2}{\sum_{n=-\infty}^{\infty}}'\int\frac{d^4k}{(2\pi)^4}log(k^2+\left(\frac{n\pi}{r_c}\right)^2) \ ,
\end{equation}
which can be solved using the techniques of dimensional regularization to give:
\begin{equation}
V^+_{\rm massless}=-\frac{3\zeta(5)}{64\pi^2}\frac{1}{r_c^4}.
\label{eq161a}
\end{equation}
Thus we can for the straightforward relation between the vacuum energy density and the size of the extra dimension;
\begin{equation}
\left< E_{\rm vac} \right> =\Lambda \propto \frac{1}{R^4}.
\end{equation}
An easier way of developing the relationship between the energy density and the expansion of space is to put things in terms of Hubble's constant H, which describes the rate of expansion of space per unit distance of space. 
\begin{equation}
H \propto \sqrt{\Lambda}, 
\end{equation}
or in terms of the radius of the extra dimension we have
\begin{equation}
H \propto \frac{1}{R^2}. 
\end{equation}
This result indicates that a sufficiently advanced technology with the ability to $\textit{locally}$ increase or decrease the radius of the extra dimension would be able to locally adjust the expansion and contraction of spacetime, creating the hypothetical warp bubble discussed earlier. A spacecraft with the ability to create such a bubble will always move inside its own local light-cone. However, the ship could utilize the expansion of spacetime behind the ship to move away from some object at any desired speed, or equivalently, to contract the space-time in front of the ship to approach any object. The possibility of the size of the compact geometry vary depending on the location in four dimensional spacetime has been explored in the context of string theory \cite{Giddings2005}, but never from the perspective of propulsion technology.

In the context of GR a similar phenomenology is produced for the case of anisotropic cosmological models, in which it is the $\textit{contraction}$ of the extra dimension that has the effect of expanding another \cite{l}. For example, consider a `toy' universe with one additional spatial dimension with the following metric
\begin{equation}
ds^2=dt^2-a^2(t)d\vec{x}^2-b^2(t)dy^2  \ .
\end{equation}
In this toy universe we will assume spacetime is empty, that there is no cosmological constant, and that all spatial dimensions are locally flat,
\begin{equation}
T_{\mu \nu}=\Lambda g_{\mu \nu}=0 \ .
\end{equation}
The action of the Einstein theory of gravity generalized to five dimensions will be
\begin{equation}
S^{(5)}=\int d^4x dy \sqrt{-g^{(5)}}\left(\frac{M_5^2 }{16\pi }R^{(5)}\right) \ .
\end{equation}
Solving the vacuum Einstein equations
\begin{equation}
G_{\mu\nu}=0 \ ,
\end{equation}
we obtain for the $G_{11}$ component
\begin{equation}
G_{11}=\frac{3\dot{a}(b\dot{a}+a\dot{b})}{a^2b} \ .
\end{equation}
Rewriting $\frac{\dot{a}}{a}=H_a$ and $\frac{\dot{b}}{b}=H_b$ where $H_a$ and $H_b$ corresponds to the Hubble constant in three space and the Hubble constant in the extra dimension respectively, we find that solving for $G_{11}=0$ yields
\begin{equation}
H_a=-H_b.
\end{equation}
This remarkable result indicates that in a vacuum, the shear of a contracting dimension is able to inflate the remaining dimensions. In other words the expansion of the 3-volume is associated with the contraction of the one-volume.  

Even in the limit of flat spacetime with zero cosmological constant, general relativity shows that the physics of the $\textit{compactified}$ space affects the expansion rate of the non-compact space. The main difference to note here is that the quantum field theoretic result demonstrates that a $\textit{fixed}$ compactification radius can also result in expansion of the three-volume as is shown in eq.\ (\ref{eq161a}) due to the Casimir effect, whereas the GR approach suggests that a $\textit{changing}$ compactifification radius results in expansion. Both add credibility to the warp drive concept presented here. 

\subsection{Energy Requirements}
\label{sec:er}

In this section, we perform some elementary calculations to determine how much energy would be required to reach superluminal speeds. We also determine an absolute speed limit based on fundamental physical limitations.

The currently accepted value for the Hubble constant is 70 km/sec/Mpsc. A straightforward conversion into SI units gives $H = 2.17\times 10^{-18} {\rm (m/s)/m}$. This tells us that one meter of space would expand to two meters of space if one were prepared to wait two billion billion seconds, or sixty five billion years. The fundamental idea behind the warp drive presented here is to increase Hubble's constant locally around the spaceship, such that space no longer expands at such a sedentary rate, but locally expands at an arbitrarily fast velocity. For example, if we want space to \textit{locally} expand at the speed of light, a simple calculation shows us by what factor we would need to increase H.
\begin{equation}
\frac{H_c}{H} \approx \frac{10^8}{10^{-18}}=10^{26}
\end{equation}
Where $H_c$ is the `modified' Hubble constant (subscript c for speed of light). This results implies that H would have to be increased by a factor of $10^{26}$ for space to expand at the speed of light. Since we know that $H \propto \frac{1}{R^2}$, we can naively form the relation
\begin{equation}
\frac{H_c}{H} = \frac{R^2}{R_c^2}=10^{26}
\end{equation}
or,
\begin{equation}
R_c=10^{-13}R
\end{equation}
Where $R_c$ is the modified radius of the extra dimension. This indicates that the extra dimensional radius must be $\textit{locally}$ reduced by a factor of $10^{13}$ to stimulate space to expand at the speed of light. In the ADD model, the size of the extra dimension can be as large as $10^{-6}{\rm m}$. If we use this number as a prototype extra-dimensional radius, this would have to be shrunk to $10^{-19}{\rm m}$ for lightspeed expansion. $\indent$An interesting calculation is the energy required to create the necessary warp bubble. The accepted value of the cosmological constant is $\Lambda \approx 10^{-47} ({\rm GeV})^4$. Converting again into SI units gives $\Lambda \approx 10^{-10} {\rm J/m}^3$ . Now, for a warp bubble expanding at the speed of light we would need to increase this again by a factor of $10^{52}$ as we have $H\propto \sqrt{\Lambda}$ . We can say
\begin{equation}
\Lambda_c=10^{52}\Lambda=10^{42}{\rm J/m}^3
\end{equation}
where $\Lambda_c$ is the \textit{local} value of the cosmological constant when space is expanding at c. Let us consider a spacecraft of dimensions
\begin{equation}
V_{{\rm craft}}=10{\rm m} \times 10{\rm m} \times 10{\rm m} =1000{\rm m}^3.
\end{equation}
If we postulate that the warp bubble must, at least, encompass the volume of the craft, the total amount of energy `injected' locally would equal
\begin{equation}
E_c=\Lambda_c \times V_{\rm craft} =10^{45}J.
\end{equation}
Assuming some arbitrarily advanced civilization was able to create such an effect, we might postulate that this civilization could utilize the most efficient method of energy production : matter-antimatter annihilation. Using $E=mc^2$ this warp bubble would require around $10^{28} {\rm kg}$ of antimatter; this is roughly the mass-energy of the planet Jupiter. This energy requirement would drop dramatically if we assumed a thin-shell of modified spacetime instead of bubble encompassing the volume of the craft.

\subsection{Ultimate Speed Limit}
\label{sec:usl}

$\indent$It is known from string theory that the absolute minimum size for an extra dimension is the Planck length, $10^{-35}{\rm m}$. This places an ultimate speed limit on the expansion of space based on the idea that there is a limit to the minimum radius of the extra dimension. From the above arguments, it is straightforward to form the relation
\begin{equation}
\frac{H_{\rm max}}{H}=\frac{R^2}{R_{\rm min}^2}=\frac{10^{-12}}{10^{-70}}=10^{58}.
\end{equation}
Here $H_{\rm max}$ is the maximum rate of spacetime expansion based on the minimum radius of the extra dimension $R_{{\rm min}}$. In this formula, we have again used a prototype extra-dimensional radius of $10^{-6}m$ which is the upper bound based on current experimental limits. Using these values and the known value of H in SI units we obtain
\begin{equation}
H_{max}=10^{58}H\approx 10^{40} {\rm (m/s)/m}
\end{equation}
A quick conversion into multiples of the speed of light reveals
\begin{equation}
V_{\rm max}=10^{32}c,
\end{equation}
which would require on the order of $10^{99}Kg$ of antimatter, more mass energy than is contained within the universe. At this velocity, it would be possible to cross the known universe in a little over $10^{-15}$ seconds. The calculations presented here are extremely `back of the envelope' and merely serve as interesting figures to contemplate, based on the formula
\begin{equation}
\left< E_{\rm vac} \right> = \Lambda \propto \frac{1}{R^4}.
\end{equation}
We emphasize again that it is not really possible to travel faster than light in a local sense. One can, however, make a round trip between two points in an arbitrarily short time as measured by an observer on board the ship. See (pending) for details on violations of the null energy conditions and causality.

\subsection{Discussion of Warp Drive Section}
\label{sec:dowds}

In this section we have explored a novel method to generate the asymmetric bubble of contracting/expanding spacetime around a spacecraft necessary for `warp drive' propulsion. We calculated the vacuum energy due to the extra dimensional scalar field contributions to the Casimir energy and associated this energy with the cosmological constant. It has been shown that this energy is intimately related to the size of the extra dimension. We have picked a generic higher dimensional model where the spacetime is simply $M^4\otimes T^1$, and similar approaches can be used for alternative models: for example, the RS1 model of warped extra dimension where a similar relation can be found.

We have proposed that a sufficiently advanced civilization could utilize this relation to generate a localized expansion/contraction of spacetime, creating a `warp bubble' in which to travel at arbitrarily high velocities. One vital aspect of future research would be $\textit{how}$ to locally manipulate an extra dimension. String theory suggests that dimensions are globally held compact by strings wrapping around them \cite{cr}. If this is indeed the case, then it may be possible to even locally increase or decrease the string tension, or even locally counter the effects of some string winding modes. This would achieve the desired effect of changing the size of the extra dimensions, which would lead to propulsion under this model. It would thus be prudent to research this area further and perform calculations as to the energies required to effect an extra dimension, and to try and relate this energy to the acceleration a spacecraft might experience. 

We have also suggested that the exciting study of exotic propulsion mechanisms is a useful tool to attract new students to the field of physics. After publishing our results, we discovered that there exists a huge amount of public interest in this field, since numerous articles discussing our concept have appeared on reputable science news websites \cite{ocMedia1, ocMedia2, ocMedia3, ocMedia5, ocMedia6, ocMedia7}, as well as discussions on well-known national radio \cite{ocMedia8}.


\chapter{Free Fermionic Models}
\label{chap:ffm}

String theory is the only framework for the unification of quantum gravity with gauge interactions. Despite the undeniable successes of the SM, field theories in which the basic object is point-like are plagued with problems as discussed in Chapter 3. The aim of free fermionic model building is to construct a realistic superstring derived standard-model which satisfies the following \cite{Faraggi:1994,fff1b,fff1c,fff1d}:   

\begin{itemize}
\vspace{-0.2in}
\begin{singlespace}
\item The gauge group must be $SU(3)\otimes SU(2)\otimes U(1)^n\otimes(hidden)$.
\item The mass spectrum must have three chiral generations and a pair of Higgs doublets that reproduce a realistic fermionic mass spectrum.
\item We must have N=1 space-time supersymmetry, ensuring that the cosmological constant vanishes, until supersymmetry is broken.
\end{singlespace}
\end{itemize}

The $E_8 \times E_8$ and SO(32) heterotic strings in ten spacetime dimensions are unique. However upon compactification their uniqueness is lost. The free fermionic formulation is applicable at a highly symmetric point in the compactification space (R=1). It is also an exact Conformal Field Theory and thus, CFT calculational tools can be utilized to calculate Yukawa couplings, for example. Also, in free fermionic models one can naturally obtain three generations with standard $SO(10)$ embedding. 

In this section we first begin by reviewing the heterotic string model building theory. Then we investigate a new realization of the $E_8$ gauge group.

\section{Basics of Fermionic Model Building}
\label{sec:bofmb}

Free fermionic heterotic models are specified by a p-dimensional basis set of vectors $V_i$. Each of these vectors has 64 components,  
which represent the different boundary conditions for the left-moving supersymmetric string and the right-moving bosonic string. The first 20 components correspond to the 20 free fermions representing world-sheet degrees of freedom for the left-moving supersymmetric string. The last 44 components specify the right-moving bosonic string. The basis vectors span a finite group which is determined via modular invariance, 
\begin{equation}
\Xi=\sum_k n_k\bold{V}_k
\end{equation}
where $n_k=0,...,N_{z_k}-1$. To fully determine a free fermionic model, a $p\times p$ dimensional matrix $\bold{k}$ of rational numbers $-1<K_{ij}<1,i,j=1,...,p,$ is required, which determines the GSO operators for physical states. One obtains physical massless states of a sector \boldmath $\alpha$\unboldmath $\in \Xi$ by acting on the vacuum with fermionic and bosonic operators and then applying the GSO projections.

The world-sheet fermions are broken up into 18 left-moving internal real fermions, $\psi ^{r=3,20}$, and 44 internal right-moving real fermions, $\overline{\psi}^{r=21,60}$, in addition to the transverse left-moving fermions, $\psi ^{r=1,2}$, in light-cone gauge. Transporting these fermions around one of the two non-contractible loops on the genus-one world-sheet results in the appearance of a phase
\begin{equation}
\psi ^r\rightarrow -e^{i\pi\alpha_r}\psi^r,\
\label{phase}
\end{equation}
for $-1<\alpha _r \leq 1$. For real complex fermions, $\alpha_r=0,1$. For complex fermions formed from pairs of real fermions, $\alpha$ is rational.  Complex fermions, $\psi_c^p\equiv \psi^{r_1}+i\psi^{r_2}$ form a charge lattice \boldmath $Q_{\alpha} \,$\unboldmath associated with these boundary vectors where
\begin{equation}
(Q_{\alpha})_p = \frac{\alpha_p}{2} + F_p.
\label{u1charges}
\end{equation}
$F_p$ is a number operator for fermion oscillator excitations with eigenvalues for complex fermions of $\{0,\pm 1\}$ and $\{0,-1\}$ for real fermions.  Pseudo-charges for non-chiral (i.e., with both left- and right-moving components) real Ising fermions can be similarly defined, with $F_\alpha(\psi)$ counting each real mode $\psi$ once.

For periodic fermions, $\alpha_p(\psi)=1$, the vacuum is a spinor representation of the Clifford
algebra of the corresponding zero modes.  For each periodic complex fermion $\psi_c$ there are two degenerate vacua ${\vert +\rangle},{\vert -\rangle}$, annihilated by the zero modes $\psi_0$ and
${{\psi_0}^*}$ and with fermion numbers $F(\psi)=0,-1$, respectively.

The masses of states can be related to the $\mathbf{Q_{\vec{\alpha}}}$ by
\begin{eqnarray}
m_{R}^2 &=& \frac{\mathbf{Q_{R}}^2}{2} - 1      \mbox{\hspace{10 mm}(non-SUSY string)}\label{massL}\\
&\mbox{and}&\nonumber\\
m_{L}^2 &=& \frac{\mathbf{Q_{L}}^2}{2} - \frac{1}{2}      \mbox{\hspace{14 mm}(SUSY string)}\label{massR}
\end{eqnarray}

Each left-moving complex fermion corresponds to a {\it global} U(1) symmetry. A complex right-moving fermion, $\psi_c^p$, is associated to a {\it local} $U(1)$ symmetry.  The massless generator of this local symmetry is produced by the simple current on the world-sheet
\begin{equation}
U_p = :\psi_c^{p*}\psi_c^{p}:\
\label{qcurrent}
\end{equation}
with normalization,
\begin{equation}
\langle U_p,U_p\rangle = 1.
\label{qcurnorm}
\end{equation}

However, not all of the fermions can always pair to form solely left-moving or right-moving complex fermions, due to possibly differing boundary conditions.  A real left-mover and a real right-mover can pair to form Ising fermions without chirality, or they can remain unpaired, forming chiral Ising fermions, both of which have boundary conditions $\alpha_r=0,1$.  Every two right-moving Ising fermions (regardless of chirality) prohibit generation of a simple current and, thus, reduce the rank of the gauge group by one.

The second object necessary to define a free fermionic model (up to vacuum expectation values of fields in the effective field theory) is an $n\times n$-dimensional matrix $\mk$ of rational numbers 
$-1<k_{i,j}\leq 1$, $i=0$, ..., $n$, that determine the GSO operators for physical states. The $k_{i,j}$ are related to the phase weights $C{{\mV_i}\choose {\mV_j}}$ appearing in a model's one-loop partition function $Z$: 
\begin{equation}
Z = \int \frac{d^{2}\tau}{[\rm{Im}( \tau) ]^2}Z_B (\tau \, \overline{\tau})\sum_{{\mbox{\boldmath ${\alpha}$\unboldmath}},\mbox{\boldmath ${\beta}$\unboldmath}}C{\mbox{\boldmath ${\alpha}$\unboldmath} \choose \mbox{\boldmath ${\beta}$\unboldmath}}Z{\mbox{\boldmath ${\alpha}$\unboldmath} \choose \mbox{\boldmath ${\beta}$\unboldmath}}
\label{partition}
\end{equation}
with phase weights,
\begin{equation}
C{{\mV_i}\choose {\mV_j}}
= (-1)^{s_i+s_j}
        {\rm exp}( \pi i k_{j,i}- \half \mV_i \cdot\mV_j), 
\label{kijpw}
\end{equation}
where $s_i$ is the 4-dimensional space-time component of $\mV_i$.  (The inner product of boundary (or charge) vectors is Lorentzian, taken as left-movers minus right-movers.  Contributions to inner products of boundary vectors, $\mV_i \cdot\mV_j$, from real fermion components are weighted by a factor of $\half$ compared to contributions for complex fermion components.)

The phase weights $C{ {\malpha}\choose {\mbeta} }$ for general sectors 
\begin{equation}
\mbox{\boldmath $\alpha$\unboldmath} = \sum_{j=0}^{n} a_j \mV_j \in \Xi , \,\,\,\,
\mbox{\boldmath $\beta$\unboldmath} = \sum_{i=0}^{n} b_i \mV_i \in \Xi 
\label{bab} 
\end{equation}
can be expressed in terms of the components in the $n\times n$-dimensional matrix $\mk$ for the basis vectors:
\begin{equation}
C{\mbox{\boldmath{$\alpha$\unboldmath}} \choose {\mbox{\boldmath {$\beta$\unboldmath}}}}
= (-1)^{s_{\malpha}+s_{\mbeta}}
        {\rm exp}\{ \pi i \sum_{i,j} b_i( k_{i,j} - \half \mV_i \cdot\mV_j) a_j \}.  
\label{kab}
\end{equation}

Modular invariance imposes constraints on the basis vectors $\mV_i$ and on the GSO projection matrix $k_{i,j}$:
\begin{eqnarray}
k_{i,j} + k_{j,i} &=& \half\, \mV_i\cdot \mV_j\, \mod{2},
\label{kseta1}\\
N_j k_{i,j} &=& 0 \, \mod{2}\,  , 
\label{kseta2}\\
k_{i,i} + k_{i,0} &=& - s_i + \fourth\, {\mV_i}\cdot {\mV_i}\, \mod{2}.
\label{kseta3}
\end{eqnarray}

The dependence upon the $k_{i,j}$ can be removed from eqns.\ (\ref{kseta1}-\ref{kseta3})
after appropriate integer multiplication, to yield three constraints on the $\mV_i$:
\begin{eqnarray}
N_{i,j} \mV_i\cdot \mV_j =  0 \, \mod{4} 
\label{ksetb1}
\end{eqnarray}

\begin{equation}
N_{i}   \mV_i\cdot \mV_i =  0 \, \mod{8}
\label{ksetb2}
\end{equation}

$N_{i,j}$ is the lowest common multiple of $N_i$ and $N_j$.  The number of simultaneous real fermions for any three $V_i$ must be even.  Thus, each basis vector must have an even number of real periodic fermions.
 
The boundary basis vectors $\mV_j$ generate the set of GSO projection operators for physical states from all sectors \boldmath $\alpha$ \unboldmath $\in \Xi$.  In a given sector, \boldmath ${\alpha}$\unboldmath, the surviving states are those that satisfy the GSO equations imposed by all $\bV_j$ and determined by the $k_{i,\malpha}$'s:
\begin{equation}
\mV_j\cdot {\mF}_{\malpha} = \left(\sum_i k_{j,i} a_i\right) + s_j
    - \half\, \mV_j\cdot {\mbox{\boldmath{$\alpha$}\unboldmath}}, \mod{2} ,
\label{gso1-a}
\end{equation}
or, equivalently,
\begin{equation}
\bV_j\cdot {\mQ}_{\balpha} = \left(\sum_i k_{j,i} a_i\right) + s_j\, \mod{2}.
\label{gso}
\end{equation}
For a given set of basis vectors, the independent GSO matrix components are $k_{0,0}$ and $k_{i,j}$, for $i>j$.  These GSO projection constraints, when combined with equations (\ref{kseta1}-\ref{kseta3}) form the free fermionic re-expression of the even, self-dual lattice modular invariance constraints for bosonic lattice models.

\section{A Non-Standard String Embedding of $E_8$}
\label{sec:anseoe8}

An algorithm to systematically and efficiently generate free fermionic heterotic string models was recently introduced \cite{fff1a, fff1b, fff1c, fff1d, fff2, fff2a} by our research group. The algorithm is being used to systematically generate the {\it complete} set of free fermionic heterotic string models with untwisted left-moving (worldsheet supersymmetric) sectors, up to continually advancing layer and order. Some interesting models have already been found. We discuss one such model here.

The standarding gauge in string models to date, is $E_8$ via an $SO(16)$ embedding, $\mbf{248} = \mbf{120} + \mbf{128}$. We explore an $SU(9)$ embedding, $\mbf{248} = \mbf{80} + \mbf{84} + \overline{\mbf{84}}$,where $80$, $84$ and $\overline{\mbf{84}}$ are the adjoint, and 2 spinor reps of $SU(9)$. This is obtained in a Layer 1, Order 6 model for which modular invariance itself dictates a gravitino sector accompany the gauge sector.

Our approach enables a {\it complete} study of all gauge group models to be generated and analyzed with extreme efficiency, up to continually increasing Layers (the number of gauge basis vectors) and Orders (the lowest positive integer $N$ that transforms, by multiplication, each basis vector back into the untwisted sector mod(2)). In this early study the models have either ${\cal{N}}=4$ or ${\cal{N}}=0$ spacetime SUSY, depending on whether the gravitino sector is or is not present, respectively.

The primary goal of this research is to systematically improve the understanding of the statistical properties and characteristics of free fermionic heterotic models, a process that is underway by a collection of research groups \cite{af0,af1,af2,af3,af4,af5,af6,af7,kd1,kd2,kd3,kd4}.              

However, as particularly interesting models appear in the course of our program, we will separately report on such models. The first of these models appears at Layer 1, Order 6 and requires a graviton sector. The intersting feature of this model is that it provides an alternative embedding of $E_8$, based not on the $E_8$ maximal subgroup $SO(16)$, but on $E_8$'s alternate maximal subgroup $SU(9)$.

\subsection{Review of $E_8$ String Models in 4 and 10 Dimension}

The $SO(16)$ realization of $E_8$ is well known: We start with the uncompactified $D=10$, ${\cal{N}}=1$ SUSY $SO(32)$ heterotic string in light-cone gauge. Free fermion construction generates this model from two basis boundary vectors: the ever-present all-periodic vector, $\bone$, and the supersymmetry generating vector $\bS$ \cite{fff1a, fff1b, fff1c, fff1d, fff2, fff2a}:
\begin{eqnarray}
\bone &=& [(1)^{8}|| (1)^{32}]\label{aps10}   \\
\bS   &=& [(1)^{8}|| (0)^{32}]\label{susy10}.
\end{eqnarray}
The $\mbf 496$ (adjoint) rep of $SO(32)$ is produced by the untwisted boundary vector $\bo = \bone+\bone$,
\begin{eqnarray}
\bo &=& [(0)^{8}|| (0)^{32}]\label{unt10}.
\end{eqnarray}

To transform the uncompactified $D=10$, ${\cal{N}}=1$ SUSY $SO(32)$ heterotic model into the $D=10$, ${\cal{N}}=1$ SUSY $E_8\otimes E_8$ model,
all that is required is the additional twisted basis boundary vector \cite{fff1a},
\beqn
\I^O =  [(0)^{8}|| (1)^{16} (0)^{16}].
\label{tsistO10}
\eeqn
The GSO projection of $\I^O$ onto $\bo$ reduces the untwisted sector gauge group to $SO(16)_O\otimes SO(16)_H$
by reducing its massless gauge states to the adjoint reps $\mbf{120}_O\otimes \mbf{1}$ + $\mbf{1}\otimes \mbf{120}_H$.
The GSO projection of $\I^O$ (or of $\bone$) on $\I^O$ results in a $\mbf{128}_O\otimes 1$ massless spinor rep of definite chirality.
Further, the GSO projection of $\bone$ onto 
\beqn
\I^{H} = \I^{O} + \bone + \bS =  [(0)^{8}|| (0)^{16} (1)^{16}],
\label{tsistH10}
\eeqn
produces a massless spinor rep $\mbf{1}\otimes \mbf{128}_H$ of $SO(16)_H$ with matching chirality.

Thus, the boundary sectors $\bo$ and $\I^O$ produce the 
 ${\mbf 248}$ (adjoint) of an observable $E_8$ via the $SO(16)$ embedding 
\beqn
\mbf{248} = \mbf{120} + \mbf{128},
\label{so16Oemb}
\eeqn
while the boundary sectors $\bo$ and $\I^H$ produce the same for a hidden sector $E_8^H$

When the $E_8\otimes E_8$ model is compactifieid down to four dimensions, without any twist applied to the compact dimensions, 
the basis vectors become,
\beqn
\bone &=& [(1)^{2}, (1,1,1)^{6}|| (1)^{44}]\label{aps4}\\
\bS   &=& [(1)^{2}, (1,0,0)^{6}|| (0)^{44}]\label{susy4}\\
\I^O  &=& [(0)^{2}, (0,0,0)^{6}|| (1)^{16} (0)^{28}].\label{tsist4}
\eeqn
Because
\beqn
\I^H  = \I^O + \bone +\bS &=& [(0)^{2}, (0,1,1)^{6} || (0)^{16} (1)^{28}],
\label{tsihst4}
\eeqn
is no longer a massless sector, the gauge group is $E_8^O\otimes SO(22)$ (with ${\cal{N}}=4$ SUSY). \newpage
An additional massless twisted sector,
\beqn
\I^{H'}  &=& [(0)^{0}, (0,0,0)^{6|}| (0)^{16}, (1)^{16}, (0)^{6}],
\label{tsihpst4}
\eeqn
is required to reclaim the second $E_8$.\footnote{In this note we we do not discuss the gauge group of the left-moving sector, since it belongs 
to the $N=4$ gravity multiplet and disappears for $N<2$.}

\subsection{$E_8$ from $SU(9)$} 

Our systematic research of free fermionic gauge models, revealed at Layer 1, Order 3 (more precisely Layer 1, Order 6 = Order(2) x Order(3))
as explained below) an intersting alternative realization of $E_8$ 
The simplest possible massless gauge sector for Order 3 is
\beqn
\I^{3}  &=& [(0)^{0}, (0,0,0)^{6|}| (\twothird)^{18}, (0)^{26}].
\label{tsi34}
\eeqn
The non-integer values in $\I^{3}$ produce a GSO projection on the untwisted sector that breaks $SO(44)$ down to
$SU(9)\otimes U(1)\otimes SO(26)$. The charges of the $SU(9)$ non-zero roots are of the form 
$\pm(1_i,-1_j)$ for $i$ and $j\ne i$ denoting one of the first 9 right-moving complex fermion. Combined with the 
zero roots of the Cartan Subalgrabra, these form the $\mbf 80$ (adjoint) rep of $SU(9)$
The $U(1)= \sum{i=1}^{9} U(1)_i$ charge is ${\rm Tr}\, Q_i$. The $SO(26)$ generators have the standard charges of
$\pm(1_r,\pm 1_s)$ with $r$ and $s\ne r$ denoting one of the last 13 right-moving complex fermion.  

However the modular invariance requirements eq.\ (\ref{ksetb1}) and eq.\ (\ref{ksetb2}) necessitate that $\I^3$ be expressed as a spacetime fermion, rather than spacetime boson. That is, the required
basis boundary vector to produce a gauge factor of $SU(9)$ in the untwisted sector in like manner to (\ref{tsi34}) is
\beqn
\I^{6}  &=& [(1)^{0}, (1,0,0)^{6|}| (\twothird)^{18}, (0)^{26}].
\label{tsi64}
\eeqn 
As an Order 6 = Order 2 x Order 3, basis boundary vector,
(\ref{tsi64}) satisfies (\ref{mi1},\ref{mi2}). $2\I^{6} = \I^{3}$ is then a massless gauge sector, as is
$4\I^{6}= -\I^{3}$. Note also that $3\I^6$ is the gravitino sector $S$. Hence $\bS$ need not, and cannot, be a 
separate basis vector.

The GSO projections of $\bone$ and $\I^{6}$ on $I^{3}$ and $-I^{3}$ yield massless gauge states from two sets of charges. Charges in the
first set have the form 
\beqn
\pm(\twothird_{i1},\twothird_{i2},\twothird_{i3},\third_{i4},\third_{i5},\third_{i6},\third_{i7},\third_{i8},\third_{i9}),
\label{set1}
\eeqn
with all subscripts different and each denoting one of the first 9 complex fermions.  
States in $I^{3}$ and $-I^{3}$ vary by their overall charge sign and form the $\mbf{84}$ and $\overline{\mbf{84}}$ reps of $SU(9)$, respectively
Thus, together the sectors $\bo$, $I^{3}$, and $-I^{3}$ contain the $\mbf{80}$, $\mbf{84}$ and $\overline{\mbf{84}}$ reps of $SU(9)$, from which
$\mbf{248} = \mbf{80} + \mbf{84} + \overline{\mbf{84}}$ emerges. Thus, here $E_8$ is obtained from its second maximal subgroup $SU(9)$. 

The second set of charges are of the form,
\beqn
\pm(\third,\third,\third,\third,\third,\third,\third,\third,\third, \pm 1_r)
\label{soe}
\label{set2}
\eeqn
with $1_r$ denoting a unit charge of one of the 13 complex fermions generating the $SO(26)$ Cartan subalgebra. 
Hence, the charges in this set are orthogonal to $E_8$, but have non-zero dot products with $U(1)= \sum_{i=1}^9 U(1)_i$ charged states, and unit dot products with the $SO(26)$ generators. Thus, this second set of states enhance $SO(26)$ to $SO(28)$.
The complete gauge group is thus $E_8\otimes SO(28)$. Since the gravitino sector is a multiple of $I^{6}$, the model has inherent ${\cal{N}}=4$ SUSY.

The whole process can be followed again with the addition of another basis boundary vector $\I^{6H}$ isomorphic with $\I^{6}$, but that has no non-zero right-moving charges in common with $\I^{6}$:
\beqn
\I^{6H}  &=& [(1)^{0}, (1,0,0)^{6|}| (0)^{18}, (\twothird)^{18}, (0)^{8}].
\label{tsi64h}
\eeqn 
$\I^{6H}$ will produce a second $E_8$ from a parallel $SU(9)$ embedding. The $SO(8)$ of the untwisted sector would be enhanced by both of the
$U(1)$'s associated with the two $SU(9)$'s to $SO(12)$, giving a standard $E_8\otimes E_8 \otimes SO(12)$ model, but with an $SU(9)\times SU(9)$ embedding for $E_8\otimes E_8$.

Heterotic models have a $SO(44)$ rotational redundancy in their charge expressions (which we are taking into account in our 
statistical analysis). In terms of solely the gauge sectors, our $E_8$ embedding from $SU(9)$ can be understood as a specific
$SO(18)\in SO(44)$ rotation of the initial charge lattice. In the $SO(16)$ basis, a set of simple roots for $E_8$ are
\beqn
E_1=(+1,-1, 0, 0, 0, 0, 0, 0) \nonumber \\
E_2=( 0,+1,-1, 0, 0, 0, 0, 0) \nonumber \\
E_3=( 0, 0,+1,-1, 0, 0, 0, 0) \nonumber \\
E_4=( 0, 0, 0,+1,-1, 0, 0, 0) \nonumber \\
E_5=( 0, 0, 0, 0,+1,-1, 0, 0) \nonumber \\
E_6=( 0, 0, 0, 0, 0,+1,-1, 0) \nonumber \\
E_7=( 0, 0, 0, 0, 0,+1,+1, 0) \nonumber \\
E_8=-(\half,\half,\half,\half,\half,\half,\half,\half),\label{e8setso16}
\eeqn
where we choose a positive chirality $\mbf{128}$ spinor. For an $SO(18)$ rotation we need 9 charge states, so we we will add
an additional zero charge onto the $E_8$ charges and include an additional $U(1)$ generator with defining charge
\beqn
U_9=( 0, 0, 0, 0, 0, 0, 0, 0, 1).\label{u1q}
\eeqn

Alternately, a simple set of roots for the $SU(9)$ basis is
\begin{eqnarray}
E_1'&=&(+1,-1, 0, 0, 0, 0, 0, 0, 0) \nonumber \\
E_2'&=&( 0,+1,-1, 0, 0, 0, 0, 0, 0) \nonumber \\
E_3'&=&( 0, 0,+1,-1, 0, 0, 0, 0, 0) \nonumber \\
E_4'&=&( 0, 0, 0,+1,-1, 0, 0, 0, 0) \nonumber \\
E_5'&=&( 0, 0, 0, 0,+1,-1, 0, 0, 0) \nonumber \\
E_6'&=&( 0, 0, 0, 0, 0,+1,-1, 0, 0) \nonumber \\
E_7'&=&( 0, 0, 0, 0, 0, 0,+1,-1, 0) \nonumber \\
E_8'&=&-(\third,\third,\third,\third,\third, -\twothird, -\twothird, -\twothird,\third).
\label{e8setsu9}
\end{eqnarray}
In the $SU(9$ basis, there is also an additonal $U(1)$ of the form
\beqn
U_9'=(\frac{1}{3}, \frac{1}{3}, \frac{1}{3}, \frac{1}{3}, \frac{1}{3}, \frac{1}{3}, \frac{1}{3}, \frac{1}{3},),\label{e8setsu9u1}
\eeqn
The SO(16) embedding of E8 can be transformed into the SU(9) embedding of E8 via a Weyl rotation that yields 
\begin{eqnarray}
E_7' &=& \frac{1}{2}(E_7-E_6)-U_9      \\
E_8' &=& \frac{2}{3}(E_8+U_9)          \\
E_9' &=& \frac{2}{3}(-E_8+\frac{1}{2}U_9)
\end{eqnarray}
It is interesting to note that this rotation between $E_8$ embeddings of maximal subgroups can be related to partition function equivalence 
involving Theta-function product identities \cite{mumford}.

\subsection{Summary} 

In this section we have presented an alternative embedding for $E_8$, involving not its maximal subgroup $SO(16)$, rather its alternate
maximal subgroup $SU(9)$. Instead of the $\mbf{248}$ (adjount) rep of $E_8$ generated as $\mbf{248} = \mbf{120} + \mbf{128}$ of $SO(16)$, we constructed a $D=4$ model in which it is generated as $\mbf{248} = \mbf{80} + \mbf{84} + \overline{\mbf{84}}$ of $SU(9)$. Interestingly, we found in this model that modular invariance requires the basis boundary vector responsibble for the pair of massless gauge sectors that yields the $\mbf{84} + \overline{\mbf{84}}$ reps to also produce the gravitino-producing sector. The model starts out with ${\cal{N}}=4$ SUSY. Thus, this alternate $E_8$ embedding cannot occur in a model without either broken or unbroken SUSY (i.e., lacking a gaugino sector).


\newpage
\vspace*{3.25in} 
 \begin{center} 
      APPENDICES
 \end{center}
 \pagestyle{plain}
  \pagebreak
\renewcommand\thesection{\Alph{chapter}.\arabic{section}}
\begin{table}[ht]
\caption{\textit{Standard Model Particles and Masses}}
\centering 
\vspace{0.2in}
\begin{tabular}{c c c} 
\hline\hline 
Particle & Mass (GeV) & Normalized Mass \\ [0.5ex] 
\hline 
$e^-$     &$0.000511$        &$5.6\time10^{-6}$\\
$\mu^-$   &$0.106$           &$1.2\time10^{-3}$\\
$\tau^-$  &$1.78$            &$2.0\time10^{-2}$\\
$\nu_e$   &$1\times 10^{-12}$&$1.0\time10^{-14}$\\
$\nu_\mu$ &$1\times 10^{-12}$&$1.0\time10^{-14}$\\
$\nu_\tau$&$1\times 10^{-12}$&$1.0\time10^{-14}$\\
$u$       &$0.003$           &$3.3\time10^{-5}$\\
$d$       &$0.006$           &$6.6\time10^{-5}$\\
$c$       &$1.3$             &$1.4\time10^{-2}$\\
$s$       &$0.1$             &$1.1\time10^{-2}$\\
$t$       &$175$             &$1.9$\\
$b$       &$4.3$             &$4.7\time10^{-2}$\\
Higgs     &$\approx 150$     &$1.64$\\[1ex] 
\hline 
\end{tabular}
\label{table:nonlin} 
\end{table}
\newpage


\bibliographystyle{ieee}
\bibliography{obousyref}



\end{document}